\documentclass[oldversion]{aa} 
\usepackage{graphicx}
\usepackage{txfonts}
\usepackage{aalongtable,lscape}
\newcommand{\gsim}{\;\lower.6ex\hbox{$\sim$}\kern-7.75pt\raise.65ex\hbox{$>$}\;}
\newcommand{\lsim}{\;\lower.6ex\hbox{$\sim$}\kern-7.75pt\raise.65ex\hbox{$<$}\;}

\begin{document}
\title{Properties of stellar generations in Globular Clusters and 
relations with global parameters
\thanks{Based on observations collected at ESO telescopes under 
programmes 072.-D0507 and 073.D-0211}
}

\subtitle{}

\authorrunning{E. Carretta et al.}
\titlerunning{Properties of stellar generations in GCs}

\author{
E. Carretta\inst{1} \and
A. Bragaglia\inst{1} \and
R.G. Gratton\inst{2} \and
A. Recio-Blanco\inst{3} \and
S. Lucatello\inst{2,4} \and
V. D'Orazi\inst{2} \and
S. Cassisi\inst{5}
}

\institute{
INAF-Osservatorio Astronomico di Bologna, via Ranzani 1, I-40127
 Bologna, Italy
\and
INAF-Osservatorio Astronomico di Padova, vicolo dell'Osservatorio 5, I-35122
 Padova, Italy
\and
Laboratoire Cassiop\'ee UMR 6202, Universit\`e de Nice Sophia-Antipolis, CNRS,
Observatoire de la Cote d'Azur, BP 4229, 06304 Nice Cedex 4, France
\and
Excellence Cluster Universe, Technische Universit\"at M\"unchen, Boltzmannstr. 2,
D-85748, Garching, Germany
\and
INAF-Osservatorio Astronomico di Collurania, via M. Maggini, I-64100 Teramo,
Italy
  }

\offprints{E. Carretta, eugenio.carretta@oabo.inaf.it}

\date{Received .....; accepted .....}

\abstract{
We revise the scenario of the formation of Galactic globular clusters (GCs) by
adding the observed detailed chemical composition of their different stellar
generations to the set of their global parameters.  We exploit the unprecedented
set of homogeneous abundances of more than 1200 red giants in 19 clusters, as
well as additional data from literature, to give a new definition of {\it bona
fide} globular clusters, as the stellar aggregates showing the presence of the
Na-O anticorrelation. We propose a classification of GCs according to their
kinematics and location in the Galaxy in three populations: disk/bulge, inner
halo, and outer halo. We find that the luminosity function of globular clusters
is fairly independent of their population, suggesting that it is imprinted by
the formation mechanism, and only marginally affected by the ensuing evolution.
We show that a large fraction of the primordial population should  have been
lost by the proto-GCs. The extremely low Al abundances found for the primordial 
population of massive GCs indicate a very fast enrichment  process before the
formation of the primordial population. We suggest a scenario for the formation
of globular clusters including at least three main phases: i) the formation of a
precursor population (likely due to the interaction of cosmological structures
similar to those that led to the formation of dwarf spheroidals, but residing at
smaller galactocentric distances, with the early Galaxy or with other
structures), ii) which triggers a large episode of star formation (the
primordial population), and iii) the formation of the current GC, mainly within
a cooling flow formed by the slow winds of a fraction of the primordial
population.  The precursor population is very effective in raising the metal
content in massive and/or metal-poor (mainly halo) clusters, while its r\^ole is
minor in small and/or metal rich (mainly disk) ones. Finally, we use Principal
Component Analysis and multivariate relations to study the phase of
metal-enrichment from primordial to second generation. We conclude that most of
the chemical signatures of GCs may be ascribed to a few parameters, the most
important being metallicity, mass, and age of  the cluster. Location within the
Galaxy (as described by the kinematics) also plays some  r\^ole, while
additional parameters are required to describe their dynamical status.}
\keywords{Stars: abundances -- Stars: atmospheres --
Stars: Population II -- Galaxy: globular clusters -- Galaxy: globular clusters:
individual: NGC~104 (47 Tuc), NGC~288, NGC~1904 (M 79), NGC~2808, NGC~3201, 
NGC~4590 (M 68), NGC~5904 (M 5), NGC~6121 (M 4), NGC~6171 (M 107), NGC~6218 (M
12), NGC~6254 (M 10), NGC~6388, NGC~6397, NGC~6441, NGC~6752, NGC~6809 (M 55),
NGC~6838 (M~71), NGC~7078 (M 15), NGC~7099 (M 30)} 

\maketitle

\section{Introduction}\label{intro}

\begin{figure*}
\centering
\includegraphics[bb=20 160 570 600, clip, scale=0.9]{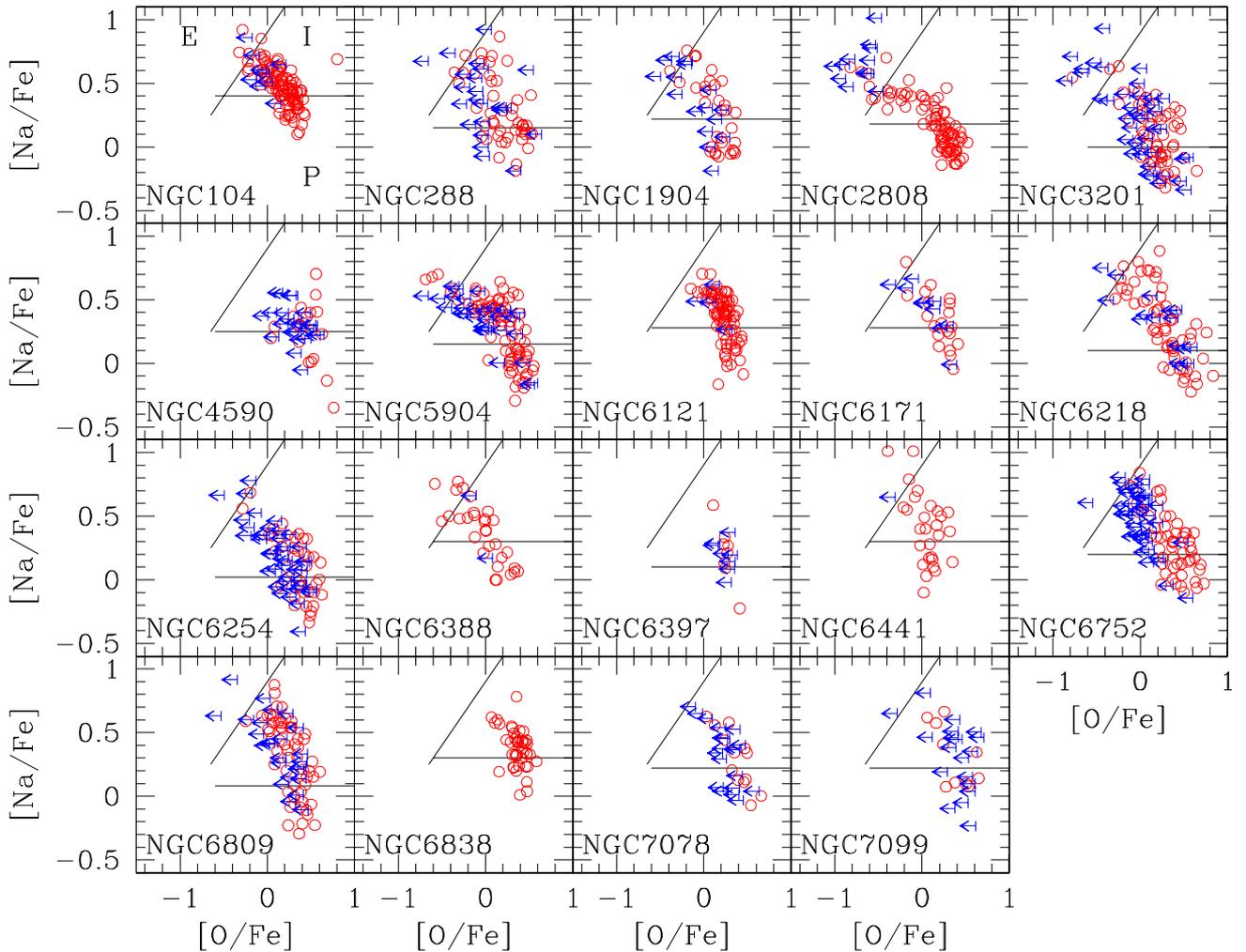}
\caption{Summary of the Na-O anticorrelation observed in the 19 globular
clusters of our sample. Arrows indicate upper limits in oxygen abundances. The 
two lines in each panel separate the Primordial component (located in the 
Na-poor/O-rich region), the Na-rich/O-poor Extreme component, and  the
Intermediate component in-between (called P, E, and I, respectively as
indicated only in the first panel). See Sect. 2 for details.}
\label{f:tutteanti}
\end{figure*}

The assembly of the early stellar populations in galaxies is one of the hottest 
open issues in astronomy. Globular Clusters (GCs) are a major component of these 
old stellar populations; they are easily detectable and can be studied in some
detail even at large distances, providing a potentially powerful link between 
external galaxies and local stellar populations. A clear comprehension of the 
mechanisms that led to the formation and evolution of GCs and
of the relations existing between GCs and field stars, is
a basic requirement to understand how galaxies assemble (see e.g. Bekki et al.
2008). Various authors proposed scenarios for the formation of GCs (Peebles 
\& Dicke 1968; Searle \& Zinn 1978;  Fall \& Rees 1985; Cayrel 1986;
Freeman 1990;
Brown et al. 1991, 1995; Ashman \& Zepf 1992; Murray \& Lin 1992; Bromm \& 
Clarke 2002; Kravtsov \& Gnedin 2005; Saitoh et al. 2006; Bekki \& Chiba 2002,
2007; Bekki et al. 2007; Hasegawa et al. 2009; Marcolini et al. 2009; 
Hartwick 2009). While very suggestive and intriguing,  these scenarios either
do not reproduce in a completely convincing way the whole  spectrum of
observations, or are likely incomplete, describing only part of the sequence of
events that lead to GC formation or only a subset of them. We still  lack the
clear understanding we would need. However, some recent progresses are  opening
new promising perspectives.

Since almost forty years ago we know that large star-to-star abundance variations 
for several light elements are present in GCs (see Gratton et al. 2004 for a 
recent review). Regarded for long time as intriguing abundance
``anomalies" restricted to some cluster stars, the observed peculiar chemical
composition only recently was explicitly understood as an universal phenomenon in
GCs, likely related to their very same nature/origin (Carretta 2006; Carretta et
al. 2006, Paper I). The observational pattern of Li, C, N, O, Na, Al, Mg in cluster
stars is currently well assessed (see e.g. the review by Gratton et al. 2004), 
thanks to several important milestones:
\begin{itemize}
\item [(i)] variations for the heavier species (O, Na, Mg, Al) are restricted 
to the denser cluster environment. The signature for other elements (Li, 
C, N) may be reproduced by assuming a mixture of primordial composition plus 
evolutionary changes. The latter are due to two mixing episodes, occurring at the end of the 
main-sequence (the first dredge-up) and after the bump on the Red Giant Branch 
(RGB), both in low mass Population II field stars and in their cluster analogues
(Charbonnel et al. 1998; Gratton et al. 2000b; Smith and Martell 2003).
\item[(ii)] the observed pattern of abundance variations is established in 
proton-capture reactions of the CNO, NeNa and MgAl chains during H-burning at 
high temperature (Denisenkov and Denisenkova 1989; Langer et al. 1993); 
\item [(iii)] the variations are found also among unevolved stars currently on 
the main-sequence (MS) of GCs (Gratton et al. 2001; Ramirez and Cohen 2002; 
Carretta et al. 2004; D'Orazi et al. 2010). This unequivocally implies that
this composition {\em has  been imprinted in the gas by a previous generation
of stars}. The necessity of this conclusion stems from the fact that
low-mass MS stars are unable to reach the high temperatures for the
nucleosynthetic chains required to  produce the observed inter-relations between
the elements (in particular the  Mg-Al anticorrelation). This calls for a class
of now extinct stars, more massive  than the low-mass ones presently evolving in
GCs, as the site for the  nucleosynthesis.
\end{itemize}

Unfortunately, we do not know yet what kind of stars produced the pollution. The 
most popular candidates are either intermediate-mass AGB stars (e.g., D'Antona 
\& Ventura 2007) or very massive, rotating stars (FRMS, e.g., Decressin et al. 
2007)\footnote{The strong objection made by Renzini (2008) on the out-flowing
of  matter from FRMS being unable to result in clearly separated  MS with
different - and {\it discrete} - He content, still applies. However, up to now
the only clear cases of several discrete MSs are the very peculiar 
\object{$\omega$ Cen} and NGC~2808. Indications  for widening of the MS have
been obtained for other clusters, like NGC~104 (Anderson et al. 2009) and 
NGC~6752,  where Milone et al. (2009b) see also  hint of a split.  On the other
hand, Renzini (2008) restricts his favourite candidate polluters,  AGB stars, to
those experiencing only a few  episodes of third dredge-up. This might appear
too much specific and at odds with  observed abundances of $s-$process elements
in some GCs.}. 

The observed abundance variations are also connected to the helium abundance, 
since He is the main outcome of H-burning (i.e., Na-rich, O-poor stars should 
also be He-rich). However,  the relation between He abundance variations and
the light element abundance pattern may be quite complicate. Multiple 
main-sequences, attributed to populations with 
different He fraction Y, have been recently found in some GCs ($\omega$ Cen, 
NGC~2808, see Bedin et al. 2004 and Piotto et al. 2007, respectively). We
have found a clear indication that Na-rich and Na-poor stars in
NGC~6218 and NGC~6752 have slightly different RGB-bump luminosities (Carretta et al.
2007b, hereinafter Paper III), as expected from models of cluster
sub-populations with different He content (Salaris et al. 2006).
In separate papers (Gratton et al. 2010; Bragaglia et al. 2010) we 
examine in more detail the relation between He and light element abundance
variations from evidence based on horizontal branch (HB) and RGB.

In summary, GCs are not exactly a Simple Stellar Population: they must harbour 
at least {\em two stellar generations}, as explained above, clearly  distinct by 
their chemistry. These populations may be separated, provided data of adequate 
quality are available. The patterns of anticorrelated Na-O, Mg-Al and, partly, 
C-N, Li-Na (and associated correlations) must be regarded as the fingerprints of 
these different sub-populations, and may be used to get insights on the early 
phases of formation and evolution of GCs, which are still obscure. The time 
scale for the release of matter processed by H-burning at high temperature is 
of the order of $10^7$~yr if it comes from FMRS, 
and a few times longer if it comes from massive AGB stars. Thus, whatever the 
candidate producers, the observed patterns were certainly already in place 
within some $10^8$~yr after the start of cluster formation. These processes 
occurred on time-scales less than 1\% of the typical total age of a GC. The 
dynamical evolution that occurred in the remaining 99\% of the 
cluster lifetime, while likely important, did not completely erase these 
fingerprints. Their fossil record is still recognisable in the chemical
composition of the low mass stars.
 
To decipher the relevant information we need large and homogeneous
data sets, like the one we
recently gathered (Carretta et al. 2009a,b). The goal of the present paper is to
exploit this wealth of data to discuss the abundance patterns of the
different populations within each GC. We will correlate them with global
cluster parameters, such as the HB morphology, and structural
or orbital parameters.  This will allow to better understand which are the main
properties of the stellar  populations of GCs, hence to get insights into
the early phases of their  evolution. Using this information as a guide, we will
sketch a quite simple scenario  for the formation of GCs, which is essentially
an updated and expanded version of  what proposed thirty years ago by Searle \&
Zinn (1978). This scenario naturally  explains the relation between GCs and
other small systems (dwarf Spheroidals), and  suggests a connection between GCs
and field stars. In fact we propose that the  primordial population of GCs might
be the main building block of the halo, although  other components are likely
present.

The present paper is organised as follows. 
In Sect. \ref{ricorda} we give a brief summary of our previous work, to set the
stage for the following discussion.
In Sect. \ref{sele} we recall some general properties of the GC population
and the division in subpopulations; we also 
present the selection criteria for our sample, discussing possible biases,
and the parameters used in the analysis.
In Sect. \ref{1gen} we discuss the
properties of the first stellar generation, 
and we present a scenario for GC formation. In Sect. \ref{2gen}
we consider the second phase of chemical enrichment in GCs, comparing the
properties of the second generation with those of the primordial one and
presenting  a number of interesting correlations. Finally, in Sect. \ref{fin}, we
discuss in a more general way the correlations with global GC parameters and
give a summary and  conclusions.  In the Appendix, we present a new
classification of all Galactic GCs, dividing them into disk/bulge, inner halo,
and outer halo ones on a kinematical basis;  we also list their metallicities
on the scale defined in Carretta et al. (2009c), their ages, re-determined
from literature using these metallicities, and a compilation of [$\alpha$/Fe]
values, that are used throughout the paper.

\section{Synopsis of previous results}\label{ricorda}

We give here  a summary of results from our 
project ``Na-O anticorrelation and HB" (Carretta et al. 2006)
propaedeutic to the present discussion.

Up to a few years ago, obtaining adequate high resolution spectroscopic
data sets was painstaking, since
stars had to be observed one-by-one. Thanks to the efforts of many researchers, 
mainly of the Lick-Texas group, spectra of some 200 stars in a dozen GCs were 
gathered using tens of nights over several years (see the reviews by Kraft
1994, Sneden 2000 and references therein). In the last few years, we used the 
spectacular data collecting capability offered by the FLAMES multi-object 
spectrograph at the ESO VLT to secure spectra for more than 1400 giant stars,
distributed over about 12\% of all known GCs. With the increase of
an order of magnitude in available data, the paradigm has  changed. We now
understand that the observed anticorrelations are
not indicative of  ``anomalies",  rather we are dealing with
{\em the normal chemical evolution of GCs}. 

Our survey has already been amply described elsewhere. Results for the first
five GCs have been presented in a series of papers (Papers I through VI:
Carretta et al. 2006, 2007a,b,c; Gratton et al. 2006, 2007),  while the 
remaining 
clusters are analysed in Carretta et al. (2009a,b: Paper VII and VIII). 
In Fig.~\ref{f:tutteanti} we show a collage of the Na-O 
anticorrelations observed in all 19 clusters in our sample. Solid lines
separate  the Primordial, Intermediate and Extreme populations, whose concept is
introduced and defined in Paper VII and recalled briefly below.

Very shortly,  we obtained GIRAFFE spectra (at R$\simeq$20000, comprising the Na
{\sc i} 568.2-568.8 nm, 615.4-6.0 nm and [O {\sc i}] 630 nm lines) of about  100
stars per cluster.  At the same time, we also
collected UVES spectra (at R$\simeq$40000, covering the  480-680 nm
region, and providing information about Mg, Al, and Si, in addition  to O and
Na) of about 10 stars (on average) per cluster. We homogeneously  determined the
atmospheric parameters for these stars using visual and near-IR  photometry and
the relations in Alonso et al. (1999, 2001). We measured Fe, O,  and Na
abundances for more than 2000 stars (more than 1200 cluster members with  $both$
O and Na detected), putting together the largest sample of this kind ever 
collected. 

This large statistics, both in the number of clusters and in stars per cluster, 
allowed us to recognise that the amount of the abundance variations among different 
clusters is related in a not trivial way to global cluster parameters (Carretta 
2006; Carretta et al. 2007a; 2009a, Paper VII: GIRAFFE data; 2009b, Paper VIII: 
UVES data). The GCs are dominated by the second (polluted) generation of stars, 
the fraction of primordial stars being roughly correlated with cluster 
luminosity. The Na-O anticorrelation has not only different extension, but also 
different $shape$, in different clusters, depending on cluster luminosity and 
metallicity. The Mg-Al anticorrelation is sometimes absent, this occurring in 
low-luminosity clusters. All these are clear indications that the polluters' 
properties change from cluster to cluster and that this change is apparently 
driven by the cluster luminosity and metallicity.

Carretta (2006) suggested to use the Interquartile Range (IQR, the difference
between the upper quartile and the lower quartile, see e.g. Tukey 1977) of the
[O/Na] ratio as a quantitative  measure of the extension of the Na-O
anticorrelation. The IQR is useful because it is less influenced by extreme
values, it refers to the range of the middle 50\% of the values and it is less
subject to sampling fluctuations in highly skewed distributions.
Statistically robust IQR  values require large enough samples of stars. Our
project was designed  to obtain Na and O abundances for a large number of RGB
stars in each  cluster, typically 100 stars per cluster,  although in some cases
only a much smaller sample of stars could be used. The number of  stars actually
measured in each cluster depends on the richness of population,  metallicity,
$S/N$ ratio, and in some cases on field stars contamination (like for  the bulge
clusters NGC~6388 and NGC~6441, or the disk clusters  NGC~6171 and NGC~6838).
The smallest sample (16 stars with both Na  and O) is for NGC~6397, the largest
(115 stars) is for 47~Tuc  (NGC~104).

In Paper VII we defined three population components in each cluster: the first 
generation stars, and two groups of second generation stars.  The lines 
separating the three components are shown in Fig.~\ref{f:tutteanti}. The 
primordial P component includes stars between the minimum 
[Na/Fe] observed and [Na/Fe]$_{min}+4\sigma$ (see Paper VII). These stars are 
defined as first generation objects since they show the same pattern of high O 
and low Na typical of galactic field stars of similar metallicity, with the 
characteristic signature of core-collapse SNe.  Since the yields of Na are
metallicity-dependent (e.g.,Wheeler et al. 1989),  the limit for the P 
component varies as a function of [Fe/H], as evident in Fig.~\ref{f:tutteanti}. 
The separation between the two sub-components of second generation stars (the 
intermediate I stars and the extreme E stars) is somewhat more arbitrary.
On the basis of the [O/Na] distributions
in our clusters, they were defined in Paper VII as those stars with [O/Na] ratios larger or 
smaller than -0.9 dex, respectively. 

Computing the fraction of stars in each component, we found that:
\begin{itemize}
\item[(i)] the Extreme 2nd generation is not present in all GCs;
\item[(ii)] the Intermediate 2nd generation constitutes the bulk (50-70\%) of 
stars in a GC;
\item[(iii)] the Primordial population is present in all GCs (at about the 30\% level).
\end{itemize}

\begin{figure}
\centering
\includegraphics[scale=0.45]{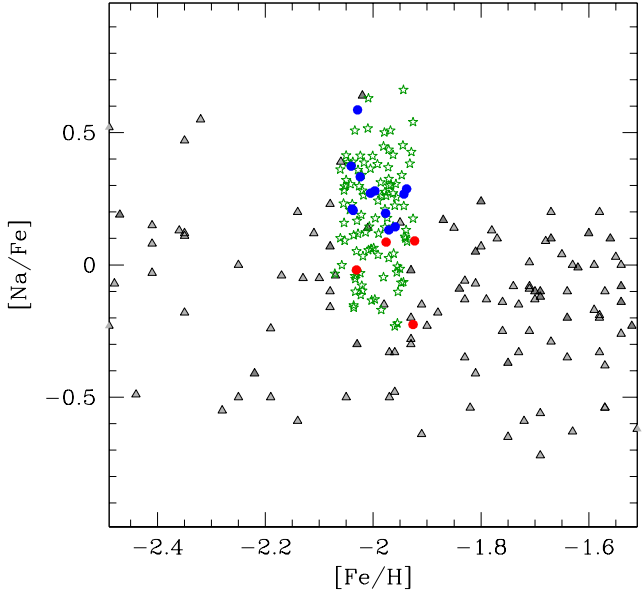}
\caption{ [Na/Fe] ratios as a function of the metallicity [Fe/H] in a range
centred on the average metal abundance of NGC~6397. Filled triangles in
grey-tones are field stars from Gratton et al. (2003a) and the compilation by
Venn et al. (2004). Filled circles are stars in NGC~6397 with determinations of
both O and Na (red: P component, blue: I component). Empty (green) star symbols
are stars in NGC~6397 with only Na abundances derived. }
\label{f:nafe6397}
\end{figure}

D'Antona and Caloi (2008) used the HB morphology to derive the fraction of stars
in two generations in GCs.  They made a claim that some clusters (including
NGC~6397) could presently host exclusively second generation stars, having
completely lost the first generation ones. However, our  abundance analysis, in
comparison with field stars, does not support that hypothesis. In
Fig.~\ref{f:nafe6397} we show the distribution of [Na/Fe] in field stars with
metallicity centred on the mean [Fe/H] value for NGC~6397. Filled circles are
our RGB stars in NGC~6397 with both Na and O measurements (Paper VII and Paper
VIII), separated in P and I components using our definition. To be very
conservative, in Paper VII we attributed to each of the three populations only
stars with both elements measured.  However, the separation between first and
second generation stars (between P and I components) only requires the knowledge
of Na abundances. In Fig.~\ref{f:nafe6397} the much more numerous stars in
NGC~6397 with Na detections are plotted as open star symbols. A good fraction of
them fall in the region populated by the Primordial component and by normal
field halo stars, unpolluted by ejecta processed in H-burning. Hence, we confirm
that {\em in all GCs analysed a primordial component of first generation stars
is still observable at present}.\footnote{While writing this paper, Lind et al.
(2009) published the analysis of an extended set of unevolved stars in NGC~6397
and we completed the analysis of Turn-Off stars in NGC~104 (D'Orazi et al. 2010),
where Na abundances show similar variations.}

\begin{figure}
\centering
\includegraphics[scale=0.45]{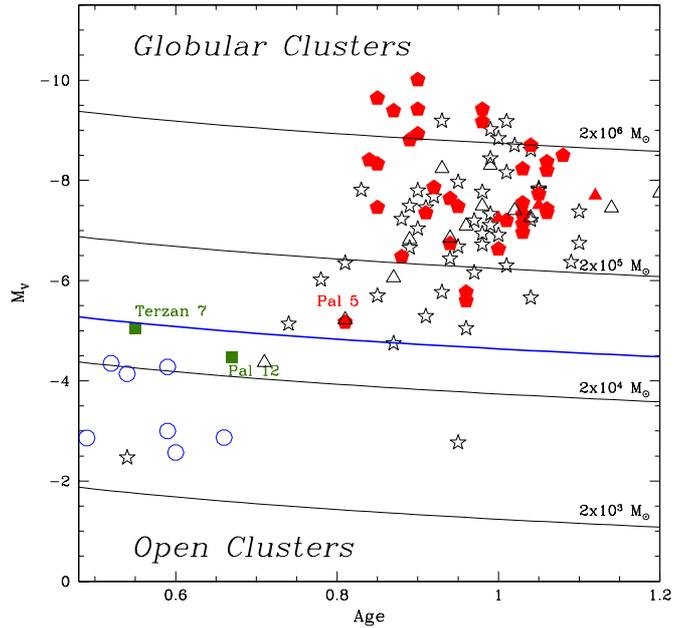}
\caption{Relative Age parameter vs absolute magnitude $M_V$ for globular and 
old open clusters (see Appendix for details). Red filled pentagons and
triangles  are GCs where Na-O anticorrelation has been observed, in the Milky
Way or the  LMC respectively; green squares are clusters which do not show
evidence  for O-Na anticorrelation, both members of Sagittarius dSph, either
of the main body (Terzan 7) or the stream 
(Pal 12). Open stars and triangles mark clusters for which not
enough data is available, in the Milky Way or the LMC respectively. Finally,
open circles are old open  clusters (data from Lata et al. 2002). Superimposed
are lines of constant mass (light solid lines, see Bellazzini et al. 2008a). The
heavy blue solid line  (at a mass of $4\times 10^4~M_\odot$) is the proposed
separation between globular and open clusters. }
\label{f:AgeMv}
\end{figure}

\section{Clusters and parameters selection}\label{sele}

\subsection{Definition of globular clusters}\label{define}

In this paper we intend to bring together the chemical properties derived
from our  in-depth study of various stellar populations in GCs with many other
global  observables.  Moreover, we present a scenario for the formation of  GCs
in the more general context of the relationship between the Milky Way and its
satellites. Thus, before entering into the  discussion, we recall a few
properties of the parent population of GCs that are  of direct interest here. 

The first point we would like to make concerns the operative 
definition of {\it bona fide} GC.
The distinction between globular and other clusters (e.g., open 
clusters) is not well drawn, and it is ambiguous in particular for
the populous clusters that are numerous in the Magellanic  Clouds. To 
better clarify this point, we plot in Figure~\ref{f:AgeMv} ages and absolute 
magnitudes for the clusters listed in the database by Harris (1996, and web 
updates). This list includes 146 GCs. Seven of these GCs are actually thought to 
be members of the Sagittarius dwarf galaxy (see van den Bergh \& Mackey 2004, 
and references therein). To this sample, we add GCs in other satellites of the 
Milky Way: 16 GCs in the Large magellanic Cloud (LMC), 8 in the Small Magellanic Cloud (SMC) and 5 in the Fornax dwarf 
Spheroidal,  dSph). Data for all these GCs are detailed in the Appendix. Age 
data are actually available for slightly more than half of the sample. We also 
add for a few old open clusters (NGC~188, NGC~6791, Collinder~261, 
NGC~1193, Berkeley~31, and Berkeley~39: data from Lata et al.  2002), that fall within the 
limits of the plot. 
From this figure, there is a clear
overlap between open  clusters and objects from the list of GCs at the faint end
of the sequence. We used filled red pentagons for those GCs where the
Na-O anticorrelation  has been found. One of the filled green squares indicates 
Terzan 7, where no spread in O abundances has been found in the (only) seven stars observed by
Sbordone et al. (2007),  and may then be the most massive cluster observed
insofar without the Na-O  anticorrelation. The other one is Pal 12, where
Cohen (2004) finds very  uniform O and Na abundances, but only
on four stars. Finally, open symbols are  for all remaining GCs, for which current
data are not adequate to state if a  Na-O anticorrelation exists or not. Similar
data are scarce for open clusters  (see Gratton 2007). However, de Silva et al. (2009) compiled 
data  for various old open clusters, finding no evidence for a Na-O anticorrelation,  and Martell \& Smith (2009) did not find
any evidence for CN variation among giants  in three open clusters (including
NGC~188). This diagram indicates that the Na-O and  related anticorrelations have
been observed in all old clusters with $M_V<-5.1$\ (which  roughly corresponds to a
mass of $\sim 4\times 10^4~M_\odot$\ for old populations),  including the vast
majority of galactic GCs, and almost all the objects with a  relative age
parameter $>0.8$. {\em We then propose to identify the GCs with those  clusters where
there is a Na-O anticorrelation}.  As we will see in Sect. \ref{proto},  this
identification corresponds to a formation scenario which clearly separates GCs 
from other clusters. Operatively, we might also define GCs either as the old 
clusters (age larger than 5 Gyr) with a $M_V<-5.1$, or those with relative age 
parameter $>0.8$. These definitions essentially include the same list
of  objects, at least in the Milky Way and its satellites. 

At the other mass limit for the GC population,  the similarity between GCs and nuclei of dwarf galaxies has been pointed out  by
many authors (see e.g. Freeman 1990; B\"oker 2008; Georgiev et al. 2009). Those 
nuclei or nuclear star clusters of dwarf galaxies that  can
be studied in good detail (like M~54 for the Sagittarius galaxy) essentially 
share the whole pattern of properties with GCs (see e.g. Bellazzini et al.
2008b; Georgiev et al. 2009, Carretta et al. 2010),  although they may have large spreads in Fe abundances, not
observed in  GCs. This occurrence suggests that also $\omega$~Cen was in the
past (in) the  nucleus of a galaxy.

\subsection{Our sample of GCs}\label{sample}

Ideally, we should have derived detailed chemical data for the whole parent
population. However, this would have required too much observing time
and we analysed only a representative subset of clusters (representing about 12\% of the total
sample). The selection procedure  was as  follows. We started from the whole
sample of galactic GCs, as listed by  Harris (1996). We then divided clusters in
different groups, according to the morphology of the HB. For each group, we
selected the two-four rich ($M_V<-5$) clusters, accessible from Paranal
($\delta<+20\degr$), with the smallest apparent distance modulus; however we did not
consider some clusters that have quite large differential reddening (like M~22:
Ivans et al. 2004\footnote{A chemical analysis similar to our has been
performed  in M~22 by Marino et al.  (2009). This data, kindly given to us
before publication, nicely fit in our relations. However, we do not include it
in the present analysis because it is not strictly homogeneous.}). 
The selected clusters were: Red HB clusters: \object{NGC 104}=47~Tuc, 
\object{NGC 6838}=M~71, \object{NGC 6171}=M~107; Oosterhoff I clusters: \object{NGC 6121}=M~4,
\object{NGC 3201}, \object{NGC 5904}=M~5; Blue Horizontal
Branch clusters: \object{NGC 6752}, \object{NGC 6218}=M~12,
\object{NGC 6254}=M~10, \object{NGC 288}, \object{NGC 1904}=M~79;
 clusters with blue, short HB's: \object{NGC 6397},
\object{NGC 6809}=M~55; Oosterhoff II
clusters: \object{NGC 7099}=M~30, \object{NGC 4590}=M~68,
\object{NGC 7078}=M~15; Clusters with very extended/bimodal
distribution of stars on the HB: \object{NGC 2808}, \object{NGC 6441}, \object{NGC 6388}.  Hence,
within each different class of HB morphology, the sample is essentially
distance limited. On the other hand, this is not true for the whole sample, 
because the adopted limits depend on the morphological classes and reddening (so
that clusters projected close to the Galactic plane are under-represented).
However, for most classes of HB's, the limit is quite uniform at about
(m-M)$_V<14.5-15.5$, that is $\sim 10$~kpc from the Sun. We needed to sample a
larger volume ((m-M)$_V<16.5$) to include GCs  with very extended/bimodal
distribution of stars on the HB, since these clusters  are rare. Due to these
choices, GCs with very extended blue HB are  over-represented in our sample
(42\% of the total).

\subsection{The Galactic GC sample}\label{mwgc}
To correctly explore the relations between the chemistry of different
stellar generations and global GC properties, it is important to assess to
what cluster population our programme GCs belong.
Zinn (1985) demonstrated that Milky Way GCs can be divided into two main groups:
disk (or bulge) GCs, and halo GCs. This separation was done according to the
metal abundance only (with the limit  at [Fe/H]$=-0.8$ dex). These two
groups correspond to the main peaks of the metallicity distribution of GCs, but
they can also be clearly distinguished from other properties (location in the
Galaxy, kinematics, etc.). According to  Searle \& Zinn (1978), halo GCs result
from the evolution of individual fragments,  while disk clusters likely formed
within the dissipational collapse. Hence this distinction is likely to play an
important r\^ole in defining the characteristics of GCs, and in particular of
their primordial population. 

Further refinements (e.g., van den Bergh \& Mackey 2004, Lee et al. 2007) 
along  Zinn's line of thought were done by using the HB morphology which,
however, is  one of the features of GCs we intend to explain.  In the following 
therefore, we will adopt a combination of location in the Galaxy and kinematics 
criteria to separate disk clusters from the halo ones\footnote{Similarly, 
Pritzl et al. (2005) adopted kinematics to assign GCs to various Galactic 
components; however they were able to do so only for 29 of the 45 GCs they
studied.}. 
Full details are given in the Appendix. Briefly, using the  Harris (1996) catalogue, we have 
first classified as 
outer halo GCs the ones currently located at distances greater than 15 kpc 
(Carollo et al. 2008)  from the Galactic  centre; clusters with Galactocentric 
distance less than 3.5 kpc were instead  considered as bulge GCs. To separate 
the inner halo clusters from the disk ones,   we used the 
rotational velocity around the Galactic centre by  Dinescu et al. (1999) and 
Casetti-Dinescu et al. (2007) whenever possible. When this datum was not available, we  used the 
differences between the observed radial velocity (corrected to  the LSR) and the 
one expected from the Galactic rotation curve (see Clemens 1985).
In the Appendix we provide the  disk/inner halo/outer halo classification 
for each cluster listed in the Harris catalogue. Finally, we consider GCs in the 
LMC, SMC, and in dSph's 
(Sagittarius and  Fornax) as separate groups.

\begin{figure}
\centering
\includegraphics[scale=0.43]{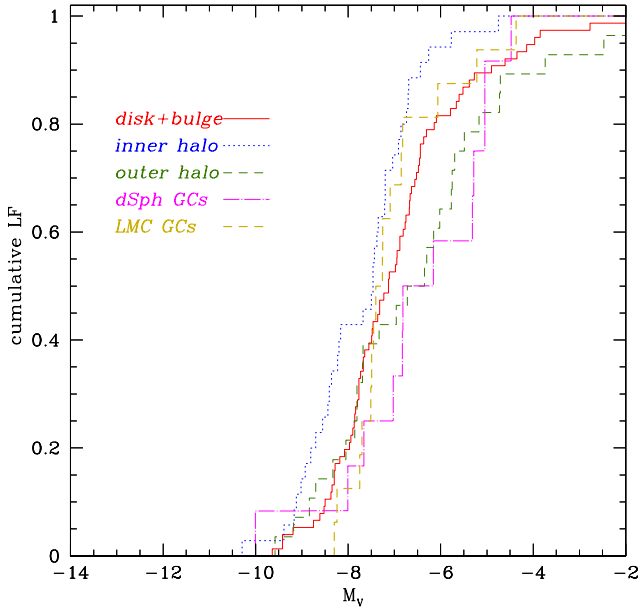}
\caption{Cumulative luminosity functions of different groups of  Galactic GCs
(from the Harris 1996 catalogue) according to our present classification
criteria, and of GCs in dwarf spheroidals (from van den Bergh and Mackey 2004;
see Appendix for references on Fornax an LMC clusters).  The red solid line
indicates disk/bulge clusters, the blue dotted line the  inner halo clusters,
the dashed green line clusters in the outer halo,  magenta dashed-dotted line
is for clusters in dSphs, and the green-gold long dashed line for GCs in LMC.}
\label{f:lfnew}
\end{figure}

The procedure to select the programme sample, described in Sect. \ref{sample}, 
results in a potential 
selection bias as a function of the distance. This shows up in a correlation 
between cluster present-day mass (as represented by the proxy of cluster total 
absolute visual magnitude, $M_V$) and distance modulus: in our sample more 
massive GCs are typically the most distant ones. This correlation is at odds 
with the total sample of  GCs in the Harris (1996) catalogue. However, since all 
programme GCs but one (NGC~1904, with $R_{GC}=18.8$ kpc) are within 15 kpc from 
the Galactic centre they belong either to the disk or to the inner halo. 
Within this sub-sample, there is a correlation of $M_V$ with 
Galactocentric distance similar to that noticed in our sample. Hence, we will 
assume that our sample is representative of the properties of the disk and inner 
halo (but not of the outer halo) GCs, and will neglect the possible bias with 
luminosity.

Other properties of the parent population of Galactic GCs relevant for
our discussion are the masses (luminosities) and the metallicities of the disk and inner 
halo GCs. These are two of the main parameters driving most of the observed properties
of GCs, as we will confirm later, and are worth a few more words. 

Inner and outer halo clusters have clearly distinct luminosity functions (LF), as
illustrated by Fig.~\ref{f:lfnew}. Small clusters (M$_{\rm V}>$-6) only exist in
the outer halo, where they make up half of the total. A Kolmogorov-Smirnov test
returns a 1\% probability that inner and outer halo LF were
extracted from the same population. Of course, this difference can be at least
in part attributed to the destruction mechanisms, which are more efficient for
clusters closer to the centre of the Galaxy. However, other mechanisms can also
be considered. In fact, all young clusters (Age parameter $<0.8$) reside in the
outer halo, and they are all faint ($M_V>-6.5$), overlapping open clusters  in
the $M_V$/age distribution (Fig.\ref{f:AgeMv}).  If we limit  to
old clusters  (Age parameter $>0.82$, essentially adopting the same definition
of  GC considered in Sect. \ref{define}), there is no clear difference between the
LF of inner halo or disk and that of outer halo clusters
(Kolmogorov-Smirnov  tests for the distribution of the GCs with age parameter
available return  significance of 16 and 32\%, respectively). 

Also the comparison between the LFs of disk and inner halo GCs is  worth more
attention. Again, we naively expected that destruction mechanisms should be 
more effective for disk clusters than for the inner halo ones. In  this case the
LF for inner halo clusters should have a fraction of small mass clusters 
intermediate between those observed in the outer halo and in the disk. However, 
while only 19\% (4\%) of the inner halo GCs have $M_V>-7$ ($M_V>-6$), this 
percentage is 41\% (15\%) for disk clusters. There are very few inner halo 
counterparts of the very frequent small disk clusters like M~71 and NGC~6397. 
Since  such clusters are more easily destroyed in the disk than in the inner
halo, this  suggests a different original mass distribution between the disk and
the inner halo (see also Fraix-Burnet et al. 2009,  who attempted a
multi-parametric classification of GCs, different from ours and leading to
different  conclusions about the properties of the different cluster
populations, see the Appendix  for further details). 

All this suggests
that the main  difference between disk, inner and outer halo clusters might be
related to their  formation (absence of young, small clusters in the inner halo)
more than to the  destruction efficiency, which is however very important for
small clusters.  This goes against a diffuse opinion, i.e., that we are now
seeing only those GCs which occupied the survival zone of parameters; however,
the notion that GCs can be formed only in a limited range of parameters is not
new, see for instance Caputo \& Castellani (1984).

It is also interesting to note that the luminosity function of the outer halo 
GCs is similar to that of GCs in dSph's (see Fig.\ref{f:lfnew}): a
Kolmogorov-Smirnov test gives a chance (73\%) that they were drawn
from the same parent population. This might depend on the fact that clusters
in the outer halo and dSph shared similar environments at birth. 

\begin{figure}
\centering
\includegraphics[scale=0.43]{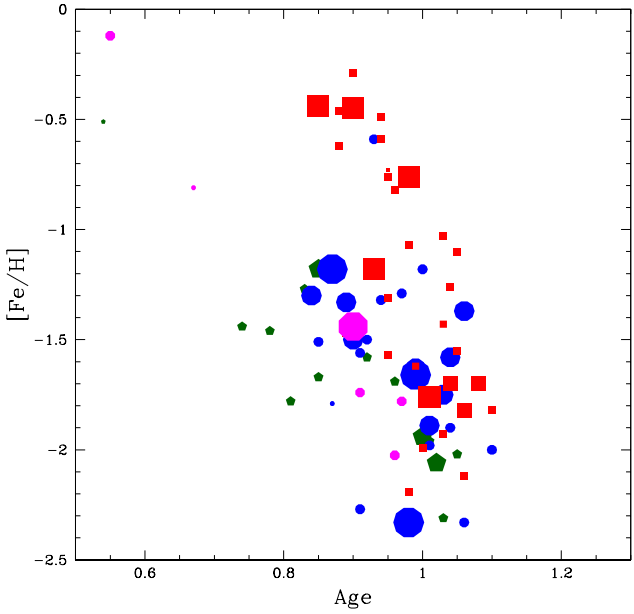}
\caption{Age-metallicity relation for different groups of globular clusters:
outer halo clusters (green pentagons), inner halo clusters (blue circles),
disk/bulge clusters (red squares). Magenta circles are GCs associated to dSphs.
Different symbol sizes are used for clusters of different luminosity.
}
\label{f:agefeh}
\end{figure}

Disk and halo GCs differ also in other important characteristics. Obviously,
the inner halo/outer halo GCs are on average more metal poor than
the disk ones (see Appendix).  
Furthermore, they seem to obey to different age metallicity relations:
metallicity increased slower in the inner halo than in the disk, and even
slower in the outer halo, see Fig.~\ref{f:agefeh}.   The age estimates were
obtained as described in the Appendix. Practically all disk/bulge GCs with
[Fe/H]$<-1$ are very old, while most of the inner halo GCs of intermediate
metallicity ($-2<$[Fe/H]$<-1$) have relative ages in the range 0.8-0.9, i.e.,
they are about 2 Gyr younger than disk GCs with the same metallicity. If the
age/metallicity calibration is correct, after 2 Gyr from
the Big Bang, the central region of the Milky Way was enriched to [Fe/H]$\sim
-0.4$ (and [$\alpha$/H]$\sim 0$), while the inner halo metallicity was still
very low\footnote{Note that this does not mean that the pace of evolution was
uniformly slower in the halo than in the disk. It is indeed possible that star
formation (and chemical evolution) in the halo actually occurred in bursts separated by
long quiescent phases, while it was characterised by prolonged phases at a
relatively low level in the disk. This might lead to the paradoxical situation
that stars in the halo have a chemical composition more appropriate to 
faster star formation than those in the disk, although the former might be
actually younger. We will come back to this point in the next Section.}. 

\begin{figure}
\centering
\includegraphics[scale=0.43]{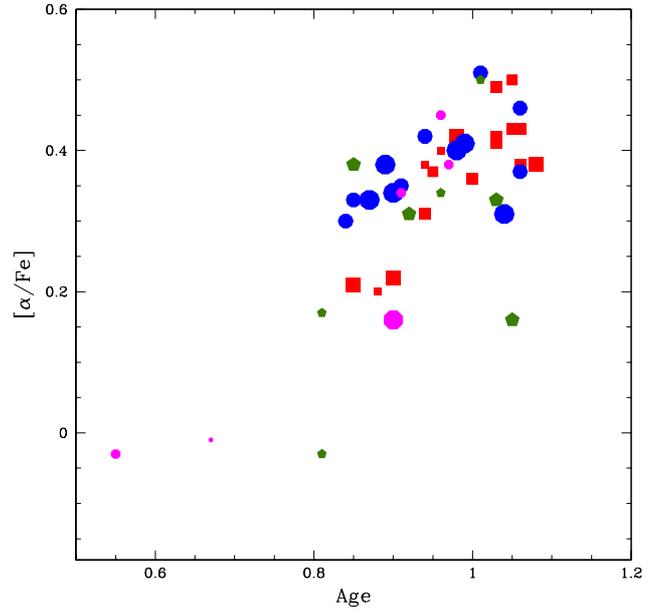}
\caption{Relation between age and excess of $\alpha-$elements for different
groups  of globular clusters (the meaning of symbols is as in
Fig.~\ref{f:agefeh}). [$\alpha$/Fe] values come from our compilation of
literature data (see Appendix). Different symbol sizes are used for clusters of
different luminosity. }
\label{f:alfaage}
\end{figure}

Finally, disk and inner halo GCs may differ in their element-to-element 
abundance ratios, as suggested by the analysis by Lee \& Carney (2002).
In part, this can be attributed to an age effect (see Fig.~\ref{f:alfaage});
however, age cannot be the only explanation. This is shown by the close 
comparison between M~4 and M~5 (NGC~6121 and NGC~5904), 
performed by Ivans et al. (2001), which is 
fully confirmed by our analysis (Carretta et al. 2009b). These GCs are 
both inner halo clusters according to our classification, although M~5 has 
much more extreme kinematics, and it is also likely younger by more than 
1~Gyr. M~5 has a smaller excess of $\alpha-$elements, and seems to be 
also deficient in nuclei produced by $s-$process nucleosynthesys (Ivans et al. 
2001; Yong et al. 2008). If confirmed, these two facts might at first look 
contradictory, since a lower excess of $\alpha-$elements is usually 
attributed to a prolonged star formation, allowing significant contribution 
by type Ia SNe. However, such a long phase of star formation should also 
allow the contribution by the intermediate and small mass AGB stars, which 
efficiently produce $s-$elements. In the next sections, we will re-examine 
this point using our extensive database, and find a solution to this conundrum.

\subsection{Cluster Parameters considered in the analysis}\label{para}

Our choice was driven by the aim of sampling the full parameter
space of GCs and to derive relations between the properties of  different
stellar generations in GCs and global parameters.

Tab.~\ref{t:tablit} lists the 19 GCs in our programme set and gives  structural
and orbital parameters taken from literature. We considered the following
parameters, mostly from Harris (1996):
\begin{itemize}
\item the apparent visual distance modulus, $(m-M)_V$
\item the reddening, E(B-V)
\item the Galactocentric distance, R$_{ GC}$
\item the total absolute visual magnitude, M$_V$
\item the HB ratio, HBR, that is the fraction of blue and red HB stars over the
total, as $(B-R)/(B+V+R)$
\item the metallicity, [Fe/H] from our Paper VIII
\item the cluster ellipticity, $ell$
\item the concentration, $c$
\item the tidal radius, $r_t$ (in pc, from Mackey and van den Bergh 2005)
\item the half light radius, $r_h$ (in pc, from Mackey and van den Bergh 2005)
\end{itemize}

We considered the following parameters related to the Galactic orbit
(from Dinescu et al. 1999;  Casetti-Dinescu et al. 2007):
\begin{itemize}
\item the total energy of orbit, E$_{tot}$
\item the period of the Galactic orbit, P
\item the apogalactic distance, R$_{apo}$
\item the perigalactic distance, R$_{per}$
\item the maximum distance from the plan, z$_{\rm max}$
\item the eccentricity of the orbit, $ecc$
\item the inclination angle of the orbit, $\psi$ 
\item the rotational velocity, $\Theta$
\end{itemize}
These orbital parameters are mean values averaged over a large number of 
orbits. Hence, they represent a way to gain knowledge on the birthplace
of clusters and on the conditions existing at their formation epoch, and
where most of their lifetime is spent in the Galaxy. Instantaneous quantities, 
such as the present Galactocentric distance of a GC, may be much less 
informative. Unfortunately, no orbital 
parameters have been determined yet for the two most massive clusters in our 
sample, NGC~6388 and NGC~6441.

 In addition, we considered  the relative age parameter, mostly
re-determined from Marin-Franch et al.
(2009) and De Angeli et al. (2005) as described in the Appendix.

\begin{table*}[bt]
\centering
\footnotesize 
\caption{Properties of the 19 GCs in our sample.}
\begin{tabular}{rlccrrrrcrrrr}
\hline
 NGC &other &(m-M)$_V^1$ &$E(B-V)^1$&R$_{GC}^1$ &Z$^1$~~ &M$_{V,tot}^1$ &HBR$^1$ &[Fe/H]$^1$   
 &ell$^1$~~ &$c^1$~~ &$r_h^1$~~ &$r_t^1$ \\  
\hline       
104 &47Tuc &13.32 &0.04& 7.4 &-3.2 &-9.42 &-0.99 &-0.768 &0.09 &2.03 &2.79 &42.86 \\
288 &	   &14.64 &0.03&12.0 &-8.8 &-6.74 & 0.98 &-1.305 &     &0.96 &2.22 &12.94 \\
1904&M79   &15.53 &0.01&18.8 &-6.3 &-7.86 & 0.89 &-1.579 &0.01 &1.72 &0.80 & 8.34 \\
2808&	   &15.59 &0.22&11.1 &-1.9 &-9.39 &-0.49 &-1.151 &0.12 &1.77 &0.76 &15.55 \\
3201&	   &14.17 &0.23& 8.9 & 0.8 &-7.46 & 0.08 &-1.512 &0.12 &1.30 &2.68 &28.45 \\
4590&M68   &15.14 &0.05&10.1 & 6.0 &-7.35 & 0.17 &-2.265 &0.05 &1.64 &1.55 &30.34 \\
5904&M5    &14.41 &0.03& 6.2 & 5.4 &-8.81 & 0.31 &-1.340 &0.14 &1.83 &2.11 &28.40 \\
6121&M4    &12.78 &0.36& 5.9 & 0.6 &-7.20 &-0.06 &-1.168 &0.00 &1.59 &3.65 &32.49 \\
6171&M107  &15.01 &0.33& 3.3 & 2.5 &-7.13 &-0.73 &-1.033 &0.02 &1.51 &2.70 &17.44 \\
6218&M12   &13.97 &0.19& 4.5 & 2.2 &-7.32 & 0.97 &-1.330 &0.04 &1.39 &2.16 &17.60 \\
6254&M10   &14.03 &0.28& 4.6 & 1.7 &-7.48 & 0.98 &-1.575 &0.00 &1.40 &1.81 &21.48 \\
6388&	   &16.49 &0.37& 3.2 &-1.2 &-9.42 &-0.65 &-0.441 &0.01 &1.70 &0.67 & 6.21 \\
6397&	   &12.31 &0.18& 6.0 &-0.5 &-6.63 & 0.98 &-1.988 &0.07 &2.50 &2.33 &15.81 \\
6441&	   &16.33 &0.47& 3.9 &-1.0 &-9.64 &-0.76 &-0.430 &0.02 &1.85 &0.64 & 8.00 \\
6752&	   &13.08 &0.04& 5.2 &-1.7 &-7.73 & 1.00 &-1.555 &0.04 &2.50 &2.34 &55.34 \\
6809&M55   &13.82 &0.08& 3.9 &-2.1 &-7.55 & 0.87 &-1.934 &0.02 &0.76 &2.89 &16.28 \\
6838&M71   &13.70 &0.25& 6.7 &-0.3 &-5.60 &-1.00 &-0.832 &0.00 &1.15 &1.65 & 8.96 \\
7078&M15   &15.31 &0.10&10.4 &-4.7 &-9.17 & 0.67 &-2.320 &0.05 &2.50 &1.06 &21.50 \\
7099&M30   &14.57 &0.03& 7.1 &-5.9 &-7.43 & 0.89 &-2.344 &0.01 &2.50 &1.15 &18.34 \\
\hline
NGC &other & E$_{tot}^2$~~ &P$^2$~~~~  
 &R$_{apo}^2$ &R$_{per}^2$ &z$_{\rm max}^2$ &ecc$^2$ &~~$\psi^2$ & $\Theta$
 &age$^3$\\ 
\hline
104 &47Tuc& -872 & 190 & 7.3 & 5.2 & 3.1 &0.17 &  29 & 161 & 0.95 \\ 
288 &	  & -787 & 224 &11.2 & 1.7 & 5.8 &0.74 &  44 & -27 & 0.90 \\ 
1904&M79  & -526 & 388 &19.9 & 4.2 & 6.2 &0.65 &  28 &  83 & 0.89 \\ 
2808&	  & -770 & 240 &12.3 & 2.6 & 3.8 &0.65 &  18 &  74 & 0.83 \\ 
3201&	  & -430 & 461 &22.1 & 9.0 & 5.1 &0.42 &  18 &-301 & 0.82 \\ 
4590&M68  & -396 & 504 &24.4 & 8.6 & 9.1 &0.48 &  30 & 300 & 0.94 \\ 
5904&M5   & -289 & 722 &35.4 & 2.5 &18.3 &0.87 &  33 & 115 & 0.85 \\ 
6121&M4   &-1121 & 116 & 5.9 & 0.6 & 1.5 &0.80 &  23 &  24 & 0.97 \\ 
6171&M107 &-1198 &  87 & 3.5 & 2.3 & 2.1 &0.21 &  44 & 151 & 0.99 \\ 
6218&M12  &-1063 & 125 & 5.3 & 2.6 & 2.3 &0.34 &  33 & 130 & 0.99 \\ 
6254&M10  &-1053 & 128 & 4.9 & 3.4 & 2.4 &0.19 &  33 & 149 & 0.92 \\  
6388&	  &	 &     &     &     &	 &     &     &     & 0.87 \\ 
6397&	  &-1017 & 143 & 6.3 & 3.1 & 1.5 &0.34 &  18 & 133 & 0.99 \\ 
6441&	  &	 &     &     &     &	 &     &     &     & 0.83 \\ 
6752&	  & -977 & 156 & 5.6 & 4.8 & 1.6 &0.08 &  18 & 199 & 1.02 \\ 
6809&M55  &-1038 & 122 & 5.8 & 1.9 & 3.7 &0.51 &  56 &  55 & 1.02 \\ 
6838&M71  & -957 & 165 & 6.7 & 4.8 & 0.3 &0.17 &   3 & 180 & 0.94 \\ 
7078&M15  & -752 & 242 &10.3 & 5.4 & 4.9 &0.32 &  36 & 128 & 1.01 \\ 
7099&M30  & -937 & 159 & 6.9 & 3.0 & 4.4 &0.39 &  52 &-104 & 1.08 \\ 
\hline 
\end{tabular}
\label{t:tablit}
\begin{list}{}{}
\item[1-] global parameters, from Harris (1996), except [Fe/H], which is from
UVES spectra (Paper VIII) and the HBR for NGC~6388, NGC~6441 calculated from
Busso et al. (2007)
\item[2-] orbital parameters, from Dinescu et al. (1999), Casetti-Dinescu et al.
(2007).  Units are: $10^2 km^2 sec^{-2}$ (E$_{tot}$), 10$^6$ yr (P), kpc
(R$_{apo}$, R$_{per}$, z$_{\rm max}$), degrees ($\psi$), km~s$^{-1}$ ($\Theta$) 
\item[3-] relative age (see Appendix).
\end{list}
\end{table*}

\begin{table*}
\centering
\caption{Quantities for the same GCs, derived by our works}
\begin{tabular}{rccrrrcccccc}
\hline
 NGC & log T$_{\rm eff}^{\rm max}$ & IQR[Na/O]$^1$ &  P$^2$ &  I$^2$ &  E$^2$ &$\langle [\alpha/{\rm Fe}] \rangle^3$ & [(Mg+Al+Si) &[Mg/Fe]$_{max}^6$&[Mg/Fe]$_{min}^6$&[Si/Fe]$_{min}^6$& [Si/Fe]$_{max}^6$\\
& (HB)            &&& && &                                                                                                 /Fe]$^6$                                     &               &               &                \\
\hline         
104   &3.756$^4$ &0.472 &  27 &  69 & 4 &0.42 &0.46 &0.60&   +0.47& 0.35&   0.43\\
288   &4.221$^5$ &0.776 &  33 &  61 & 6 &0.42 &0.41 &0.55&   +0.41& 0.30&   0.41\\
1904  &4.352$^4$ &0.759 &  40 &  50 &10 &0.31 &0.31 &0.40&   +0.16& 0.25&   0.34\\
2808  &4.568$^4$ &0.999 &  50 &  32 &18 &0.33 &0.29 &0.42& $-$0.30& 0.22&   0.38\\
3201  &4.079$^4$ &0.634 &  35 &  56 & 9 &0.33 &0.32 &0.45&   +0.27& 0.25&   0.41\\
4590  &4.041$^4$ &0.372 &  40 &  60 & 0 &0.35 &0.40 &0.48&   +0.28& 0.30&   0.48\\
5904  &4.176$^4$ &0.741 &  27 &  66 & 7 &0.38 &0.36 &0.55&   +0.31& 0.21&   0.39\\
6121  &3.968$^5$ &0.373 &  30 &  70 & 0 &0.51 &0.54 &0.65&   +0.50& 0.45&   0.64\\
6171  &3.875$^4$ &0.522 &  33 &  60 & 7 &0.49 &0.53 &0.60&   +0.46& 0.45&   0.63\\
6218  &4.217$^4$ &0.863 &  24 &  73 & 3 &0.41 &0.43 &0.60&   +0.46& 0.22&   0.43\\
6254  &4.400$^5$ &0.565 &  38 &  60 & 2 &0.37 &0.37 &0.58&   +0.33& 0.18&   0.36\\ 
6388  &4.255$^4$ &0.795 &  41 &  41 &19 &0.22 &0.30 &0.40&   +0.16& 0.20&   0.46\\
6397  &3.978$^4$ &0.274 &  25 &  75 & 0 &0.36 &     &0.55&   +0.40& 0.25&   0.43\\
6441  &4.230$^4$ &0.660 &  38 &  48 &14 &0.21 &0.34 &0.45&   +0.20& 0.15&   0.45\\
6752  &4.471$^5$ &0.772 &  27 &  71 & 2 &0.43 &0.44 &0.60&   +0.36& 0.28&   0.49\\
6809  &4.153$^5$ &0.725 &  20 &  77 & 2 &0.42 &0.43 &0.60&   +0.18& 0.30&   0.51\\
6838  &3.763$^4$ &0.257 &  28 &  72 & 0 &0.40 &0.44 &0.60&   +0.39& 0.30&   0.51\\
7078  &4.477$^4$ &0.501 &  39 &  61 & 0 &0.40 &0.46 &0.68& $-$0.01& 0.28&   0.60\\
7099  &4.079$^4$ &0.607 &  41 &  55 & 3 &0.37 &0.44 &0.60&   +0.44& 0.20&   0.45\\
\hline 
\label{t:tabnoi}
\end{tabular}
\begin{list}{}{}
\item[1-] IQR([Na/O]) comprises stars with GIRAFFE and UVES data
\item[2-] P,I,E are fraction of stars of Primordial, Intermediate, 
Extreme populations from Paper VII
\item[3-] [$\alpha$/Fe] is the average of [Mg/Fe]$_{\rm max}$, [Si/Fe]$_{\rm min}$ and [Ca/Fe]
\item[4-] log T$_{\rm eff}^{\rm max}$(HB) taken from Recio-Blanco et al. (2006)
\item[5-] log T$_{\rm eff}^{\rm max}$(HB) derived in the present work (see Section 5.2)
\item[6-] from Paper VIII
\end{list}
\end{table*}
 
In Tab.~\ref{t:tabnoi} we report a few of the parameters derived by our works,
related to the chemistry of first and second generation stars in GCs and their
link with primordial abundances existing at the epoch of their formation.
Other parameters derived, but not listed, in Paper VII and VIII, are also given
in this Table.

Among these, we considered parameters related to the chemistry of first 
generation stars:
\begin{itemize}
\item the maximum O abundance, [O/Fe]$_{\rm max}$
\item the minimum Na abundance, [Na/Fe]$_{\rm min}$
\item the maximum Mg abundance, [Mg/Fe]$_{\rm max}$
\item the minimum Al abundance, [Al/Fe]$_{\rm min}$
\item the minimum Si abundance, [Si/Fe]$_{\rm min}$
\item the total Mg+Al+Si content, where the average is done in number, not in logarithm,
[(Mg+Al+Si)/Fe]
\item the overabundance of $\alpha-$elements, [$\alpha$/Fe], as given
by the average of [Mg/Fe]$_{\rm max}$, [Si/Fe]$_{\rm min}$ and [Ca/Fe] (see
Sect. \ref{alfa} for an explanation of the choice).
\end{itemize}
Coupled with [Fe/H], these parameters essentially describe the starting 
composition of the proto-GCs. The elements here considered are mainly produced
by core collapse SNe, with some contribution by thermonuclear SNe for what
concerns Fe (and marginally Si). The sum Mg+Al+Si essentially
describes the primordial abundance of elements with $24<{\rm A}<28$ for two
reasons. First, this quantity does not differ between different stellar
generations in a cluster. As an example, in NGC~2808, the average value of the ratio
[(Mg+Al+Si)/Fe] is $+0.29\pm0.01$ dex (rms=0.03 dex) for the 9 stars in the P, I 
components and  $+0.28\pm0.02$ dex (rms=0.04 dex) for the three stars with sub-solar Mg values, belonging to the E
component (see Paper VIII, Al was measured only on UVES spectra). 
Second, the only way to get significant
modifications of this primordial ratio is to produce  the dominant species
$^{24}$Mg or $^{28}$Si from SN nucleosynthesis.
Hence, in the following we
will adopt this ratio essentially as another indicator of the $\alpha-$element
level in a cluster.\footnote{Of course Al is not an $\alpha-$element. However, in the primordial
populations, Al abundance is always negligible with respect to that of Mg and Si;
hence, for these stars the sum of Mg+Al+Si is essentially the sum of Mg+Si.
Within the GC, when some stars are very rich in Al, this comes from p-captures on
$^{24}$Mg, $^{25}$Mg, $^{26}$Mg. This Al results then from material originally
produced as $\alpha-$rich, and the total of Mg+Al+Si is conserved throughout
these reactions. For this reason, we may use  this sum as an indicator of the
abundance of the $\alpha-$elements.}.

Parameters related to the internal chemical evolution within the clusters are:
\begin{itemize}
\item the minimum O abundance, [O/Fe]$_{\rm min}$
\item the maximum Na abundance, [Na/Fe]$_{\rm max}$
\item the minimum Mg abundance, [Mg/Fe]$_{\rm min}$
\item the maximum Al abundance, [Al/Fe]$_{\rm max}$
\item the maximum Si abundance, [Si/Fe]$_{\rm max}$
\item the relative fraction of stars in Primordial (P), Intermediate (I),
and Extreme (E) groups
\item the interquartile range of the [O/Na] ratio, IQR[O/Na]
\end{itemize}
Minimum and maximum abundances of different elements were estimated as 
discussed in Papers VII and VIII, using a dilution model that reproduces the 
run of the observed Na-O, Mg-Al and Mg-Si anticorrelations in GCs. 

Finally, to  explore the connection between chemical patterns of light elements,
He  abundances, and HB morphology we considered  the maximum temperature $\log
T_{\rm eff}^{\rm max}(HB)$ reached on the blue tail of the HB (taken by
Recio-Blanco et al. 2006 or computed by us for programme clusters  not listed in
that study).

\section{First generation stars, primordial abundances, and scenarios for
cluster formation}\label{1gen}

The chemical pattern in GC first generation stars is strictly related to the
pre-enrichment established in the precursors of GCs, an issue for which we only
have, at the very best, indirect evidence. In this section we discuss what
evidence can be obtained from our data on the scenario of formation of GCs.

\subsection{The masses of proto-GCs and the relation between the
primordial population of GCs and the field}\label{proto}

The scenario we are devising assumes that practically $all$ globular 
clusters started their evolution as large cosmological fragments.
To put cluster formation in a broader context, we try to establish the  order
of magnitude of the mass involved in cluster formation, and discuss the possible
link between GCs and field stars. Using different lines of thought,
several authors (Larson 1987;  Suntzeff \&
Kraft 1996; Decressin et al. 2008; D'Ercole et al. 2008) suggested that
present-day GCs are only a fraction (likely small) of the original  structures
where they originated. Large amounts of mass should be lost by  proto-clusters
during the early phases of formation (a few $10^7$~yr), mainly  for two reasons.
First, the efficiency of transformation of gas into stars is unlikely to be 
larger than 50\%, and it is more likely between 20 to 40\% (Parmentier et  al. 
2008). The interaction  with 
the high velocity winds from massive stars and by their SN explosions
expels the residual gas from the cluster; 
probably ram pressure contributes to the loss. Second, massive stars lose a large fraction of their 
mass before they become collapsed remnants. Several tens per cent of the 
initial mass of the cluster  may be lost by these stars, depending on the 
stellar Initial Mass Function (IMF).
 
Due to this huge mass loss, the clusters experience a violent relaxation 
(Lynden-Bell 1967), with a considerable expansion - beyond the tidal radius -
and  ensuing loss of stars. As shown by Baumgardt et al. (2008), the gas loss
may destroy  as much as 95\% of the clusters; this is a basic difficulty in
forming bound star  clusters. Only clusters with very large mass and initial
concentration may  survive. Clusters with a relatively flat stellar mass 
spectrum would
be disrupted  by this mass loss (Chernoff \& Weinberg 1990). A bell-shaped
cluster mass function, not too dissimilar from the observed one, can be
reproduced by a proper tuning of parameters (efficiency of star formation,
initial central concentration, original mass distribution, initial stellar mass
function: see e.g.,  Parmentier \& Gilmore 2007; Kroupa \& Boily 2002). However,
given the uncertainties  existing in these parameters, the exact fraction of
primordial mass lost by  the proto-GCs is not well determined.

\begin{table*}
\centering
\caption{Original/final mass of GCs required to produce the observed
ratios between number of stars in first and second generations.}
\begin{tabular}{lcccccc}
\hline
Prim/2nd gen&	IMF		& M$_{\rm min}$&M$_{\rm max}$&Prim/2nd gen&	Original/Current&	Field/GCs\\
Current		&	Slope	&		 &		&Original &&\\
(1)                       & (2)                   & (3)           &  (4)        & (5)         & (6) & (7) \\
\hline
\multicolumn{7}{c}{Massive AGB Stars (HBB)}\\
\hline
0.5			&	1.35	& 4		& 8	 	&11.1	&	~7.4	&	6.4 \\
0.5			&	1.85	& 4		& 8	 	&~9.7	&	~6.5	&	5.5 \\
0.5			&	2.35	& 4		& 8		&15.8	&	10.6	&	9.6 \\
0.5			&	MS		& 4		& 8	 &~7.3	&	~4.9	&	3.9 \\
\hline
\multicolumn{7}{c}{Fast Rotating Massive Stars}\\
\hline
0.5			&	1.35	& 12	&50	 	& 1.8	& 1.2	&	0.2 \\
0.5			&	1.85	& 12	&50	 	& 3.3	&	 2.2	&	1.2 \\
0.5			&	2.35	& 12	&50	 	&11.2	&	 7.5	&	6.5 \\
0.5			&	MS		& 12	&50	 &~3.0	&	 2.3	&	1.3 \\
\hline
\label{t:tabratio}
\end{tabular}
\begin{list}{}{}
\item[(1) ] Current ratio between primordial and second generation stars
\item[(2) ] Slope $\alpha$ of the IMF ($N(M)=M^{-\alpha}$; Salpeter (1955) IMF has 
$\alpha=2.35$; MS means that the IMF by Miller \& Scalo (1979) is adopted.
\item[(3) ] Minimum polluter mass (in $M_\odot$).
\item[(4) ] Maximum polluter mass (in $M_\odot$).
\item[(5) ] Original ratio between primordial and second generation stars.
\item[(6) ] Ratio between the number of low-mass stars in the proto-cluster and in 
the current GC.
\item[(7) ] Current ratio of field stars (=primordial population lost by the cluster)
and of GC stars. This is the value that can be compared with observational data.
\end{list}

\end{table*}

On the other hand, it is currently well assessed that the second generation
stars (that  presently make up some 2/3 of the stars of a typical GC, see Paper
VII) should have formed from the ejecta of only a fraction of the first
generation, primordial stars (Prantzos  \& Charbonnel 2006). In order to explain
the present GC mass, we should then  assume that (i) the clusters originally had
a much larger number of stars  of the primordial generation than we currently
observe; and ii) that they  selectively lost most of their primordial
population, while retaining most  of the second generation stars. D'Ercole et
al. (2008) presented a viable  hydrodynamical scenario that meets both these
requirements. In this scenario, a cooling flow channels the material, ejected 
as low velocity winds from massive AGB stars of the first generation, to the
centre of  the potential well.  The first generation stars were at the epoch
expanding due to the violent relaxation caused by the mechanisms  cited above.
Given their very different kinematics, first and second generation  stars are
lost by the cluster at very different rates (at least in the early phases), 
leaving a kinematically cool, compact cluster dominated by second generation 
stars. This selective star loss may continue until two-body relaxation
redistributes  energy among stars: this takes a few relaxation times, that is
some  $10^8-10^9$~yr in  typical GCs. After that, the effect could even be
reversed if He-rich second  generation stars are less massive than first
generation ones (see D'Ercole et al. 2008; Decressin et al. 2008). 

We may roughly estimate the initial mass of the primordial population needed
to provide enough mass for the second generation by the following procedure:
\begin{itemize}
\item [(i)] We assume an IMF for both the first and second generation. For simplicity, 
we assume that the two populations have the same IMF. We considered both 
power-law (like the Salpeter 1955 one) and the Miller \& Scalo (1979, MS) IMF's. 
As often done,
in the first case we integrated the IMF over the range 
0.2-50~$M_\odot$\footnote{ Had we integrated the IMF over the range
0.1-50~$M_\odot$, which clearly leads to an overestimate of the fraction of
small mass stars (see Chabrier 2003), the values in  Columns (5), (6), and (7)
of Tab.~\ref{t:tabratio} should be increased by  $\sim 50$\%.}, 
while in the second case we considered the range 0.1-100~$M_\odot$.
\item [(ii)] We also assume an initial-final mass relation. In practice, we assumed a
linear relation, with final mass ranging from 0.54 to 1.24~$M_\odot$, over the
mass range from 0.9 to 8~$M_\odot$ (Ferrario et al. 2005). A second linear
relation with final mass ranging from 1.4 to 5~$M_\odot$ was assumed
for the mass range from 8 to 100~$M_\odot$. We note that the
latter relation is not critical, since
massive stars lose most of their mass.
\item [(iii)] We assume that the second generation is made of the ejecta of stars
in the mass range between $M_{\rm min}$\ and $M_{\rm max}$. The adopted ranges
were 4-8~$M_\odot$ for the massive AGB scenario, and 12-50~$M_\odot$ for the
FRMS. Note that second generation stars likely result
from a dilution of these ejecta with some material with the original cluster
composition. A typical value for this dilution is that half of the material 
from which second generation stars formed was polluted, and half has the
original  composition. The origin of this diluting material is likely to be
pristine gas (not included into primordial stars, see Prantzos and Charbonnel
2006).
\item [(iv)] We finally assume that none of the second generation stars is lost, while
a large fraction of the primordial generation stars evaporate from the clusters.
Of course, this is a schematic representation.  
\end{itemize}

With these assumptions, the original population ratio between primordial and
second generation stars in GCs depends on the assumed IMF, as detailed in 
Tab.~\ref{t:tabratio}. 
To reproduce the observed ratio between primordial and second generation stars
(33\%/66\%=0.5), the original cluster population should have been much larger
than the current value. If the polluters were massive AGB stars, larger by
roughly an order of magnitude if a Salpeter (1955) IMF is adopted (the same
result was obtained by Prantzos \& Charbonnel 2006), and by a smaller value
(about 7) if   the Miller \& Scalo (1979) IMF  is adopted. If this is correct,
we may expect to find in the field many stars coming  from this primordial
population. According to Tab.~\ref{t:tabratio} and in the extreme hypothesis
that all field stars formed in the same episode that led to the formation  of
present-day GCs, the ratio between field and GC stars ranges between 4 and 10 
if the polluters were massive AGB stars. It may be lower  by more than a factor
of 2 if the polluters were instead FRMS. Of course,  this is likely an
underestimate, since we neglected various factors: also second  generation stars
may be lost by GCs; small mass stars are selectively lost; a  significant
fraction - even the majority - of field stars may have formed in smaller 
episodes of star formation. 

We conclude that during the early epochs of dynamical evolution, a proto-GCs should
have  lost $\sim 90$\% of its primordial stellar population. A GC a few
$10^7$~yrs old should  have then appeared as a compact cluster immersed in a
much larger loose association of stars and an even more extended expanding cloud
of gas; objects with such  characteristics have been observed in galaxies with
very active star formation (see e.g. Vinko et al. 2009).

\begin{figure}
\centering
\includegraphics[bb=30 160 560 680, clip,scale=0.45]{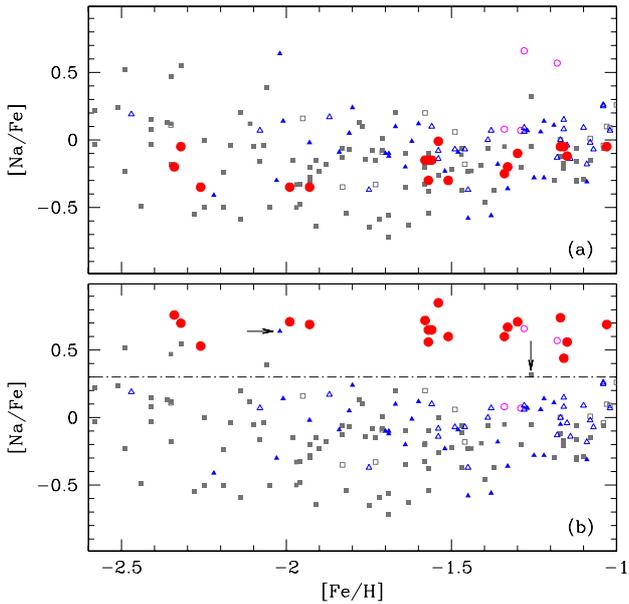}
\caption{Comparison of [Na/Fe] values between field and GC stars  
as a function of metallicity. In both panels the filled 
red circles are for our sample of GCs, indicating [Na/Fe]$_{\rm min}$ in panel
(a)  and  [Na/Fe]$_{\rm max}$ in panel (b), with the dot-dashed line at
[Na/Fe]=0.3 (see text). [Na/Fe] ratios for field stars are the same in both
panels and are taken 
from Fulbright et al. (2007: magenta open circles, bulge stars), Gratton et al.
(2003a: blue filled triangles for accreted and open triangles for dissipation
components, respectively), and Venn et al. (2004: grey filled squares for halo
and open squares for thick disk stars, respectively). Small arrows in panel (b)
indicate the two field stars that seem genuine second generation stars evaporated
from GCs (see text)Ä.}
\label{f:napop}
\end{figure}

Observational constraints to the ratio between primordial and second 
generation stars may be obtained by comparing the number of stars within the GCs 
with that of the related field population. To have an estimate of the amount
of mass lost by GCs during their evolution, we may use the peculiar composition 
of GC stars, namely the large
excesses of Na that are often observed in GC stars, to trace these lost stars in
the field. Fig.~\ref{f:napop} shows the run of [Na/Fe] among field stars,
comparing it to the extremes of the distributions for GC stars. To this purpose,
we collected data for field stars  from three different sources: Gratton
et al. (2003a), Venn et al. (2004),  and Fulbright et al. (2007). Beside the
abundance ratios, they indicated the  population of each star on the basis of
the kinematics. Gratton et al. divided  stars into "accretion" and
"dissipation", while Venn et al. used the more common  separation between halo
and disk (the correspondence accretion/halo and dissipation/disk  is largely
true for the stars in common); finally all stars from Fulbright et al. are bulge
ones. In Fig.~\ref{f:napop} we plot field stars -taken only once if they are
present in more than one source- between metallicity $-2.5$  (to fit the lower
limit of the GC metallicity range) and $-1$ (to avoid  thin disk stars). In the
upper panel we also plot  [Na/Fe]$_{\rm min}$ for our GCs, i.e. the original,
first generation value that sits in the middle of the field stars distribution.
In the lower panel we plot instead  [Na/Fe]$_{\rm max}$ for our GCs, i.e. the
second generation value, well above the bulk of field stars. 

Examining this plot, we find that while most of the field stars roughly have
[Na/Fe]$\la 0$, there are a few objects with rather large excess of Na,
comparable to that observed in second generation stars of GCs. In the sample of
144 field stars with [Fe/H]$<-1$, there  are six stars with [Na/Fe]$>0.3$; for
comparison, 50\% (735 over 1483) of the stars in our  survey of GC stars have
such large Na excesses\footnote{We only considered those GCs with  [Fe/H]$<-1$.
This ratio does not change significantly if we compute the fraction of  Na-rich
stars in each cluster, and then average this value. Note also that in this way
we underestimate  the fraction of second generation stars, but we use this value
here for consistency with the field stars.}. 
However, only two  of these field stars are likely to be second generation stars
evaporated from GCs: they are  \object{HD74000} and \object{HIP37335} (=G112-36). 
These two stars are also moderately depleted in Li  (Hosford et al. 2009;
Pilachowski et al 1993), as expected for second generation stars  in GCs (see
Pasquini et al. 2005). The remaining four Na-rich stars are extremely 
metal-poor stars residing in binary systems, and the Na excess may be attributed
to mass  transfer.  Two of them (\object{CS22898027} and \object{CS22947187}) 
are C-rich stars. \object{G246-38} is extremely Li-poor (Boesgaard et al. 2005).
Finally, also \object{HD178443} is a giant in a binary system. While the
 statistics is poor, we may conclude that some 1.4\% of the
field metal-poor stars are likely Na-rich stars evaporated from GCs. Since these
are half of the GC stars, we may conclude that stars evaporated from GCs make up
2.8\% of the metal-poor component of the Milky Way. We may compare this value
with the current fraction of stars in GCs, that is 1.2\% using the Juric et al (2008)
in situ star counts, and 5\% using the Morrison (1993) ones (we neglect the impact of
selective loss of small mass stars; this is not too bad an approximation because
spectroscopic data are only available  for stars with typically the current TO
mass). We conclude that the GCs should have made up some 4\% of the original
mass of metal-poor stars, if Juric et al.  star counts  are used, and as much as
7.8\% adopting the Morrison  ones. Note that these values may still be
underestimates. In fact, if the cooling flow scenario is correct, second 
generation stars were originally a very kinematically cold population; this means
that they evaporated  from the GCs only after dynamical relaxation led to energy
equipartition,  $\sim 1$~Gyr after the GC formation, i.e., much later
than the formation phase. Then, there should be many more primordial stars of
GCs now in the field, lost during the  early phases. As discussed above, these
values should be increased by an order of magnitude.

The conclusion is that precursors of GCs likely had a baryonic mass $\sim
20$~times  larger than the current mass (if both the efficiency of star
formation  and the huge star loss factors are taken into account). If they also 
contained dark matter, they were likely two orders of magnitudes larger than
they currently are, with total masses up to a few $10^8~M_\odot$,  that is the
size of dSph's (see also Bekki et al. 2007).

We propose that the fraction of the primordial population lost by GCs is a 
major building block of the halo, although we do not exclude other minor 
contributors. This is supported by many other arguments, including their 
total mass, the metallicity distribution, and the location within the  Milky 
Way; all these are discussed in a separate paper (Gratton et al. in 
preparation). GCs might have played a similar role in the formation of the 
metal-poor  component of the thick disk ([Fe/H]$<-1$), while the specific 
frequency is much lower (by an order of magnitude) for the metal-rich component 
([Fe/H]$>-1$), and they are obviously absent from the thin disk. 

The formation phase of GCs may be very important to understand star formation in
the early phases of the Milky Way (and likely of other galaxies). Any
information on the composition of the primordial population  would help to shed
light on the formation mechanism of GCs. We will now examine what evidence can
be obtained from our data.

\subsection{The chemical evidence: the primordial abundance ratios and
the scenario for GC formation}\label{forma}

\begin{figure}
\centering
\includegraphics[bb=25 160 480 700, clip,scale=0.48]{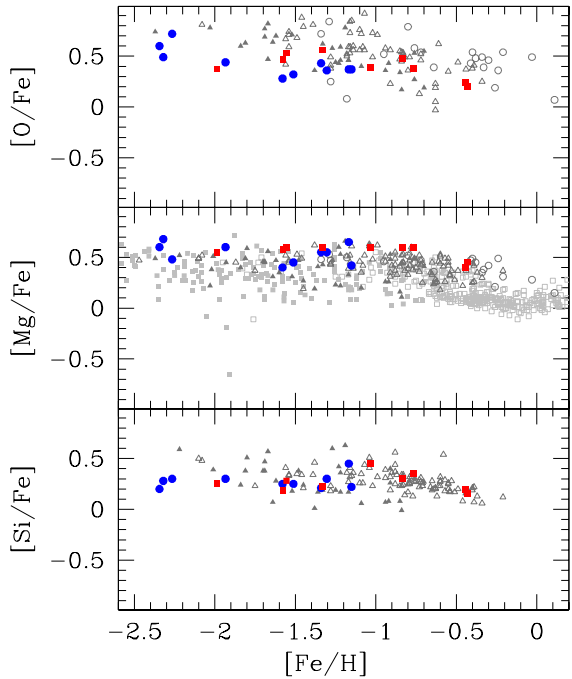}
\caption{[O/Fe]$_{\rm max}$, [Mg/Fe]$_{\rm max}$, [Si/Fe]$_{\rm min}$ for our
programme clusters (large filled red squares for disk/bulge clusters and blue 
circles for inner halo clusters, here and in the following figures) as a function of
the  metallicity [Fe/H], superimposed to field stars  from
literature (grey symbols).}
\label{f:omaxmgmaxsimin}
\end{figure}

\subsubsection{$\alpha-$elements}\label{alfa}

\begin{figure}
\centering
\includegraphics[scale=0.42]{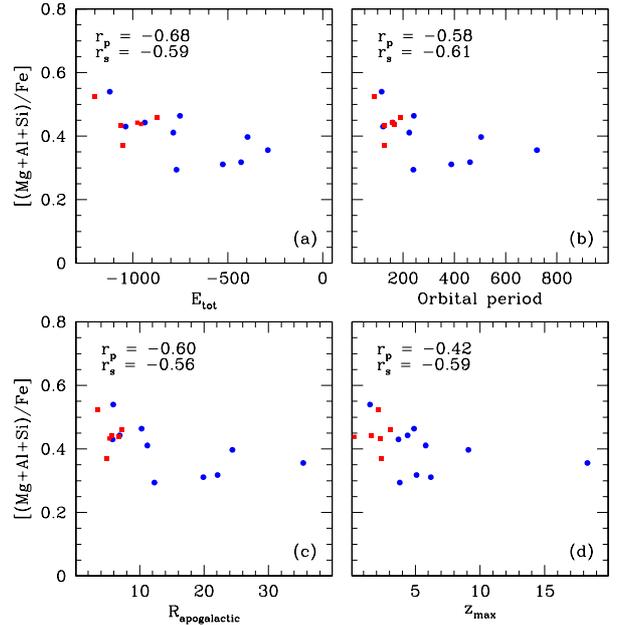}
\caption{Relation of the ratio [(Mg+Al+Si)/Fe] as a function of
several orbital parameters: (a) total energy of the orbit, (b) orbital period, 
(c) apogalatic distance, and (d) maximum distance above the galactic plane. 
Also indicated are the Spearman and Pearson correlation coefficients.}
\label{f:mgalsifeorb}
\end{figure}

As demonstrated by several authors (Gratton et al. 1996, 2000, 2003a,b; 
Fuhrmann 1998, 2004), the [$\alpha$/Fe] ratio is a good population
discriminator. Nissen \& Schuster (1997) found that halo subdwarfs have an
[$\alpha$/Fe] ratio on average lower and with more scatter than that typical of
the thick disk  population at the same metallicity, a result later confirmed by
several other  investigations (see e.g. Gratton et al. 2003b).

While several of the
$\alpha$-elements  are involved in the nuclear cycles related to high
temperature H-burning, there  are many possible indicators of the primordial
[$\alpha$/Fe] ratio that we can  obtain from our data. A short list includes the
maximum O and Mg abundances  ([O/Fe]$_{\rm max}$ and [Mg/Fe]$_{\rm max}$), the
minimum Si abundance  ([Si/Fe]$_{\rm min}$), the total Mg+Al+Si content
[(Mg+Al+Si)/Fe] (see Sect. \ref{para} and footnote there), and the average  
of [Mg/Fe]$_{\rm max}$, [Si/Fe]$_{\rm min}$, and the typical $\alpha-$element Ca. All these indicators 
give  concordant results. In the following, we 
will  mainly use the total [(Mg+Al+Si)/Fe] content.            

In Fig.~\ref{f:omaxmgmaxsimin} we plotted the run of the original abundance
ratio  of various $\alpha-$elements with [Fe/H]: [O/Fe]$_{\rm max}$,
[Mg/Fe]$_{\rm max}$,  and [Si/Fe]$_{\rm min}$. In the same Figure, we also
plotted the distribution of field stars from Fulbright (2000), Gratton et al.
(2003a), Reddy et al. (2006), Venn  et al. (2004), Fulbright et al. (2003,2006).
GCs lie close to the location of  the field stars, specially considering
possible offsets among different analyses.

We do not have an {\it a priori} idea of what we should expect for the 
exact form of most of the relations between the new parameters we are introducing 
and global cluster parameters, hence we will adopt the simpler one, a linear 
relation. These relations are evaluated using the Pearson coefficient for 
linear regressions and the Spearman coefficient of rank correlation, that can be 
used to characterise the strength and direction of a relationship of two given 
random variables (e.g. Press et al. 1992).

Tight relations are obtained between the overabundance of $\alpha-$elements
(represented e.g., by  Mg+Al+Si) and orbital parameters
(confirming  earlier findings by Lee and Carney 2002). In
Fig.~\ref{f:mgalsifeorb} we show the  relations of the ratio [(Mg+Al+Si)/Fe]
with total energy of the orbit, orbital  period, apogalactic distance, and the
maximum distance above the galactic plane.  The correlation coefficients,
represented by the Spearman and Pearson coefficients ($r_s$ and $r_p$), are high;
they would further increase excluding NGC~5904 (M~5), the cluster affected
by the largest uncertainties in the orbit. 

Similar trends are seen when plotting the overabundance of Ca or the average 
between Ca and Ti {\sc i}. It seems that clusters populating large-sized, more
eccentric orbits, with large apogalacticon distances (i.e., mostly the inner
halo GCs, in our classification), also have a proclivity
to have a lower abundance of elements produced in $\alpha-$capture processes. 
We consider these results as an indication that the initial position affected 
the chemical enrichment of GCs

Is there a risk of a bias introducing spurious trends among orbital and chemical
parameters? We believe that the correlation existing in our sample between
absolute magnitude $M_V$ and distance (Sect. \ref{mwgc}) is not a source of concern.
We found that the total Mg+Al+Si sum is anticorrelated (with moderate
significance, between 90 and 95\%) with $M_V$: the [(Mg+Al+Si)/Fe] ratio is
lower in more massive clusters. However, we found that  there is a slight trend for 
orbital parameters to be $correlated$ with the cluster mass (luminosity) for our distance 
limited sample. Albeit scarcely significant
from a statistical point of view, this trend is present also in the control
sample of GCs with $D_{sun}<12.9$ kpc and known orbital parameters.
Thus, when taken together, these opposite trends should combine in such a way to
erase any dependence of the total Mg+Al+Si sum on orbital parameters, whereas
we find good and significant relations. We conclude that these trends
are likely significant, and should be considered when discussing scenarios
for cluster formation.

We conclude that disk and halo GCs share the same [$\alpha$/Fe] ratio of 
thick disk and halo stars respectively. As observed in the field, halo GCs
have on average smaller excess of $\alpha-$elements, and a scatter larger
than that observed for the (thick) disk populations.
From the [$\alpha$/Fe] ratios collected in the Appendix we
found that below a metallicity of [Fe/H]$=-1$, the average values are 
$+0.42\pm0.01$ dex ($rms=0.05$ dex from 15 GCs) for disk/bulge clusters and 
$+0.36\pm0.02$ dex ($rms=0.07$ dex from 14 GCs)  for inner halo ones.

The explanation that we propose here is not the classical one requiring the
contribution of SNe Ia to raise the iron content, hence lower the [$\alpha$/Fe]
ratio.  
We propose the possibility 
that the contribution of core-collapse SNe to metal enrichment
is weighted towards higher-mass SNe for the precursors of lower-mass clusters: the most
kinematically energetic products (rich in particular in $\alpha-$elements)
might have been lost in more massive GCs, due to a powerful wind.
Evidence for such a wind is found around very massive and young star clusters,
like that observed in NGC~6946 (Sanchez Gil et al. 2009).
Note also that in the Milky Way these massive GCs are preferentially found in the
inner halo. This explanation is substantiated by the comparison of M~5
and M~4, providing a solution to the conundrum described in Sect. \ref{mwgc} . 

\subsubsection{Aluminium}\label{al}

In Paper VIII we presented the run of [Al/Fe]$_{\rm min}$\ with [Fe/H] in GCs.
[Al/Fe]$_{\rm min}$\ is expected to represent the Al abundance of the
primordial  population. We underlined there the large scatter observed in this
diagram, which  exceeds the observational uncertainties by far; we also noticed that
there is a group  of clusters, mainly belonging to the inner halo, characterised
by very low values  of [Al/Fe]$_{\rm min}$. Similar results were obtained
previously for  individual clusters (see e.g. Melendez \& Cohen 2009), but our
extensive survey shows  that this is a widespread property of GCs. 

However, Fe is not the best reference element for Al, because it has a very
different nucleosynthesis. Cleaner insight can be obtained considering Mg as
reference. Mg and Al may both be produced by massive stars exploding as core
collapse SNe. While Mg is an $\alpha-$rich element, whose production is primary,
Al requires the existence of free neutrons for its synthesis, and therefore its
production is sensitive to the initial metallicity 
(Arnett \& Truran 1969; Truran \& Arnett 1971; Woosley \& Weaver 1995). The
exact  dependence of the ratio of Al to Mg as a function of overall metallicity
has  never been defined satisfactorily by theory, but it should likely show
some  sort of secondary behaviour with respect to Mg. 

\begin{figure}
\centering
\includegraphics[scale=0.45]{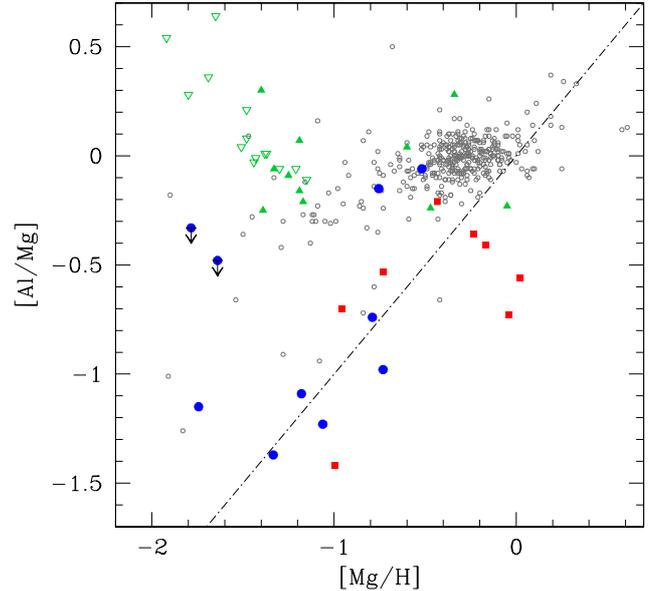}
\caption{Run of the [Al/Mg] ratio as a function of the [Mg/H] ratio in field
stars (grey-tone open circles), globular cluster in our sample (red and blue
filled circles and squares, respectively) and stars in dSphs (green triangles,
the open ones are  upper limits in Al, see text for references). The [Al/Mg]
ratio in our sample refers to primordial abundances, namely [Al/Fe]$_{\rm min}$
and  [Mg/Fe]$_{\rm max}$. The dot-dashed line indicate the line of secondary
production [Al/Mg]=[Mg/H].}
\label{f:mgh}
\end{figure}

Fig.~\ref{f:mgh} shows the run of [Al$_{\rm min}$/Mg$_{\rm max}$] (that is the 
ratio for the primordial population) with [Mg$_{\rm max}$/H] for our programme 
cluster. We also plot, with different symbols, the run of [Al/Mg] with [Mg/H] 
for metal-poor field stars in the Milky Way (Fulbright et al. 2007; Reddy et al.
2003; Gehren et al. 2006; Jonsell et al. 2005; Fulbright 2000), as well as for
stars in  dSph galaxies (Koch et al. 2008; Geisler et al. 2004; Shetrone et al.
2001, 2003; Sbordone et al. 2004). As shown by Gehren et al. (2006), the local
subdwarfs have  markedly different [Al/Mg], depending on their kinematics: halo
subdwarfs have much  lower [Al/Mg] than thick disk ones, so much that Gehren et
al. proposed to use this  ratio as a population diagnostics. The locus occupied
by primordial populations in  most GCs is clearly distinct from that for the
thick disk subdwarfs, and close to  that defined by halo subdwarfs. Several
clusters, including the most massive ones,  have very low primordial
Al abundances and lie
close to the line of secondary production:  [Al/Mg]=[Mg/H]. The exception
consists in a few small clusters (NGC~288, NGC~6121, NGC~6171,  NGC~6838), that have
large primordial Al abundances. A multivariate analysis, using  [Mg$_{\rm
max}$/H] and $M_V$\ as independent variables, and  [Al$_{\rm min}$/Mg$_{\rm
max}$] as dependent one, yields:
\begin{eqnarray}
[{\rm Al_{\rm min}}/{\rm Mg_{\rm max}}]&=(-0.01\pm 0.24) + (0.66\pm 0.14) [{\rm Mg_{\rm max}/H}] + \nonumber\\ 
&(0.13\pm 0.06) ({\rm M_V}+7.5),
\end{eqnarray}
with a highly significant linear correlation coefficient of r=0.80 (16 clusters,
13 degrees of freedom). This behaviour is different from what is observed in dSph's, that
are  typically characterised by rather large Al abundances. 

\subsubsection{A scenario for the formation of GCs}\label{scenario}

How may we explain this behaviour? First, the {\it primary-like} run of Mg and
Al  observed in disk subdwarfs requires that most of the Al in these stars was
produced  in a site different from massive stars (where the production is
secondary).  Second,  we observed very high Al$_{max}$ abundances within second 
generation stars in GCs (Paper VIII, and Table~\ref{t:tabnoi}) even exceeding
the values observed in disk subdwarfs,  and similar to those observed in dSph's
stars. This suggests an obvious astrophysical  site where these large amounts of
Al can be produced: the same stars  responsible for the Mg-Al
anticorrelation (either fast rotating massive stars - FRMS - or 
intermediate mass AGB stars). In this case the very low Al abundances, 
characteristic of the
primordial population of massive GCs, may be explained  {\it if the  raise in
metal content in the environment where these stars formed was so fast  that no
star with intermediate metallicity could form}. On the other hand, 
pre-enrichment should have had to be more gradual for small mass clusters, with 
different generations having little difference in the Mg content, and an
efficient re-processing of Mg into Al in the next  generation. A similar
conclusion was drawn for the case of M~71 (NGC~6838) by Melendez \& Cohen (2009). However,
we find that this is a general feature of GCs, even of clusters reputed younger
with respect to the disk ones. This is unlikely to be a
coincidence, rather is more probably related to the typical sequence of events
that led to cluster formation. 

This consideration suggests a (still qualitative) sketch for the formation of
typical  massive GCs, which is a more elaborated and updated version of what
proposed  more than thirty years ago by Searle \& Zinn (1978) and
later elaborated by many other authors (see e.g. B\"oker 2008 and references
therein). 

\begin{enumerate}
\item Consider a cosmological fragment/satellite of $10^6-10^9~M_\odot$\, i.e.
the same range  of masses of dSph's, but which is near the Milky Way
(R$_{GC}\sim 10$~kpc) at a very early epoch ($<2$~Gyr from
Big Bang).  In a cold dark matter (CDM) scenario, we expect many such satellites to have
existed (see e.g. Bromm \& Clarke 2002; De  Lucia \& Helmi 2008). At this very
early epoch, this satellite is still made of dark  matter and gas
($10^5-10^8~M_\odot$), with  negligible/small stellar contribution  and metal
pre-enrichment, depending on its age, i.e., on the time allowed for an
isolated evolution before the phases described in the following. 
\item Likely due to its motion, which brings the cluster in proximity of the
denser  central region of the Milky Way, this fragment has a strong interaction,
possibly with  the same early disk of the Milky Way or with another substructure
(Bekki 2004). This  strong interaction triggers an early star formation
(Whitmore \& Schweizer 1995). In  a short timescale (a few million years)
$10^4-10^5~M_\odot$\ of gas are transformed  into stars; the most massive of
these stars explode as SNe after $\sim 10^7$~yr.  Hereinafter, we will call this
population $precursor$, because while needed to form the  GC itself, it is
unlikely that we will find any representative of this population within  the
present GC (see below).
\item The precursor core collapse SNe have two relevant effects: i) they enrich
the  remaining part of the fragment/satellite of metals, raising its metallicity
to the value  currently observed in the GC\footnote{This enrichment
should be very uniform,  suggesting a super-wind from the precursor association
(Mac Low \& McCray 1988). There might also be a selection effect against
the most massive and energetic SNe,  possibly reducing the typical [$\alpha$/Fe]
ratio of the next generation stars.}; and  ii) efficiently trigger star
formation in the remaining part of the cloud, before the  intermediate mass
stars can efficiently contribute to nucleosynthesis. This second  episode (or
phase, since it is not clear that there should not be a continuum in star
formation) forms a few $10^5-10^6~M_\odot$\ of stars, in a large association.
These associations have mass and size ($\sim$ 100 pc) comparable to the 
knots commonly observed in luminous and ultra-luminous infrared galaxies
(see e.g. Rodriguez Zaurin et al. 2007).
\item The strong wind from massive stars and core collapse SNe of this huge
association disperses the remaining primordial gas on a timescale of 
$\sim 10^7$~yr (see the case of the super star cluster in NGC~6946, Sanchez Gil
et al. 2009).
\item While the large association is expanding, the low velocity winds from FRMS
or, perhaps more likely\footnote{Given the very fast
evolutionary lifetimes of FRMS it is possible that the SN
explosions from this component would hamper the formation of an efficient
cooling-flow.}, from the more  massive intermediate mass stars feed a cooling
flow, that forms a kinematically cool population at the centre of the
association (D'Ercole et al. 2008). Possible examples of objects in this phase
are Sandage-96 in NGC~2403 (Vinko et al. 2009) or the super star cluster
in NGC~6946 (Hodge 1967, Larsen \& Richtler 1999, Larsen et al. 2006).
A fraction of the  primordial population
stars (but very few precursors if any, since they are much rarer  and possibly
were at some distance from the newly forming cluster) remains trapped  into the
very compact central cluster formed by this second generation stars. This is 
the GC that may survive over a Hubble time, depending on its long-scale
dynamical  evolution, and that we observe at present.
\item Core collapse SNe from this second generation sweep the remaining gas
within the  cluster, terminating this last episode of star formation. This
occurs earlier in more  massive clusters:  these clusters will be then enriched
by stars over a restricted range of  mass (only the most massive among the potential polluters), 
leading
to very large He abundances.  Hence, there should be correlations between He
enrichment, cluster mass, and fraction of  primordial stars. Note however that
this may occur naturally only in a cooling flow scenario,  where second generation
star formation is well  separated from the evolution of individual stars. In the
original FRMS scenario of Decressin et al. (2008),
second generation stars form within the  individual equatorial
disks around the stars, as a consequence of the large mass loss rate
and fast rotation. Within this scenario, it is difficult (although not strictly
impossible) to link  properties of individual stars to global cluster
properties.
\item  At some point during these processes or just after, 
the DM halo is stripped from the GC and merges with the
general DM  halo (see e.g. Saitoh et al. 2006; Maschenko \& Sills 2005). It is
not unlikely that  the loss of the DM halo is due to the same interaction
causing the formation of the GC.  The cluster has now acquired the typical
dynamical characteristics that we observe at  present, and has hereinafter
essentially a passive evolution (see e.g. Ashman \& Zepf  1992).
\end{enumerate}

Small and/or metal rich clusters (mainly residing in the disk) differ in many 
respects:  forming within a disk (not necessarily the disk of the main 
galaxy), they have a  considerable pre-enrichment of metals. In the smaller 
clusters, the precursor population (if it exists) does not enrich significantly 
the next generation, which will share the  chemical composition of the field.
Moreover, as we recall in Sect. \ref{mwgc}, the typical luminosity of disk GCs is 
smaller than that of halo GCs. A large fraction of disk stars form in small 
clusters and association, which are easily  disrupted due to infant mortality 
and dynamical evolution (this might be closely  related to the low specific 
frequency of clusters in spirals: Harris and Racine 1979;  Harris 1988). 
In addition, 
dynamical interaction with the disk or the presence of other nearby star forming 
regions might delay or reduce the efficiency of cooling flows  in forming second 
generation stars.  Observations of many clusters with multiple  turn-off's in the Magellanic Clouds  
(Mackey et al. 2008; Milone et al. 2009a)  might 
indicate that this effect is not so important; see however Bastian \& De Mink 
(2009) for  a different interpretation of these observations.
Finally, disk GCs all formed very early in the galactic history (Fall \& Rees 
1988;  Zinn 1988). After these very early phases, conditions within the Milky
Way disk were 
never again  suitable for formation of very massive GCs, likely due to the low 
pressure characteristic  of quiet disks (Ashman \& Zepf 2001).

As noticed by Zinn (1985), within this scheme the different chemical history of 
disk  and halo GCs may easily explain the most obvious characteristics of GCs, 
systematically observed in virtually all GC systems,  
such as: i) the bimodal colour and metallicity distribution , because blue clusters 
are essentially self-enriched, while red clusters form from pre-enriched material in 
the early phases of dissipational collapses (note however that 
there should be a few blue and metal-poor disk clusters); ii) the absence of discernible 
metallicity trends with R$_{GC}$ for halo GCs (see Searle \& Zinn 
1978); and iii) the presence of relatively young GCs in the halo (Zinn 1985).

This scenario also unifies the view of GCs and dSph's. In fact, according to
this  scheme, both GCs and dSph's start as DM-dominated cosmological
structures  with masses in the range $10^6-10^9~M_\odot$. The main difference is
their location  with respect to the Galaxy. GCs formed from DM halos closer to
the centre of the  Galaxy or to other structures (even NGC~2419, the farthest
known  GC). They had only a limited significant independent  chemical evolution
prior to their interaction with the Galaxy, which occurred quite  early, when
the structures were still gas-rich. The age-metallicity relation for  halo GCs
(see Fig.~\ref{f:agefeh}) suggests that pre-clusters clouds that had more  time
to evolve, actually produced part of their metals during this pre-cluster phase.
When they interacted with other structures or with the main Galaxy body itself,
their independent evolution was interrupted by the sequence of phases we
described above (see Melendez \& Cohen 2009 for a similar view). 

On the other hand, dSph's formed much farther, at typical distances of several
hundred kpc. They could have a long independent evolution before interaction (if
any) with the Milky Way. Their evolution is essentially determined by their
initial mass, so that they obey mass-luminosity and mass-metallicity relations.
If any interaction with the Milky Way occurred (as is the case for Sagittarius),
this happened once the dSph had become gas poor;  this did not lead to any
further GC formation (although more massive dSph's could include a few already
formed GCs).

As noticed by the referee, the scenario we considered implies that significant 
GC formation may still occur in gas-rich environments insofar the high pressure
needed is available, e.g. due to galaxy-to-galaxy interaction. This is possibly
the case of the Antennae (Whitmore \& Schweizer 1995) or even of the Magellanic 
Clouds, where there is evidence of formation of six young clusters in the 
Magellanic Bridge (Irwin et al. 1985).

\section{The second phase of cluster self-enrichment}\label{2gen}

In our scenario, the early phases of the evolution of the structures that will
ultimately form GCs are linked to the composition of the primordial population;
the late phases determine the cluster self-enrichment processes.
We expect that such processes are linked to some global characteristic of GCs. In Paper VII and VIII we showed that very important r\^oles are played by mass and metallicity
of  the cluster. In this Section, we revisit this issue, considering many
other parameters; a more global approach will be applied in next Section.

\subsection{The chemistry of second generation stars}\label{chem2}

The chemical properties of second generation stars in GCs were discussed
quite extensively in Papers VII and VIII. They may be described by the fraction 
of the intermediate I and of the extreme E components, and by the extreme values
associated to the polluted stars using a dilution model, namely [O/Fe]$_{\rm
min}$, [Na/Fe]$_{\rm max}$, [Mg/Fe]$_{\rm min}$, [Al/Fe]$_{\rm max}$. In Paper
VIII we demonstrated that the ratio [Si/Fe]$_{\rm max}$ is an additional marker of
second generation stars, because of the leakage of the Mg-Al cycle which
produces $^{28}$Si at temperatures in excess of about 65 million K (Carretta et
al. 2009b; Arnould et al. 1999; Yong et al. 2005).

\begin{table*}
\begin{center}
\small
\caption{ Pearson's correlation coefficients and level of significance for relations in P, I, and E stars }
\begin{tabular}{lllll}
\hline
Par	             & IQR		&P		   &I		      & E              \\
\hline     		
IQR[O/Na]  	     &	          	  &  +0.25 $<$90\%&$-$0.49 95-98\% &   +0.65 $>$99\% \\
P	             &  +0.25, $<$90\%&	               &$-$0.91, $>$99\% &   +0.56, $>$99\% \\
I	             &$-$0.49, 95-98\%&$-$0.91, $>$99\%&	    	     & 
$-$0.85, $>$99\% \\
E	             &  +0.65, $>$99\%&  +0.56, 98-99\%&$-$0.85, $>$99\% &	     	  		    \\
$[$(Mg+Al+Si)/Fe$]$  &$-$0.56, $>$99\%&$-$0.59,  99\%   &  +0.69,  $>$99\% &
$-$0.73,  $>$99\% \\
$[$Ca/Fe$]$ 	     &  +0.03, $<$90\%&$-$0.43, 90-95\%&  +0.52, ~98\%   &
     $-$0.51, 95-98\% \\
$[$O/Fe$]_{\rm min}$ &$-$0.79, $>$99\%&$-$0.37, $<$90\%&  +0.62, $>$99\% &
$-$0.78, $>$99\% \\
$[$O/Fe$]_{\rm max}$ &$-$0.23, $<$90\%&$-$0.12, $<$90\%&  +0.40, 90-95\% &
$-$0.67, $>$99\% \\
$[$Na/Fe$]_{\rm min}$&$-$0.03, $<$90\%&  +0.13, $<$90\%&$-$0.20, $<$90\% &         +0.27, $<$90\% \\
$[$Na/Fe$]_{\rm max}$&$-$0.26, $<$90\%&$-$0.26, $<$90\%&  +0.21, $<$90\% &
     $-$0.10, $<$90\% \\
$[$Al/Fe$]_{\rm min}$&$-$0.46, 95\%   &$-$0.26,  $<$90\%&  +0.36,  $<$90\% &
     $-$0.35,  $<$90\% \\
$[$Al/Fe$]_{\rm max}$&  +0.10,  $<$90\%&  +0.49, 95-98\%&$-$0.38, , $<$90\% &        +0.10,  $<$90\% \\
$[$Mg/Fe$]_{\rm max}$&$-$0.42, $~$95\%&$-$0.55, 98-99\%&  +0.74, $>$99\% &
$-$0.80, $>$99\% \\
$[$Mg/Fe$]_{\rm min}$&$-$0.48, 95-98\%&$-$0.61, $>$99\%&  +0.67, $>$99\% &
     $-$0.56, 98-99\% \\
$[$Si/Fe$]_{\rm max}$&$-$0.47, 95-98\%&$-$0.21, $<$90\%&  +0.32, $<$90\% &
      $-$0.38, ~90\%  \\
$[$Si/Fe$]_{\rm min}$&$-$0.44, 95\%   &$-$0.32, $<$90\%&  +0.40, 90-95\% &
     $-$0.40, 90-95\% \\
${\rm [Fe/H]}$       &  +0.18, $<$90\%&  +0.01, $<$90\%&$-$0.28, $<$90\% &         +0.58, 98-99\% \\
M$_V$                &$-$0.46,95-98\% &$-$0.39, $~$95\%&  +0.55, 98-99\% &
$-$0.61, $>$99\% \\
age                  &$-$0.33, $<$90\%&$-$0.40, $~$95\%&  +0.59, $>$99\% &
$-$0.71, $>$99\% \\
T$_{\rm eff}^{\rm max}$(HB) &  +0.69, $>$99\%  &  +0.45, $~$95\% &$-$0.47, 95-98\%	&  +0.36, $<$90\% \\
\hline 
\label{t:corrtab}
\end{tabular}
\begin{list}{}{}
\item  Note - The number of degrees of freedom is generally 17, except for the cases involving
[(Mg+Al+Si)/Fe], [Al/Fe]$_{min}$, 
and [Al/Fe]$_{max}$, where it is 16.
\end{list}
\end{center}
\end{table*}

In Tab.~\ref{t:corrtab} we list the Pearson's correlation coefficients, the
number of degrees of freedom, and the statistical significance level of the
correlations involving the extremes of the abundance distributions and the
fraction of P, I, and E stars. In addition, we also considered the interquartile
of the Na-O distribution IQR[O/Na]\footnote{We verified that the corresponding
quantity for the Mg-Al anticorrelation IQR[Mg/Al] shows a general correlation
with IQR[O/Na], with some scatter. However, IQR[Mg/Al] could be estimated only
for the comparatively few stars observed with UVES. Furthermore, we could not
compute this index for four GCs of our sample, three  having  only 5-7 UVES
spectra per cluster (NGC~6171, NGC~6388, and NGC~6441) and one without Al
determination (NGC~6397). Hence, IQR[Mg/Al] has large uncertainties and will not
be used in the remaining of this paper. On the other hand, the IQR[O/Na]
for NGC~6397, based on only 16 stars, is quite indiscernible from the others
(based on more stars) in all relations shown in the following. This supports the
robustness of this indicator also for the GC in our sample with the smallest
size of sampled stars.}.  We find several correlations with high
statistical significance level (better than 99\%), as can be seen from the
Table. The most interesting correlations are shown in Figs. 11 to 16.

\begin{figure}
\centering
\includegraphics[bb=40 160 450 700, clip, scale=0.6]{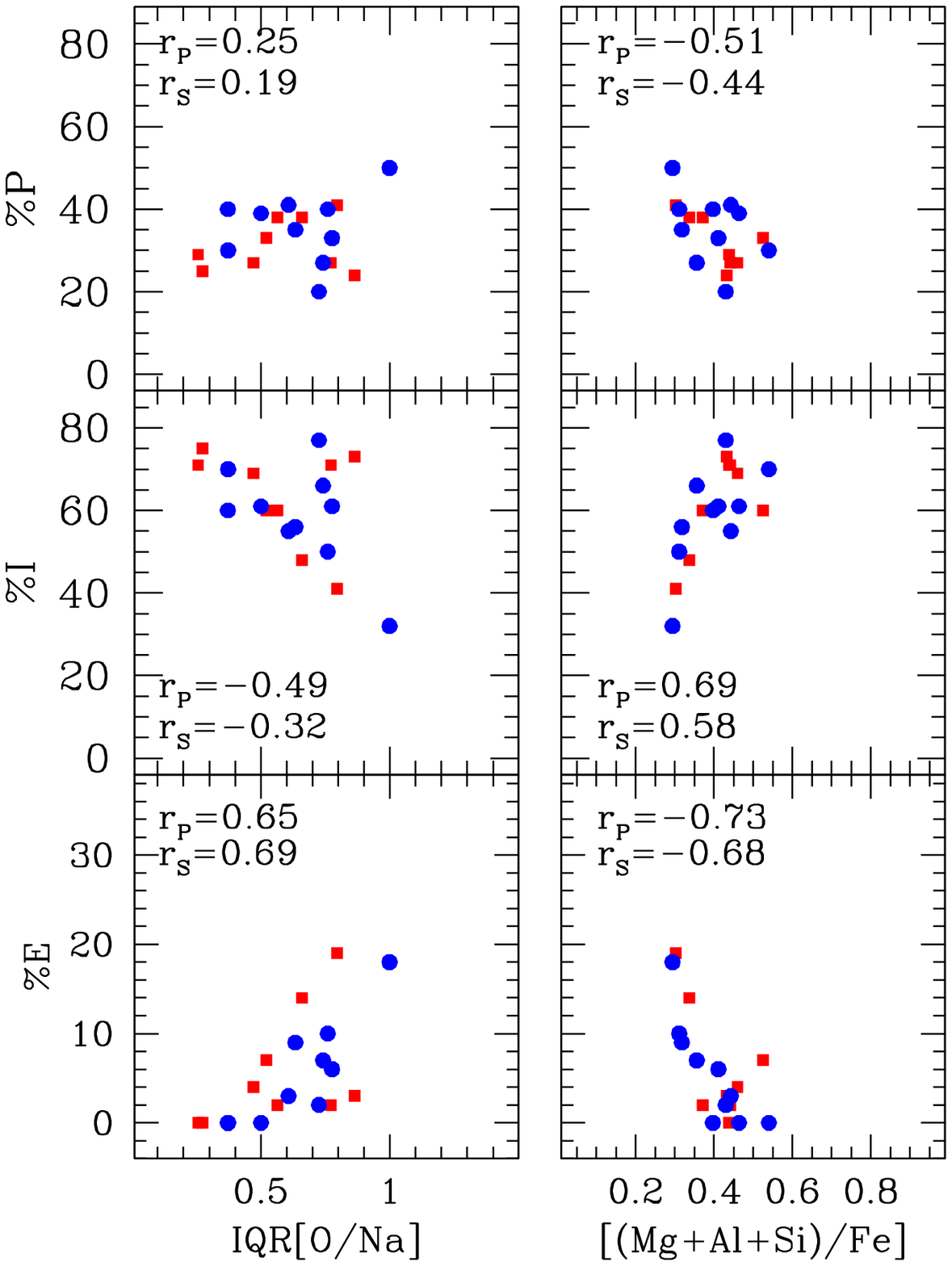}
\caption{Left: Fraction of stars in the primordial P (upper panel), intermediate 
I (middle panel) and Extreme E (lower panel, with different scale on the y-axis) 
components of GCs as a function of the IQR[O/Na]. Right: the same,  as a 
function of [(Mg+Al+Si)/Fe]. The  Pearson and the Spearman Rank Correlation 
coefficients are shown in the box, here and in the next figures.}
\label{f:iqrPIE}
\end{figure}

From Tab.~\ref{t:corrtab} and the left panels in Fig.~\ref{f:iqrPIE} we see
that there is a tight correlation between the E fraction and the extension of
the Na-O anticorrelation, IQR[O/Na]. This does not come as a surprise, since
IQR[O/Na] is driven by the stars with extreme chemical modifications. The I
fraction, which is the dominant group in all GCs, is anticorrelated with
IQR[O/Na], while the primordial P component is not related to the extension of
the Na-O anticorrelation.

\begin{figure}
\centering
\includegraphics[scale=0.47]{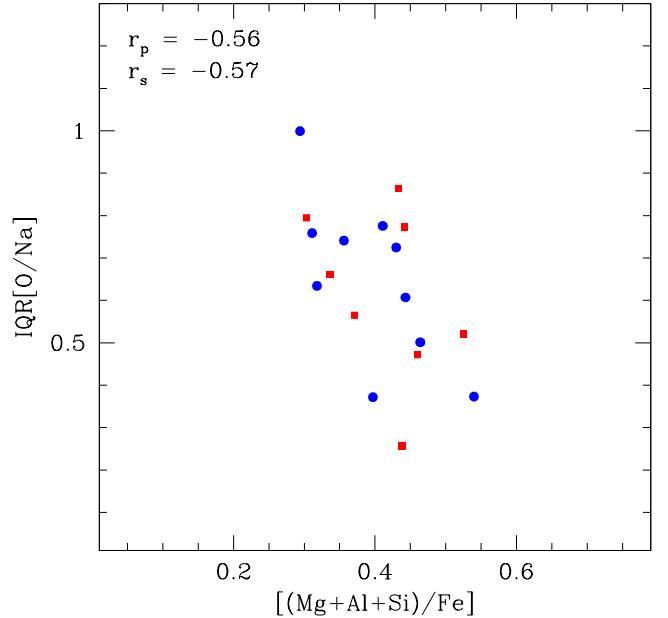}
\caption{Extension of the Na-O anticorrelation, measured by the IQR[O/Na], as a
function of the total sum of Mg+Al+Si atoms for our sample of clusters.}
\label{f:iqrmgalsife}
\end{figure}

\begin{figure}
\centering
\includegraphics[bb=40 160 450 700, clip,scale=0.6]{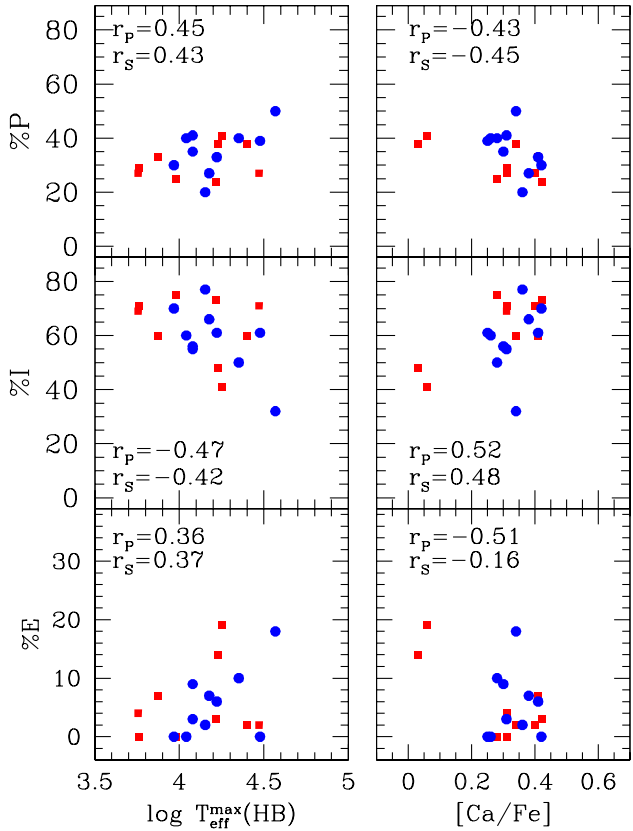}
\caption{Left: fraction of stars of the P, I and E components (upper, middle and
lower panels, respectively) as a function of log T$_{\rm eff}^{\rm max}$(HB).
Right: the same, as a function of the mean [Ca/Fe] ratio.}
\label{f:piealsi}
\end{figure}

The run of the P, I and E fractions as a function of the total Mg+Al+Si sum is
illustrated in the left panels of Fig.~\ref{f:iqrPIE}. The fraction of stars in
the P population is higher when this total sum is lower. Even tighter relations
do exist for the I and E components. The sign of these  relations is opposite
to those with IQR[O/Na]. The combination of the left and right panels in
Fig.~\ref{f:iqrPIE} results in the statistically significant anticorrelation
between IQR[O/Na] and total Mg+Al+Si sum shown in Fig.~\ref{f:iqrmgalsife}:
clusters where the Na-O anticorrelation is more extended have a lower value of
the total Mg+Al+Si sum. This result
is confirmed by the relations existing between the P, I, and E fractions and
[Ca/Fe] (right panels in Fig.~\ref{f:piealsi}). However, we also notice that
these findings are strongly influenced by the two bulge clusters NGC~6388 and
NGC~6441.

The E population is the only one showing a significant correlation with [Fe/H].
However, this is not surprising, because we found (Carretta et al. 2009a) that
O$_{min}$, whose value is related to the presence of the E component, is well
represented by a linear combination of M$_V$ and [Fe/H]. We interpreted this
result as a proof that the mass of the average polluters varies regularly as a
function both of cluster mass and metal abundance.  

\begin{figure}
\centering
\includegraphics[bb=40 160 450 700, clip,scale=0.6]{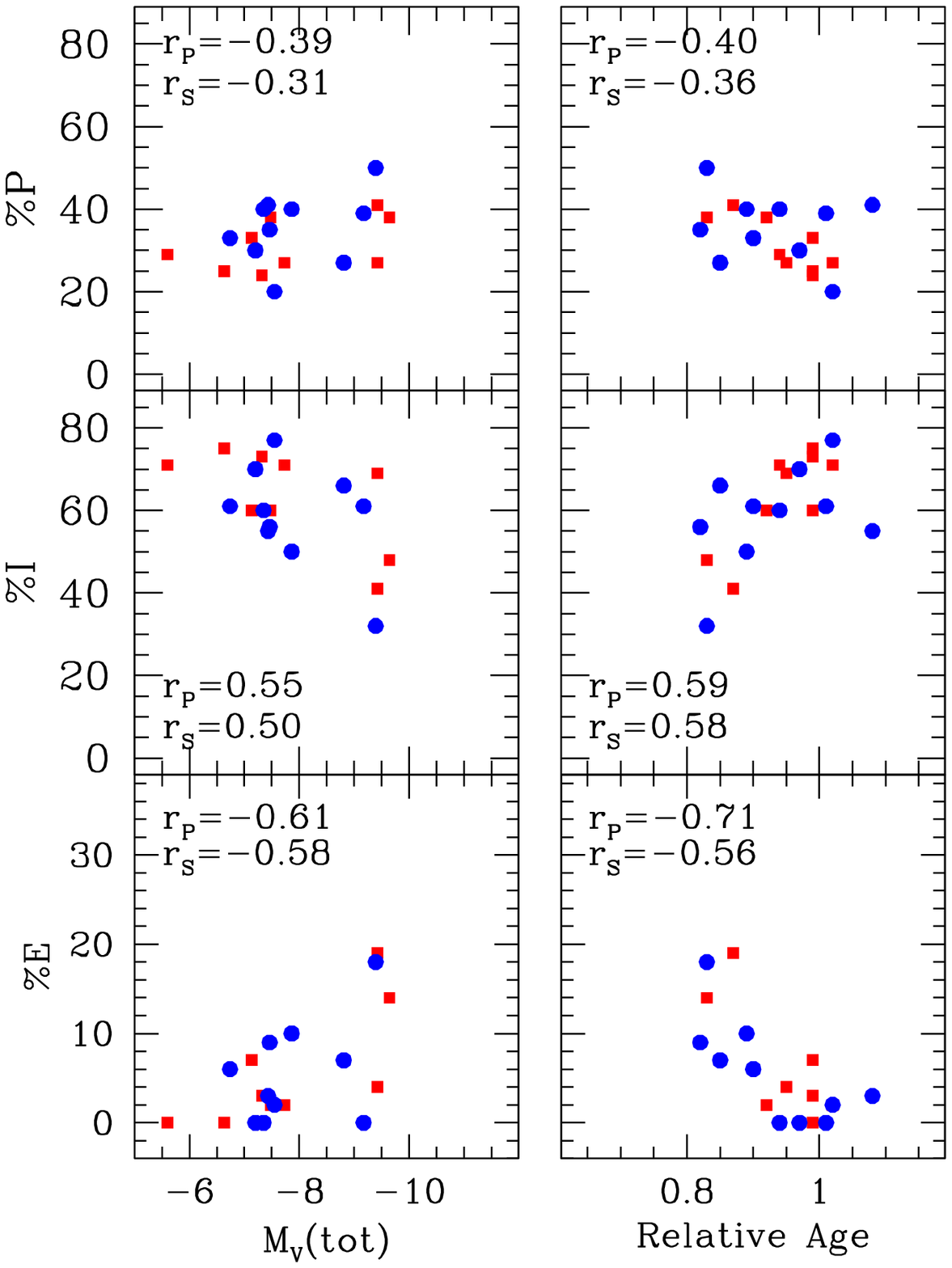} 
\caption{Left panels: The fraction of stars in the P, I and E components of
second generation stars as a function of the total absolute magnitude (hence,
mass) of clusters.  Right: the same, as a function of relative age. }
\label{f:popsMvrh}
\end{figure}

In Fig.~\ref{f:popsMvrh} (right panels) the fractions of first and second
generation stars are shown as a function of the relative GC ages. These
relations are statistically significant, the P and E components being
anticorrelated and the I fraction correlated with age. This result might be an
artefact due to the relations with $M_V$ (left panels of the same figure). As
seen in Sect. \ref{sele}, the more massive clusters in our sample are those in
the inner halo, where the younger clusters reside. However, we cannot exclude
other factors. For instance, the presence of a population with enhanced He
might alter the derivation of ages; this is discussed in detail in Gratton et
al. (2010).

In summary:
\begin{itemize}
\item[(i)] a P component of first generation stars is present in each GC; its incidence does not depend on the cluster metallicity, being approximatively constant (at a level of about one third of the cluster stars, Paper VII) in each GCs. This component does not affect the extension of the Na-O anticorrelation, but it is less conspicuous in GCs with higher abundances of $\alpha-$elements. 
\item[(ii)] the bulk of second generation stars is composed of stars with moderate
alterations in the light elements Na, O, Mg, Al participating to proton-capture reactions in H-burning at high temperature (see Paper VII). This I component is lower in GCs with longer Na-O anticorrelations, and this fraction is higher in GCs with larger $\alpha-$element ratio.
\item[(iii)] the E component, with extreme changes in light element abundances with respect to first generation stars, is not present in all clusters (Paper VII). It is anticorrelated to the complementary I component, and is the key component driving the extension of the Na-O anticorrelation, showing a very tight and significant correlation with IQR[O/Na]. The E component decreases with increasing overabundance of $\alpha-$elements.
\end{itemize}

In the following, we will analyse the relations between the chemistry of different sub-populations and global cluster parameters.

\subsection{Extension of the Na-O anticorrelation and (blue) HB}\label{naohb}

\begin{figure}
\centering
\includegraphics[scale=0.45]{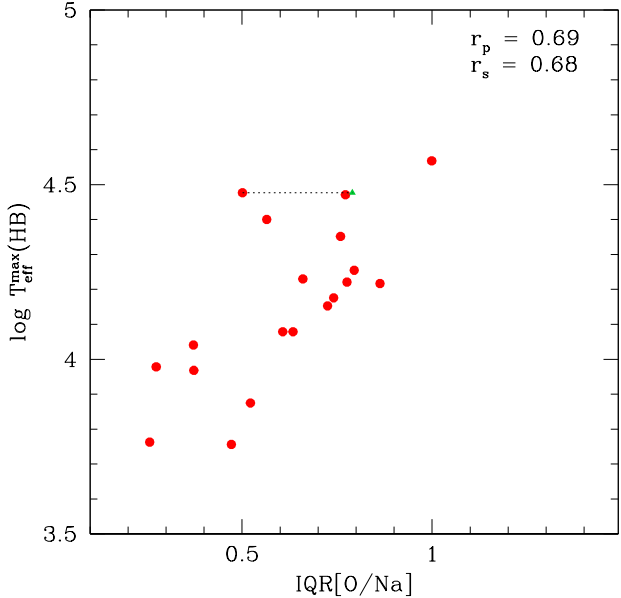}
\caption{Interquartile range of the [O/Na] ratio as a function of the maximum
temperature reached on the HB (taken from Recio-Blanco et al. 2006 or derived
here). The line connects the value for M~15 based on our data and the one
derived from literature (open green triangle, see text).}
\label{f:iqrteff}
\end{figure}

We demonstrated in Carretta et al. (2007d) that there is a tight
correlation between IQR[O/Na] (derived from part of our sample or from the
literature) and the maximum temperature of HB stars,  $\log{\rm T}_{\rm
eff}^{\rm max}$(HB) (Recio-Blanco et al. 2006). To verify that this relation
holds also for the more extended sample of clusters considered here, we
complemented the values of $\log{\rm T}_{\rm eff}^{\rm max}$(HB) obtained by
Recio-Blanco et al.  with new ones. We   employed good quality CMDs (Bellazzini
et al. 2001 for NGC~288; Marino et al. 2008 for NGC~6121; Rosenberg et al. 1999
for NGC~6254; Momany et al. 2003 for NGC~6752; Vargas Alvarez \& Sandquist 2007
for NGC~6809) with the same HB models (Cassisi et al. 1999) used by those
authors. Furthermore, we replaced the values of IQRs of Carretta et al. (2007d)
that were derived from the limited samples of stars with UVES spectra available
at the time, with new ones obtained from the full sample of stars with Na and O
determinations from GIRAFFE and UVES in all the 19 GCs in our sample
(Tab.~\ref{t:tabnoi}). We fully confirm our previous findings, as seen in
Fig.~\ref{f:iqrteff}\footnote{This results would not change had we used older
parameters measuring the length of the HB, such as $L_{tail}$ or BT from Fusi
Pecci et al. (1993)}.

The position of M~15 (NGC~7078) in Fig.~\ref{f:iqrteff} is somewhat uncertain.
The value from literature lies very well on the relation defined by all other
clusters, while the value we derive is more offset from the main trend. Since
M~15 is among the most metal-poor GCs and our GIRAFFE resolution is worse than
the one in Sneden et al. (1997), it is   possible that we missed some very
O-poor or Na-poor stars. However,  the O abundance of two most O-poor stars in
Sneden et al. is flagged as uncertain.  With these caveats in mind, we adopt
here  the value of the IQR[O/Na] for M~15 derived in our analysis for
homogeneity. The correlation is anyhow very tight: the Spearman test returns
coefficients $r_S$ of 0.69 (with our value for M~15), 0.82 (with M~15 from
literature), or 0.81 (without M~15). The probability to get such a tight
relation by chance is negligible: the (one-tailed) $t-$test returns values of
2$\times$10$^{-4}$,  1$\times$ 10$^{-6}$, or 4$\times$ 10$^{-6}$ in the three
cases,  respectively.

We confirm that this is a real, very strong relation: the first conclusion we
can draw is that the same mechanism drives/affects the extent of the pollution
on the RGB and the morphology of the bluest end of the HB.

However, this can be only considered $a$ second parameter: the global
distribution of stars on the HB [as indicated e.g., by the HB  ratio
HBR=(B-R)/(B+R+V) -see Harris 1996 and web updates] is $not$ correlated with 
IQR[O/Na], i.e. with the extension of the Na-O anticorrelation, as already
discussed by Carretta (2006) and Carretta et al. (2007d). The relation between
the distribution of stars along the HB and the Na-O anticorrelation is further
discussed in a separate paper (Gratton et al. 2010).

Finally,  the relations of  the three individual stellar components P, I, E
with  log T$_{\rm eff}^{\rm max}$(HB), shown in the right panels of 
Fig.~\ref{f:piealsi}, simply reflect the correlations and anticorrelations of 
the latter with IQR[O/Na].

\subsection{Extension of the Na-O anticorrelation and total cluster mass}\label{naomv}

\begin{figure}
\centering
\includegraphics[scale=0.45]{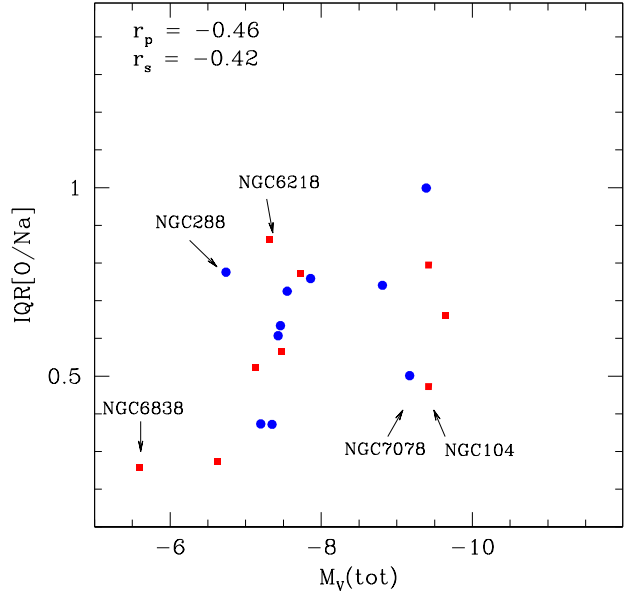}
\caption{The extension of the Na-O anticorrelation (measured using 
IQR[O/Na]) as a function of the clusters' absolute magnitude M$_V$ from Harris
(1996). The Pearson correlation coefficient is given
and five GCs discussed in the text are indicated.}
\label{f:iqrMv}
\end{figure}

Possibly a major result of our analysis is the good correlation we found between
the IQR[O/Na] and the present-day total mass of the GCs (using the absolute
magnitude M$_V$ as a proxy for the mass), see Table~\ref{t:corrtab} and
Fig.~\ref{f:iqrMv}. A large mass seems to be a requisite for an extended
anticorrelation. This is not unexpected, because Recio-Blanco et al. (2006)
already found a good correlation between total mass and highest temperature on
the HB, and in the previous Section we found that the latter is strictly related
to the extent of the Na-O anticorrelation.

The correlation found in Recio-Blanco et al. is also a good test for the 
reliability of our sample against a possible bias related to selection criteria.
We do not have values of IQR[O/Na] for all clusters in the Harris catalogue:
however, our distance-limited sample of 19 GCs shows a correlation between $M_V$
and log T$_{\rm eff}^{\rm max}$(HB), with a Pearson correlation coefficient $r_p
=-0.47$, 17 degrees of freedom, significant at better than 95\% level of
confidence. The same correlation is present in all clusters in the Harris
catalogue with  available log T$_{\rm eff}^{\rm max}$(HB) ($r_p =-0.43$, 54 GCs)
and for clusters in the control sample restricted to distances less than
12.9 kpc from the Sun ($r_p =-0.52$,  42 GCs), both significant to more than
99\%.

Thus this correlation is not due to selection bias. Moreover, since we included
GCs with all morphologies of HB in our sample, it is not biased against HB type
or log T$_{\rm eff}^{\rm max}$(HB). It follows that the correlation we found
between IQR[O/Na] and $M_V$ is real. 

The explanation of this correlation is likely quite complex. On one hand, it is
clear that only massive enough cluster can have a second generation. However, i)
(almost) all GCs are massive enough (while open clusters likely are not, see
Fig.~\ref{f:AgeMv}), and ii) IQR[O/Na] does not depend on the fraction of second
generation stars (that is on I+E, or P=1-(I+E)). Rather, IQR[O/Na] is correlated
with E, and anticorrelated with I (see left panels of Fig.~\ref{f:iqrPIE}). This
indicates that what is related to $M_V$ is the extreme of the  anti-correlation,
not its median value. This has been already found in Paper VII and VIII, where
we showed that O$_{\rm min}$\ and Al$_{\rm max}$\ are strongly correlated with a
combination of $M_V$\ and [Fe/H]. In these papers we connected this fact to the
typical mass of the polluters: the larger this mass, the higher the H-burning
temperature, the lower is O$_{\rm min}$ and the higher is Al$_{\rm max}$. 

Hence, available data indicate that the typical polluter mass is related to
$M_V$ (i.e., total cluster mass). To explain this relation, either the maximum
or the minimum mass (or both) defining the range of polluters is changing. This
might be understood quite easily in the case of AGB polluters. The maximum mass
may be reduced if the cooling flow is delayed due to the presence of adjacent
regions where massive stars are still present, as expected e.g. from formation
of clusters in spiral arm. The minimum mass may be larger in more massive
clusters: the larger the cluster mass, the faster the critical mass needed for
second generation star formation is reached, before the earliest second
generation SNe explode, halting further star formation. In the case of massive
stars polluters, the arguments are more complex but similar: only stars formed
by the ejecta of the most massive stars may contribute if the cluster mass is
very large, perhaps due  to devastating effect of super-winds blowing in very
massive clusters.

However, the scatter we observe in Fig.~\ref{f:iqrMv} suggests that high mass
alone is not a sufficient condition. 

Clusters lying off the global relation, at the left edge, namely NGC~288,
NGC~6218, (M~12) and NGC~6838 (M~71) might be reconciled with the bulk of other
clusters if we assume that they lost a higher than average  fraction of their
original mass (we already mentioned in Sect. \ref{proto} arguments suggesting
that a noticeable fraction of the original mass of GCs is lost after  the
formation phase). We searched the literature for observational evidence.  De
Marchi et al. (2006) argue for severe tidal stripping in NGC~6218, and they
estimate that the present mass of this cluster might even be only one fifth of
the original one. The flat mass function of NGC~6218 is also consistent with a
large fraction of stars lost by evaporation or tidal stripping. A tidal tail is
also suspected to be associated to NGC~288 (Leon et al. 2000), although other
studies (Kiss et al. 2007) found no extended extra-tidal structures. The most
intriguing evidence of large mass loss in these later phases is maybe the one
for M~71. It comes from a totally  independent line of thought. With its total
absolute magnitude of $M_V=-5.6$ (Harris 1996), this cluster is the less
massive and leftmost object in Fig.~\ref{f:iqrMv}. Elsner et al. (2008)
published a study based on $Chandra$ X-ray observations of M~71; they found
that there is an excess of sources for the present cluster mass, with respect
to the relation defined by 47~Tuc, M~4 and NGC~6397. Another way to state the
problem is that in order to bring M~71 on the relation given by the other
clusters one must assume that 50 to 70\% of its original mass was lost in the
past. Elsner et al. use scaled values of the mass $M_h$ inside the half-mass
radii and a relation by Kong et al. (2006) linking $M_h$ to the cluster
absolute visual magnitude. Using the same relation, but assuming that the
scaled mass of M~71 is 1 instead of 0.3, we get a value of $M_V=-7.2$ which
would bring this cluster perfectly on the relation between IQR[O/Na] and $M_V$.

On the other hand, even if accounting for less massive clusters is perhaps
possible, some problems are left also at the high mass end. A large mass is not
always matched by a very extended Na-O anticorrelation. NGC~104 (47~Tuc) is a
notable  example: this  cluster simply does not show very O-poor stars and
presents a short/normal Na-O anticorrelation even if it is a very massive
object. Some other factors must be involved. We reconsider the case of 47~Tuc
in Gratton et al. (2010)

\section{The next level of the game and conclusions}\label{fin}

\begin{figure*}
\centering
\includegraphics[bb=30 180 550 695, clip,scale=0.44]{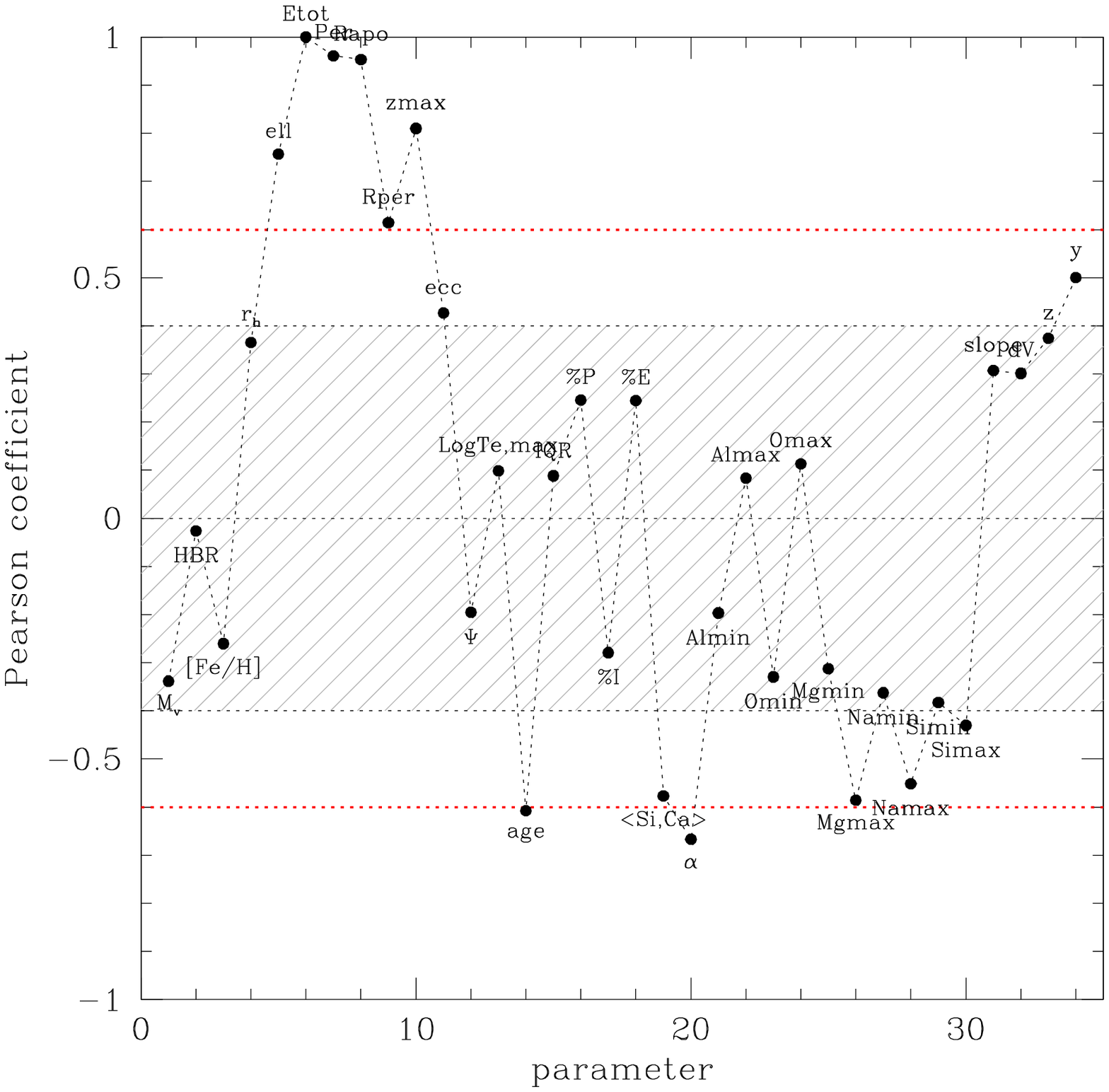} 
\includegraphics[bb=30 180 550 695, clip,scale=0.44]{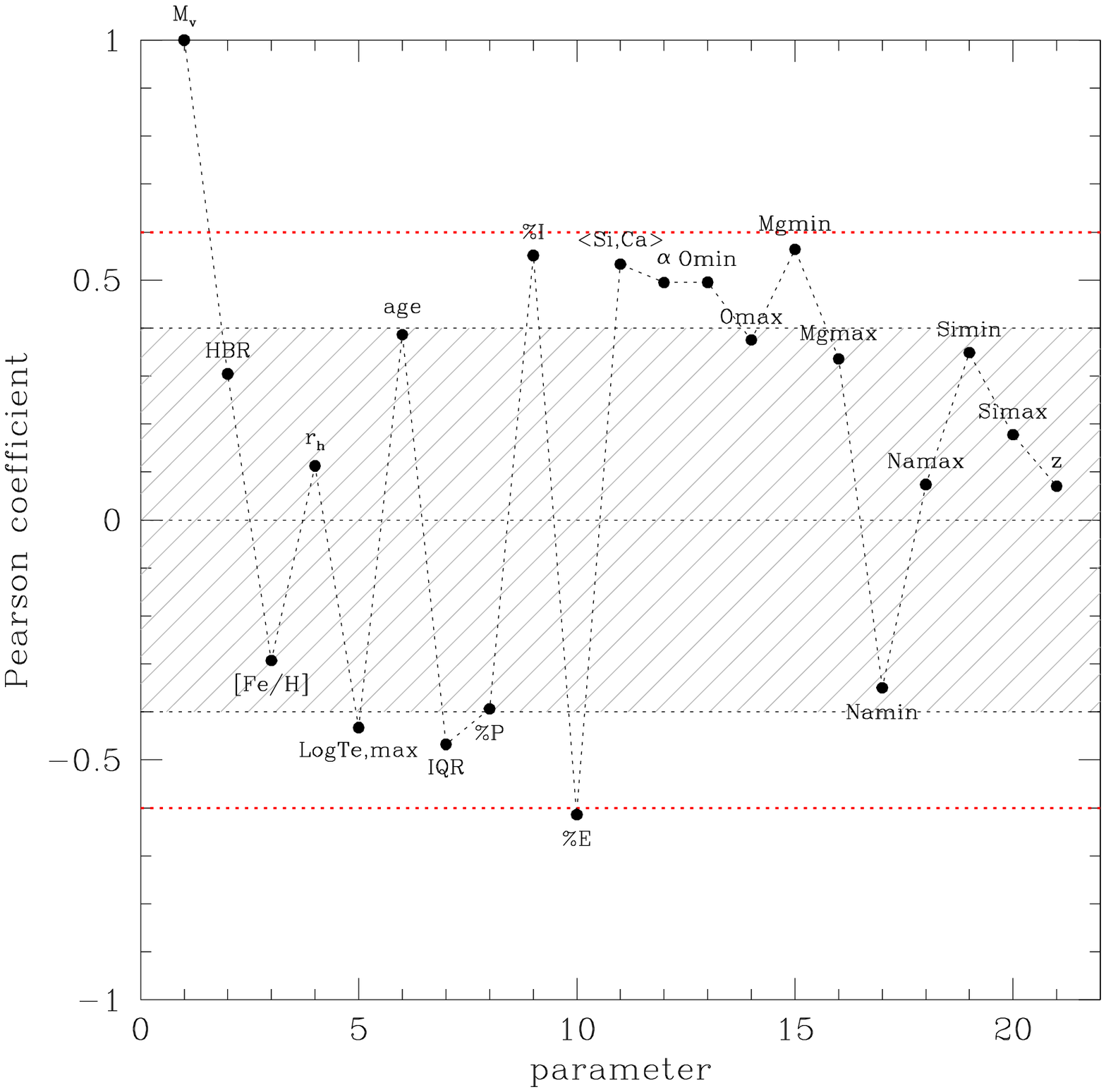}
\includegraphics[bb=30 180 550 695, clip,scale=0.44]{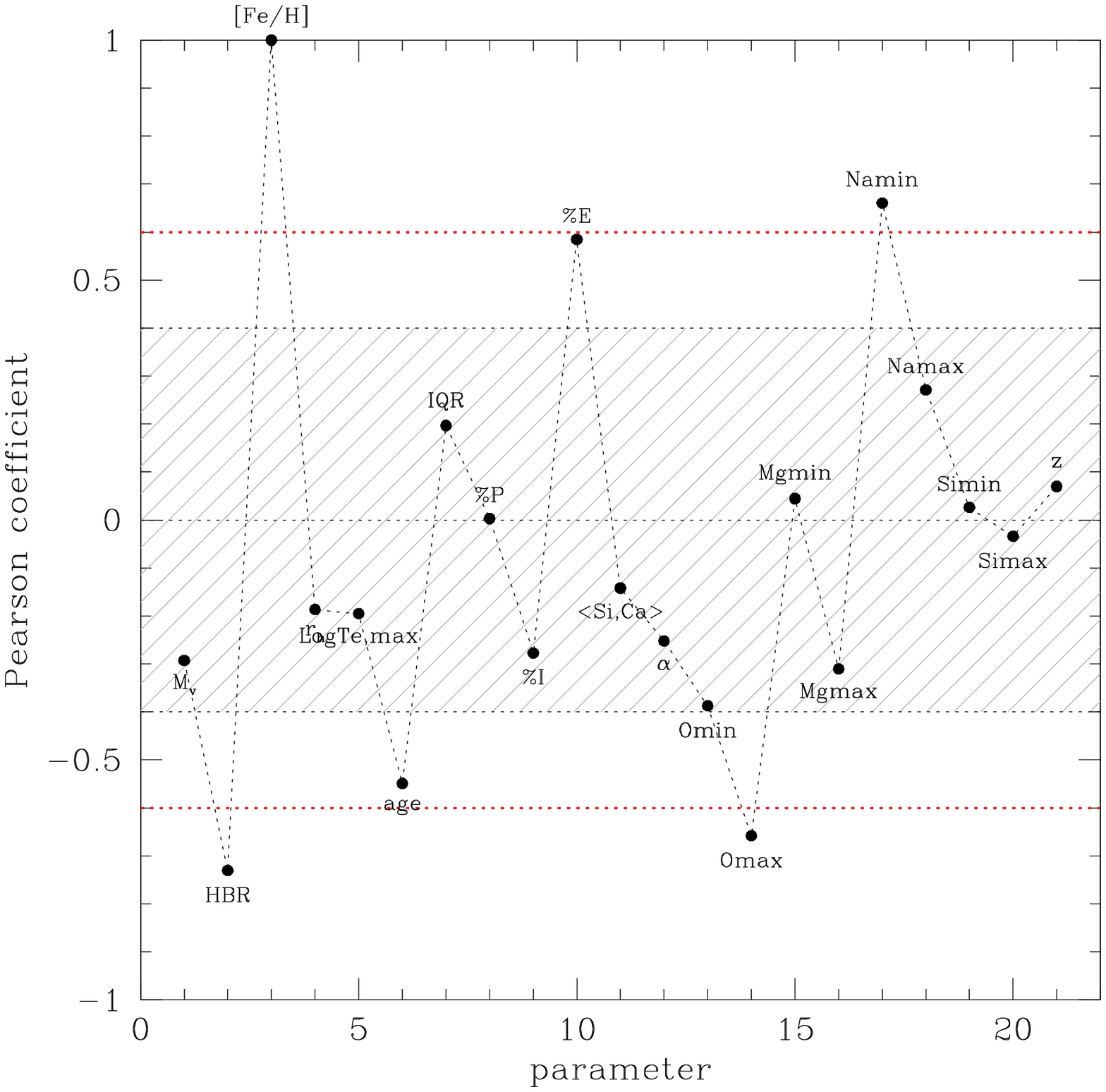}
\includegraphics[bb=30 180 550 695, clip,scale=0.44]{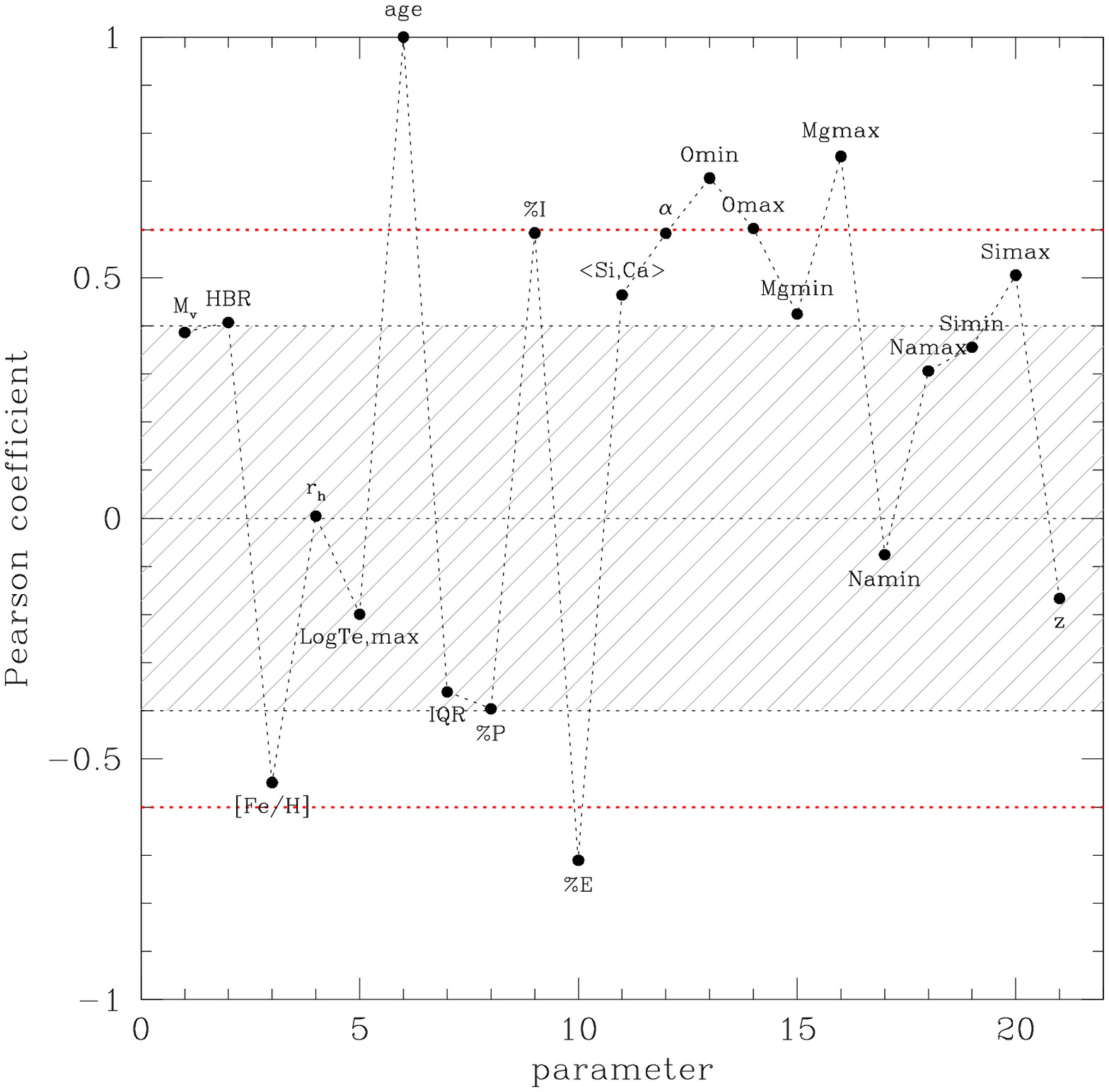} 
\caption{Example of the  linear monovariate correlations found for  E$_{tot}$
(left upper panel), $M_V$ (right upper panel), [Fe/H] (left lower panel), and
age (right lower panel). The first is for 15 GCs, the three other for 19 GCs.
The grey zone indicates that there is no significant correlation, while the
horizontal lines are drawn at (anti-)correlations larger than 0.4 and 0.6, which
represent limits for a significance  better than about 90\% and 99\% for our
samples. }
\label{f:pca1}
\end{figure*}

In the previous sections, as well as in Paper VII and VIII, we have seen the key
r\^ole of the main parameters affecting the formation and early evolution of
GCs, namely (i) the mass, (ii) the metallicity, and (iii) the galactic
population/region to whom they belong (inner/outer halo vs disk). These main
parameters well summarise the history of a GC, in the context of the evolution
of the Galaxy itself. The correlations with chemical properties suggest
that both the initial conditions at cluster's birth and the following evolution
while orbiting the Galaxy determine the resulting ratio of first and second
generation stars and several other properties.

The next step is to systematically explore all the possible relations
between the chemical signatures (in particular of second generation stars) of
GCs and global cluster parameters, to see whether they are independent relations
or may be explained with a combination of the three main parameters seen above.
We can divide the parameters in five broad groups: i) structural parameters
(including  HBR, concentration $c$, ellipticity, $r_h$, $r_t$, and  $\log$
T$_{\rm eff}^{max}$(HB)); ii) orbital parameters and/or parameters depending on
the location in the Galaxy (R$_{GC}$, $|Z|$, E$_{tot}$, age, $M_V$,
[Fe/H]\footnote{[Fe/H] was included in this group to take into account that
inner halo GCs are on average more metal-poor than disk clusters.}); iii)
primordial abundances ([(Mg+Al+Si)/Fe] and average [$\alpha$/Fe]) and first
generation ones (Mg$_{max}$, Si$_{min}$, Al$_{min}$, Na$_{min}$, O$_{max}$, plus
fraction of P stars); iv) chemical parameters of second generation stars
(Mg$_{min}$, Si$_{max}$, Al$_{max}$, Na$_{max}$, O$_{min}$, and fractions of I
and E stars); and  v) parameters linking first and second generation stars
(slope of the [Al/Fe] vs [Na/Fe] ratios and IQR[O/Na])\footnote{We used here all
the possible indicators of chemistry in first and second generation stars
because some are affected by saturation (e.g., Na$_{min}$ and O$_{max}$) and
others only show a small variation from the original value (e.g., Si).}.

The GC family is a complex one, with properties depending on many parameters,
often interconnected (see e.g., Djorgovski and Meylan 1994). While finding
dependencies of a particular cluster property on others may be a viable
approach, a more comprehensive and detailed analysis should take into account
all clusters' parameter using a multivariate approach, so we applied the
Principal Component Analysis (PCA) on our set. This method has been used many
times to understand the properties of the GC system (e.g., Fusi Pecci et al.
1993; Djorgovski and Meylan 1994; Recio-Blanco et al. 2006) but this is the
first time that it is also extended to the properties of the different stellar
generations in GCs. The method finds the correlation matrix of the whole
parameter set 
and defines the dimensionality of the data set, i.e., the minimum number of
dimensions required to fully explain the data. 

Unfortunately, a  rigourous study using the PCA is hampered by  i) the limited
sizes of the sample, 19 GCs only, since this is the only set for which detailed
and homogeneous abundances of first and second generation stars are available at
present; and ii) the fact that not all parameters are available for all (neither
for the same) clusters.  We first tried it on the maximum sample for which all
the 34 above mentioned  parameters are available. This means 15 GCs: NGC~6388
and NGC~6441 were excluded because they do not have orbital information,
NGC~6397 because we did not measure Al, and NGC~288 because it lacks a measure
of ellipticity in Harris (1996).  However, excluding four clusters could result
in loss of significance in some cases. For instance, $\alpha$-elements show a
good correlation with $M_V$ in our complete sample, but not in the reduced one,
possibly because two very massive clusters are excluded. We repeated the
analysis using the whole GC sample, excluding those parameters that were absent
for at least one of the GCs; this was possible for 21 parameters. 

We give in Table~\ref{t:pca2} (available only on-line) the complete list of the 
correlation coefficients among the 21 parameters. We show in Fig.~\ref{f:pca1} 
four examples, one taken from the first computation (E$_{tot}$) and three from
the second  one ($M_V$, [Fe/H], and age); the correlation coefficients are for
linear relations. Table~\ref{t:pca}  gives information on the dimensionality 
(the number of eigenvalues larger than 1, the limiting value usually assumed in 
this kind of analysis),  and on the fractional and cumulative contributions of 
the components.  This Table refers to both cases: 34 parameters for 15 GCs
(first four columns) and 21  parameters for 19 GCs (last four columns). In the 
first case, the dimensionality is eight, the first three eigenvalues account 
only for 62\% of the total variance, while eight components account for 93 \%. 
In the second case,  the dimensionality is five (or six), the first three 
eigenvalues account  for 66\% of the total variance, while five (six)
components  account for 83 (87) \%.

\begin{table}
\setcounter{table}{5}
\centering
\caption{Eigenvalues, fractional, and cumulative contributions for
the two cases discussed in the text.}
\begin{tabular}{rrrrrrrrccrrrrrrrr}
\hline
 \multicolumn{4}{c}{34 parameters, 15 GCs} && 
&\multicolumn{4}{c}{21 parameters, 19 GCs}\\
$i$~ &$e_i$~~ &$V_i$~~ &$C_i$~~ &&&$i$~ &$e_i$~~ &$V_i$~~ &$C_i$~~ &\\
\hline
 1 & 11.35 & 33.39 & 33.39 & & &  1 & 8.25 & 39.30 & 39.30 \\ 
 2 &  5.19 & 15.28 & 48.67 & & &  2 & 3.61 & 17.18 & 56.48 \\ 
 3 &  4.52 & 13.28 & 61.95 & & &  3 & 2.11 & 10.05 & 66.53 \\ 
 4 &  2.99 &  8.79 & 70.73 & & &  4 & 1.86 &  8.83 & 75.36 \\ 
 5 &  2.40 &  7.06 & 77.79 & & &  5 & 1.56 &  7.51 & 82.87 \\ 
 6 &  2.06 &  6.17 & 83.96 & & &  6 & 0.96 &  4.57 & 87.44\\ 
 7 &  1.76 &  5.17 & 89.13 & & &  7 & 0.69 &  3.20 & 90.74 \\ 
 8 &  1.16 &  3.42 & 92.55 & & &  8 & 0.61 &  2.91 & 93.65 \\ 
\hline
\end{tabular}
\label{t:pca}
\end{table}

Besides monovariate relations, we tried a semiempirical approach to explore the
dependencies of cluster parameters from the fundamental quantities. 
In particular, we wanted to test whether the key parameters mass, metallicity,
age, and orbital elements (the last two representing the position in the Galaxy)
are actually those giving the most significant relations with structural and
chemical parameters of GCs.  
We derived the correlations for parameters chosen in all the groups defined above
as a function of all bi-variate combinations of $M_V$, [Fe/H], age, and
$E_{tot}$; then, we counted the frequency of the most significant correlations
(larger than 99\%) for each combination in each group.
The results are are given in 
Table~\ref{t:c} (on-line only) and are summarised in Table~\ref{t:ria}.
The best combinations are essentially two: $M_V-$age and [Fe/H]-age. 

\begin{table}
\setcounter{table}{7}
\centering
\caption{Number of significant correlations with different class of parameters.}
\setlength{\tabcolsep}{1.5mm}
\begin{tabular}{lcccccr}
\hline
 &Struct. &Orbit./ &Primord. &Chem. &Chem. &Total \\
 &          &posit. &abund. &of FG     &of SG     &      \\
\hline
$M_V$        	     &0     &0     &0	   &0	 &1	& 1 \\
${\rm [Fe/H]}$       	     &1     &0     &0	   &2	 &0	& 3 \\
age          	     &0     &1     &2	   &2	 &3	& 8 \\
E$_{tot}$    	     &0     &1     &2	   &0	 &0	& 3 \\
R$_{GC}$             &0     &1     &0      &0    &0     & 1 \\
$M_V$,${\rm [Fe/H]}$         &1     &0     &0	   &2	 &2	& 5 \\
$M_V$,age            &1     &1     &2	   &2	 &4     &10 \\
$M_V$,E$_{tot}$      &3     &1     &2	   &1	 &0     & 7 \\
$M_V$,R$_{GC}$       &0     &1     &0	   &0	 &3     & 4 \\
${\rm [Fe/H]}$,age           &1     &1     &2      &3    &3     &10 \\   
${\rm [Fe/H]}$,E$_{tot}$     &2     &1     &2      &1    &0     & 6 \\
${\rm [Fe/H]}$,R$_{GC}$      &0     &1     &0      &2    &1     & 4 \\
E$_{tot}$,age        &2     &0     &2      &2    &2     & 8 \\
R$_{GC}$,age         &0     &1     &2      &2    &3     & 8 \\
R$_{GC}$,E$_{tot}$ &1     &1     &2      &1    &0     & 5 \\
\hline
\end{tabular}
\label{t:ria}
\end{table}

Obviously, this  is meaningful only
if the main parameters we are using are independent. From the monovariate
correlations we see that $M_V$ is not correlated with age.
A significant correlation does instead exist between  [Fe/H] and age.
This is  evident both in our sample (Fig.~\ref{f:pca1}) and in the global one of galactic GCs
(Fig.~\ref{f:agefeh}), and it 
is stronger when individual sub-populations (inner halo vs disk/bulge) are
considered. However, the age-metallicity relation is a physical one, intrinsic
to the system of GCs, and not a spurious bias affecting our sample.
We may safely conclude that most of the ``phenotypes" of GCs can be explained
rather well by the variations of three fundamental parameters: total mass,
metallicity, and age, linked by the origin and the interaction of GCs in, and with, our Galaxy.

In summary, in this paper we combine the results of our extensive survey of
abundances in RGB stars in 19  GCs (Carretta et al. 2009a,b,c)
with previous knowledge of globular 
cluster, in order to discuss scenarios for their formation. The novelty of our
approach is to fully take into account the fact that GCs cannot
anymore be regarded as Simple Stellar Populations. For the first time it is
possible to include among the properties of GCs a quantitative estimate of the
ratio and of the chemical composition of different stellar generations.
Our main findings are:

\begin{itemize}
\item[(i)] We first analyse the definition of globular clusters. The presence of 
the Na-O anticorrelation may be very well used to separate GCs 
from smaller (open) clusters. Second, we divide GCs according to 
their kinematics and location in the Galaxy in three populations: disk/bulge, 
inner halo, and outer halo. We find that the LF of bona fide 
GCs (that is,  those which exhibit the Na-O anticorrelation) is 
fairly independent on their population. This suggests that it is imprinted by the 
formation mechanism, and only marginally affected by the following evolution.

\item[(ii)] We then use the evidence of different generations within GCs given 
by their chemistry, and consider separately the composition of 
the primordial population and of the second generation. A large fraction of the 
primordial population should  have been lost by the proto-globular clusters. We 
propose that the fraction of  primordial population stars lost by GCs make up 
the main component of  halo field stars. Arguments in favour 
include the total number of stars, the metallicity, kinematic and density 
distribution, and the chemistry.

\item[(iii)] In addition, we argue that the extremely low Al abundances found 
for the primordial  population of massive GCs is an indication of 
a very fast enrichment  process before the formation of the primordial 
population. We then suggest a scenario for the formation of GCs 
including at least three main phases: a) the formation of a precursor population 
(likely due to the interaction  with the early Galaxy or with other  cosmological 
structures similar to those that led to the formation of dwarf spheroidals,
but residing at smaller Galactocentric distances), b) which 
triggers a large episode of star formation (the primordial population), and c) then 
the formation of the current GC, mainly within a cooling flow 
formed by the slow winds of a fraction of this primordial population. Some stars 
of the primordial population remains trapped in the newly forming cluster, producing the primordial component still observed in GCs. 
The precursor population is very effective in raising the metal content in 
massive and/or metal-poor (mainly halo) clusters, while its r\^ole is minor in 
small and/or metal rich (mainly disk) ones.

\item[(iv)] We then re-examine the second phase of metal-enrichment (from 
primordial to second generation): we consider monovariate, bivariate relations, 
and an explorative  PCA. Our conclusion is that most of 
the ``phenotypes" of globular clusters (and in particular their detailed 
chemical characteristics) may be ascribed to a few parameters, the most 
important being metallicity, mass, and age of  the cluster. Location within the 
Galaxy (as described by the kinematics) also plays some  r\^ole, while 
additional parameters are required to describe their dynamical status. 
\end{itemize}

The proposed scenario for the origin of GCs offers a first framework to
interpret the increasing information coming from spectroscopy and
photometry of GCs, and implicitly suggests several tests that can be performed.
Some are currently undertaken by our and other groups. Is it possible to
reproduce the observed features of the HBs in term of the fundamental parameters
mass, metallicity and age (see e.g., Gratton et al. 2010)? Is the proposed scenario
(precursor-present GC hosting multiple stellar generations)  valid also outside
the typical mass range of GCs? We are presently exploring both the very low-mass end, 
with  the old open cluster \object{NGC~6791}, and the high-mass end,
with  \object{M~54} (NGC~6715); results will be published in the next future.

While corroborating evidence is coming from these studies,  several issues are
still dramatically open and/or poorly explored. A full hydrodynamical treatment
of the formation and early evolution of GCs is still missing, the relation
between GCs and past and present dSphs needs to be well studied, the nature and
the precise yields of candidate first generation polluters has still to be
definitively assessed, the r\^ole of binaries has to be well understood,
to mention only a few open problems. Although in the past few years the road
to a deeper understanding of the globular clusters has been opened, the way is
still long.

\begin{acknowledgements}
We warmly thank Michele Bellazzini for useful discussions and a careful reading 
of the manuscript. The comments of the referee were very useful to produce a
clearer manuscript. AB thanks the  Observatoire de la C\^ote d'Azur for the 
hospitality during the preparation of the paper. This research has made use of 
the SIMBAD database, operated at CDS, Strasbourg, France, and of NASA's 
Astrophysics Data System. This work was partially funded by the Italian MIUR 
under PRIN 2003029437 and PRIN 20075TP5K9, and by INAF by the grant INAF 2005 
``Experimenting nucleosynthesis in clean environments". SL acknowledges the 
support by the DFG cluster of excellence ``Origin and Structure of the 
Universe".
\end{acknowledgements}

\begin{appendix}

\section{Classification of Galactic globular clusters}

Since the work by Zinn (1985) a few progresses were done with respect to his
criteria for the separation of the sub-populations of Galactic globular clusters.
Many classification schemes rest on the appearance of the clusters' CMD and  related
parameters (namely metallicity, age and HBR index). However, one of  the main
aim of our project is to explain the HB morphology and its  relations with
chemical signatures of stellar generations in GC, thus we cannot use  the
distribution of stars along the HB as a separation criterion. In 
Tab.~\ref{t:class} we list the quantities used in Sect. \ref{mwgc} to separate
disk/bulge clusters from the halo ones according to the combination of their
location in the Galaxy and their kinematics.

\begin{figure}
\centering
\includegraphics[scale=0.40]{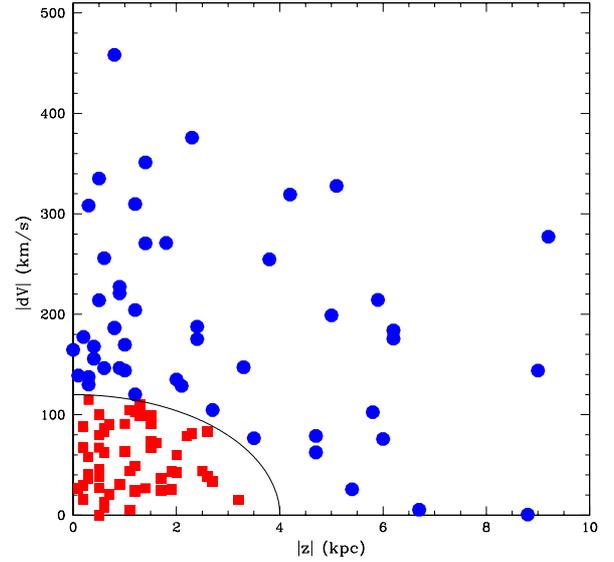}
\caption{Classification of disk (red squares) and inner halo (blue dots)
clusters. The curve is the discriminating line as obtained from our selection
criteria.}
\label{f:dvz}
\end{figure}

\begin{figure}
\centering
\includegraphics[scale=0.40]{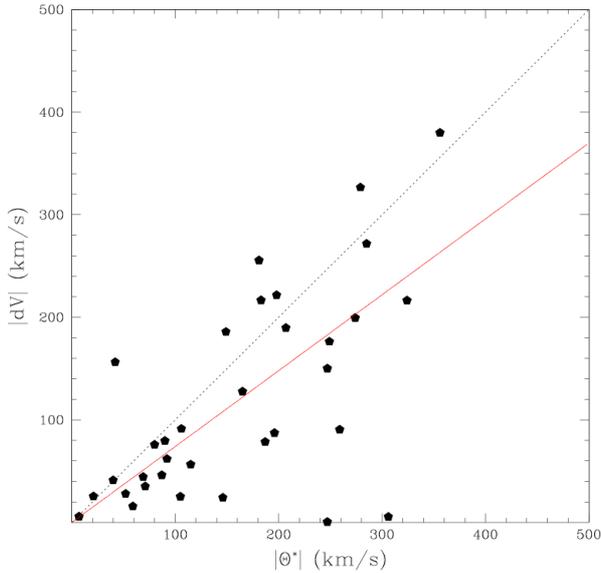}
\caption{Absolute values of the difference $dV$ between the observed radial
velocity of globular clusters and the one expected from the galactic rotation
curve, as a function of the rotational velocity given by Dinescu et al. (1999) 
and Casetti-Dinescu et al. (2007). The dotted line indicates one-to-one 
correlation, while the red solid lines indicates the linear regression.
}
\label{f:vdinescu}
\end{figure}

As said in Sect. \ref{mwgc},  outer halo clusters were simply classified as those
currently located at more than 15 kpc from the centre of the Galaxy (see Carollo
et al. 2008). Clusters with R$_{GC}$ below 3.5 kpc were
considered as bulge GCs, even though some of them  might be halo clusters on
very elongated orbits presently close to the pericentres.  To separate the
remaining clusters  in inner halo and disk GCs, we computed the differences
($dV$, column 8  in Tab.~\ref{t:class}) between the observed radial velocity 
(corrected to  the LSR) and the one expected from the Galactic rotation curve 
(see Clemens 1985). In the $dV-Z$\ plane (where $Z$ is the clusters' distance
from the Galactic plane, in kpc, see Col. 10 of Tab.~\ref{t:class}) we defined
an ellipse with equation:
$$y=\left(\frac{|dV|}{120}\right)^2+\left(\frac{|z|}{4}\right)^2.$$ 
Clusters with y$<$1 were classified as disk GCs, while we considered as halo GCs
the ones with y$>$1, see Fig~\ref{f:dvz}. Of course, a better estimate of GC
kinematic is possible when the whole orbit is available. Given the quite good
correlation (see Fig.~\ref{f:vdinescu}) between $dV$  and the rotational velocity
(Col. 9 of Tab.~\ref{t:class}, where $\Theta^{\star}$=$\Theta$-220) given by
Dinescu et al. (1999) and Casetti-Dinescu et al. (2007), we replaced, when
available, our $dV$ values with the ones provided from those studies. Since we
find that $|dV|=(0.74\pm 0.06)|\Theta^{\star}|$, the definition of $y$ was in
this case: 
$$y=\left(\frac{|\Theta^{\star}|}{162}\right)^2+\left(\frac{|z|}{4}\right)^2.$$
It should however be clear that $dV$ is not at all a synonymous of
($\Theta$-220), and the existing correlation has only a statistical meaning.

Column 11 in Tab.~\ref{t:class} shows our classification for the population of
the Galactic globular clusters in disk/bulge (D/B), inner halo (IH), outer halo
(OH) and dwarf Spheroidal's clusters (dSph). For each cluster we  report the
integrated magnitude (M$_{\rm V}$, Col. 2), the horizontal branch morphology
parameter (HBR, Col. 6), and the Galactocentric distance (R$_{GC}$, Col. 7)
as directly retrieved from Harris' catalogue. The metallicity  values ([Fe/H] ,
Col. 3)  were instead replaced  with
the determinations by Carretta et al. (2009c). Additionally, we list in Col. 4
the [$\alpha$/Fe] ratios (the corresponding references are given in the table
notes). 
 Column 5 displays the age parameter: more in detail, we computed an
average value between the two different estimates by Marin-Franch et al. (2009)
and De Angeli et al. (2005), after applying a correction of 0.08 to the second
ones for GCs with [Fe/H] ranging from -1.8 to -1.1 dex, as suggested by a
cluster-to-cluster comparison.  When neither of these estimates were available,
we adopted the ones calculated by Vandenberg (2000), normalised to the
Marin-Franch scale by assuming that 13.5 Gyr = 1.00. We then corrected the values
so obtained for the difference between the metallicities considered in those
papers and those listed in Carretta et al. (2009c), transformed into [M/H]
using an average [$\alpha$/Fe] of +0.4 (see Table~\ref{t:tabnoi}). This
correction was made using the sensitivity of age on metallicity given by
Marin-Franch et al. (2009).

First, we decided to compare our new classification with the previous ones
relying only on metallicity and HB morphology (see Sect.~\ref{mwgc}): as
representative of this approach, we choose the work by Mackey \& van den Bergh
(2005).  Briefly, they defined as disk component all the GCs with
[Fe/H]$>$$-$0.8 dex; the so-called ``old" halo and ``young" halo clusters were
then divided following Zinn (1993),  namely by computing the offset in HB type
-at a given metallicity- with respect to the fiducial line of the inner halo 
clusters\footnote{Note that both Zinn (1993) and Mackey \& van den Bergh (2005) 
labelled inner halo GCs the ones more metal-poor than [Fe/H]=$-$0.8 dex
and located at Galactocentric distance less than 6 kpc}.   GCs with an offset
larger and smaller than $-$0.3 in HB type were classified as old halo and young
halo, respectively. The Table ~\ref{t:matrix} schematically shows the comparison
and emphasises the different nature of the two classifications. As (partially) 
expected, the matrix is not diagonal, i.e. there is not a one-to-one correlation
between old halo (young halo) and inner halo (outer halo) subgroups.  More in
detail, the metallicity criterion is largely responsible for such a discrepancy:
had we adopted the requirement  of [Fe/H]$>$$-$0.8 dex,  35 of the 36
clusters that we classified as disk/bulge and Mackey \& van den Bergh as old
halo, should be moved into  inner halo+old halo cell;  while all the 5 GCs
placed in the disk+young halo box should become inner halo+young halo clusters
thanks to their low metallicity. 
However, and most important, even taking into
account these changes in the relative population  of the matrix cells, the
resulting correspondence is not yet one-to-one: as to the inner halo GCs, 
80\% constitute the old halo and the remaining 20\% the young halo clusters; 
for the outer halo GCs, the promiscuity is even larger resulting in 40\% and
60\%, respectively for old and young halo.     This is the direct evidence of
the strong difference between kinematics (and/or positional) criteria and the
ones based only on metallicity and HB type.  

One last word on the
classification. While this work was in preparation a paper by
Fraix-Burnet et al. (2009) appeared; they used a cladistic technique to divide a sample of
54 GCs into three sub-samples (called Group 1,  2, and 3 and  later identify
with inner halo, outer halo, and disk, respectively) on the basis of [Fe/H],
M$_V$, T$_{eff}^{max}(HB)$, and age. We cross-checked the assignments for the
clusters in common and found a good agreement only for disk clusters, and also
this in a limited sense. When they classify a cluster as disk, we agree (in 17
cases out of 18), but we have many other disk clusters that they instead
classify in the halo sub-samples. In particular, for the 19 GCs in our FLAMES
sample, the two classifications agree for seven clusters and disagree for seven
others (five GCs are not present in their data set). We think that the main
factor producing this difference is in the fact that they ignored the
kinematics, although the information is present for their sample, while our
method rests on that. 

To conclude, we report in  Table~\ref{t:class2} the analogous of
Table~\ref{t:class}  but for the LMC, SMC and Fornax globular cluster system; as
in the previous case, the number in  brackets corresponds to the reference (for
[$\alpha$/Fe], [Fe/H], age and HBR) whose decoding is given in the Notes.  The
integrated magnitudes $M_V$ for the LMC and Fornax GCs are taken from van
den Bergh \& Mackey (2004),  while for the SMC were computed from the apparent
magnitudes {\it UBV} by van den Bergh (1981) along with the distance moduli as 
given in those papers providing the clusters' age. (ref. 15, 16, 17 see
Table~\ref{t:class2}).  As to age, for LMC and SMC GCs, since absolute values
were available  (see ref. given in Table~\ref{t:class2}) we report them to our
relative scale  adopting the previous conversion of 13.5 Gyr = 1.00.  For the
Fornax clusters, the ages were instead derived starting from the cluster-to-cluster  relative differences ($\Delta$Age) obtained   by Buonanno et al. (1998,
1999) and assuming that 1.05 = 14.2 Gyr.

\begin{scriptsize}
\begin{longtable}{lccccccccccl}
\caption{}\\
\hline
\hline
     Name   &	 $M_V$   & [Fe/H] & [$\alpha$/Fe]  &  $<$Age$>$ &  HBR &  R$_{GC}$ & dV 	&	$\Theta^{\star}$	&  Z  & 	Pop.\\ 
           &                   &                  &                  &             &     & (kpc) & (km/s)  &  (km/s)  & (kpc) &   \\ \hline
\endfirsthead
\caption{continued.}\\
\hline
\hline  
     Name   &	 $M_V$   & [Fe/H] & [$\alpha$/Fe] &   $<$Age$>$ & HBR &   R$_{GC}$ & dV 	&	$\Theta^{\star}$	&  Z  &  Pop.\\ 
           &                   &                  &                  &             &     & (kpc) & (km/s)  &  (km/s)  & (kpc) &   \\ \hline
\endhead
\hline\endfoot
   NGC 104    &  -9.42  & -0.76  &   ~0.42(1) &   0.95 & -0.99  & ~~7.4  & ~-15  &  ~59 & ~-3.2  & D/B\\       
   NGC 288    &  -6.74  & -1.32  &   ~0.42(1) &   0.90 & ~0.99  & ~12.0  & ~~~1  &  247 & ~-8.8  & IH\\        
   NGC 362    &  -8.41  & -1.30  &   ~0.30(2) &   0.80 & -0.87  & ~~8.9  & -176  &  249 & ~-6.2  & IH \\       
   NGC 1261   &  -7.81  & -1.27  &   ---      &   0.79 & -0.71  & ~18.2  &  ---  &  --- & -12.9  & OH \\       
   Pal 1      &  -2.47  & -0.51  &   ---      &   0.52 & -1.00  & ~17.0  &  ---  &  --- & ~~3.6  & OH \\       
   AM1	      &  -4.71  & -1.84  &   ---      &    --- & -0.93  & 123.2  &  ---  &  --- & -91.3  & OH\\        
   Eridanus   &  -5.14  & -1.44  &   ---      &   0.71 & -1.00  & ~95.2  &  ---  &  --- & -59.6  & OH \\       
   Pal 2      &  -8.01  & -1.29  &   ---      &    --- &  ---	& ~35.4  &  ---  &  --- & ~-4.3  & dSph \\     
   NGC 1851   &  -8.33  & -1.18  &   ~0.38(3) &   0.81 & -0.36  & ~16.7  &  ---  &  87  & ~-6.9  & OH \\       
   NGC 1904   &  -7.86  & -1.58  &   ~0.31(1) &   0.89 & ~0.89  & ~18.8  &  ---  &  137 & ~-6.3  & OH \\       
   NGC 2298   &  -6.30  & -1.96  &   ~0.50(4) &   1.01 & ~0.93  & ~15.7  &  ---  &  246 & ~-3.0  & OH \\       
   NGC 2419   &  -9.58  & -2.20  &   ~0.20(5) &    --- & ~0.86  & ~91.5  &  ---  &  --- & ~35.9  & OH \\       
   Pyxis      &  -5.75  & -1.33  &   ---      &    --- & -1.00  & ~41.7  &  ---  &  --- & ~~4.8  & OH \\       
   NGC 2808   &  -9.39  & -1.18  &   ~0.33(1) &   0.83 & -0.49  & ~11.1  & ~-26  &  146 & ~-1.9  & IH \\       
   E3	      &  -2.77  & -0.73  &   ---      &   0.92 & ---	& ~~7.6  &  ---  &  --- & ~-1.4  & D/B \\      
   Pal 3      &  -5.70  & -1.67  &   ---      &   0.83 & -0.50  & ~95.9  & ---   &  72  & ~61.8  & OH \\       
   NGC 3201   &  -7.46  & -1.51  &   ~0.33(1) &   0.82 & ~0.08  & ~~8.9  & -458  &  521 & ~~0.8  & IH \\       
   Pal 4      &  -6.02  & -1.46  &   ---      &   0.76 & -1.00  & 111.8  &  ---  &  --- & 103.7  & OH\\        
   NGC 4147   &  -6.16  & -1.78  &   ~0.38(6) &   0.96 & ~0.55  & ~21.3  &  ---  &  174 & ~18.8  & dSph\\      
   NGC 4372   &  -7.77  & -2.19  &   ---      &   1.00 & ~1.00  & ~~7.1  & ~-91  &  106 & ~-1.0  & D/B  \\     
   Rup 106    &  -6.35  & -1.78  &  -0.03(7)  &   0.79 & -0.82  & ~18.5  & ---   &  --- & ~~4.3  & OH \\       
   NGC 4590   &  -7.35  & -2.27  &  ~0.35(1)  &   0.94 & ~0.17  & ~10.1  & ~~76  &  -80 & ~~6.0  & IH\\        
   NGC 4833   &  -8.16  & -1.89  &   ---      &   1.01 & ~0.97  & ~~8.7  & -221  &  198 & ~~5.0  & IH\\        
   NGC 5024   &  -8.70  & -2.06  &   ---      &   1.04 &  ~0.81 & ~18.3  & ---   &  -18 & ~17.5  & OH \\       
   NGC 5053   &  -6.72  & -2.30  &   ---      &   1.01 &  ~0.52 & ~16.9  & ---   &  --- & ~16.1  & OH\\        
   NGC 5139   & -10.29  & -1.64  &    ---     &    --- &   ---  & ~~6.4  & -271  &  285 & ~~1.4  & IH\\        
   NGC 5272   &  -8.93  & -1.50  &  ~0.34(8)  &   0.88 & ~0.08  & ~12.2  & ~-56  &  115 & ~10.2  & IH  \\      
   NGC 5286   &  -8.61  & -1.70  &   ---      &   1.02 & ~0.80  & ~~8.4  & ~-60  &  --- & ~~2.0  & D/B\\       
   AM 4       &  -1.60  & -2.07  &   ---      &    --- &  ---	& ~25.5  & ---   &  --- & ~16.5  & OH\\        
   NGC 5466   &  -6.96  & -2.31  &   ~0.33(9) &   1.07 & ~0.58  & ~16.2  &---	 &  280 & ~15.2  & OH\\        
   NGC 5634   &  -7.69  & -1.93  &   ---      &    --- &   ---  & ~21.2  & ---   &  --- & ~19.1  & OH\\        
   NGC 5694   &  -7.81  & -2.02  &  ~0.16(10) &   1.05 & ~1.00  & ~29.1  & ---   &  --- & ~17.5  & OH \\       
   IC 4499    &  -7.33  & -1.62  &   ---      &    --- & ~0.11  & ~15.7  &---	 &  --- & ~-6.6  & OH \\       
   NGC 5824   &  -8.84  & -1.94  &  ---       &   1.00 & ~0.79  & ~25.8  &---	 &  --- & ~12.0  & OH \\       
   Pal 5      &  -5.17  & -1.41  &  ~0.17(11) &    --- & -0.40  & ~18.6  &---	 &  178 & ~16.7  & OH\\        
   NGC 5897   &  -7.21  & -1.90  &  ---       &   1.03 & ~0.86  & ~~7.3  & -184  &  149 &  ~6.2  & IH \\       
   NGC 5904   &  -8.81  & -1.33  &  ~0.38(1)  &   0.85 & ~0.31  & ~~6.2  & ~~26  &  105 & ~~5.4  & IH \\       
   NGC 5927   &  -7.80  & -0.29  &   ---      &   0.91 & -1.00  & ~~4.5  & ~~~8  &  ~-7 & ~~0.6  & D/B   \\    
   NGC 5946   &  -7.20  & -1.29  &  ---       &   0.93 &  ---	& ~~5.8  & -186  &  --- & ~~0.8  & IH\\        
   BH 176     &  -4.35  &  ---   &   ---      &   0.55 & -1.00  & ~~9.7  & ~~-3  &  --- & ~~1.2  & Open Cluster\\ 
   NGC 5986   &  -8.44  & -1.63  &   ---      &   0.97 & ~0.97  & ~~4.8  & -188  &  207 & ~~2.4  & IH	\\     
   Lynga 7    &  -6.37  & -0.68  &   ---      &    --- & -1.00  & ~~4.2  & -115  &  --- & ~-0.3  & D/B \\      
   Pal 14     &  -4.73  & -1.63  &   ---      &    --- & -1.00  &  69.0  &  ---  &  --- & ~49.6  & OH  \\      
   NGC 6093   &  -8.23  & -1.75  &  ~0.24(12) &   1.01 & ~0.93  & ~~3.8  & -147  &  247 & ~~3.3  & IH  \\      
   NGC 6121   &  -7.20  & -1.98  &  ~0.51(1)  &   0.97 & -0.06  & ~~5.9  & ~-87  &  196 & ~~0.6  & IH \\       
   NGC 6101   &  -6.91  & -1.18  &  ---       &   1.01 & ~0.84  & ~11.1  & -319  &  --- & ~-4.2  & IH  \\      
   NGC 6144   &  -6.75  & -1.82  &  ---       &   1.08 & ~1.00  & ~~2.6  & -376  &  356 & ~~2.3  & D/B \\      
   NGC 6139   &  -8.36  & -1.71  &  ---       &   0.89 & ~0.91  & ~~3.6  & -120  &  --- & ~~1.2  & IH \\       
   Terzan 3   &  -4.61  &-0.755  &  ---       &    --- &  ---	& ~~2.4  & ~-23  &  --- & ~~1.2  & D/B\\       
   NGC 6171   &  -7.13  & -1.03  &   ~0.49(1) &   0.99 & -0.73  & ~~3.3  & ~-44  &  ~69 & ~~2.5  & D/B \\      
   1636-283   &  -3.97  &-1.525  &   ---      &    --- & ---	& ~~2.0  & ---   &  --- & ~~1.6  & D/B \\      
   NGC 6205   &  -8.70  & -1.58  &   ~0.31(8) &   1.02 & ~0.97  & ~~8.7  & -199  &  274 & ~~5.0  & IH  \\      
   NGC 6229   &  -8.05  & -1.33  &   ---      &    --- & ~0.24  & ~29.7  & ---   &  --- & ~~2.7  & OH  \\      
   NGC 6218   &  -7.32  & -1.43  &   ~0.41(1 )&   0.99 & ~0.97  & ~~4.5  & ~~78  &  ~90 & ~~2.2  & D/B   \\    
   NGC 6235   &  -6.44  & -1.38  &   ---      &    --- & ~0.89  &  ~4.1  & -105  &  --- & ~~2.7  & IH \\       
   NGC 6254   &  -7.48  & -1.57  &   ~0.37(1) &   0.92 & ~0.98  & ~~4.6  & ~-36  &  ~71 & ~~1.7  & D/B \\      
   NGC 6256   &  -6.52  & -0.62  &   ---      &    --- &  ---	& ~~1.8  & ~-79  &  --- & ~~0.5  & D/B   \\    
   Pal 15     &  -5.49  & -2.10  &   ---      &    --- & ~1.00  & ~37.9  &  ---  &  --- & ~18.3  & OH\\        
   NGC 6266   &  -9.19  & -1.18  &   ---      &   0.89 & ~0.32  & ~~1.7  & ~-31  &  --- & ~~0.9  & D/B \\      
   NGC 6273   &  -9.18  & -1.76  &   ---      &   0.99 & ~0.96  & ~~1.6  & -351  &  --- & ~~1.4  & D/B \\      
   NGC 6284   &  -7.97  & -1.31  &   ---      &   0.91 & ~0.88  & ~~7.6  & ~-39  &  --- & ~~2.6  & D/B \\      
   NGC 6287   &  -7.36  & -2.12  &  ~0.38(13) &   1.07 & ~0.98  &  ~2.1  &~~271  &  --- & ~~1.8  & D/B\\       
   NGC 6293   &  -7.77  & -2.01  &  ~0.42(13) &    --- & ~0.90  & ~~1.4  & ~-49  &  --- & ~~1.2  & D/B   \\    
   NGC 6304   &  -7.32  & -0.37  &   ---      &    --- & -1.00  & ~~2.2  & ~~63  &  --- & ~~0.6  & D/B  \\     
   NGC 6316   &  -8.35  & -0.36  &   ---      &    --- & -1.00  & ~~3.2  & -105  &  --- & ~~1.1  & D/B  \\     
   NGC 6341   &  -8.20  & -2.35  &  ~0.46(14) &   1.11 & ~0.91  & ~~9.6  & ~~79  &  187 & ~~4.7  & IH  \\      
   NGC 6325   &  -6.95  & -1.37  &   ---      &    --- &  ---	& ~~1.1  & ~~-5  &  --- & ~~1.1  & D/B\\       
   NGC 6333   &  -7.94  & -1.79  &   ---      &    --- & ~0.87  & ~~1.7  & ~-99  &  --- & ~~1.5  & D/B \\      
   NGC 6342   &  -6.44  & -0.49  &  ~0.38(15) &   0.92 & -1.00  & ~~1.7  & ~~74  &  --- & ~~1.5  & D/B \\      
   NGC 6356   &  -8.52  & -0.35  &   ---      &    --- & -1.00  & ~~7.6  & ~-33  &  --- & ~~2.7  & D/B  \\     
   NGC 6355   &  -8.08  & -1.33  &   ---      &    --- &   ---  & ~~1.8  & ~147  &  --- & ~~0.9  & D/B \\      
   NGC 6352   &  -6.48  & -0.62  &  ~0.20(16) &   0.87 & -1.00  & ~~3.3  & ~~21  &  --- & ~~0.7  & D/B  \\     
   IC 1257    &  -6.15  &-1.725  &   ---      &    --- & ~1.00  & ~17.9  &  ---  &  --- & ~~6.5  & OH  \\      
   Terzan 2   &  -5.27  & -0.29  &   ---      &    --- & -1.00  & ~~0.9  & -308  &  --- & ~~0.3  & D/B \\      
   NGC 6366   &  -5.77  & -0.59  &   ---      &   0.91 & -0.97  & ~~5.0  & ~144  &  --- & ~~1.0  & IH	\\     
   Terzan 4   &  -6.09  & -1.62  &  ~0.50(17) &    --- & ~1.00  & ~~1.3  & ~-89  &  --- & ~~0.2  & D/B\\       
   HP 1	      &  -6.44  & -1.57  &   ~0.34(18)&    --- & ~0.75  & ~~6.1  & ~-67  &  --- & ~~0.5  & D/B  \\     
   NGC 6362   &  -6.94  & -1.07  &   ---      &   0.94 & -0.58  & ~~5.1  & ~-81  &  --- & ~-2.3  & D/B \\      
   Liller 1   &   ---	&  0.40  &   ---      &    --- & -7.63  & ~~1.8  & -165  &  --- & ~~0.0  & D/B  \\     
   NGC 6380   &  -7.46  & -0.40  &   ---      &    --- &  ---	& ~~3.2  & ~-83  &  --- & ~-0.6  & D/B \\      
   Terzan 1   &  -4.90  & -1.29  &   ---      &    --- &   ---  & ~~2.5  & -139  &  --- & ~~0.1  & D/B\\       
   Ton 2      &  -6.14  &-0.525  &   ---      &    --- &   ---  & ~~1.4  & ~~~0  &  --- & ~-0.5  & D/B   \\    
   NGC 6388   &  -9.42  & -0.45  &   ~0.22(1) &   0.87 &  -0.65 & ~~3.2  & -204  &  --- & ~-1.2  & D/B  \\     
   NGC 6402   &  -9.12  & -1.39  &   ---      &    --- & ~0.65  & ~~4.1  & ~175  &  --- & ~~2.4  & IH	\\     
   NGC 6401   &  -7.90  & -1.01  &   ---      &    --- &  ---	& ~~2.7  & ~~90  &  --- & ~~0.7  & D/B  \\     
   NGC 6397   &  -6.63  & -1.99  &  ~0.36(1)  &   0.99 & ~0.98  & ~~6.0  & ~-46  &  ~87 & ~-0.5  & D/B \\      
   Pal 6      &  -6.81  & -1.06  &  ~0.40(19) &    --- & -1.00  & ~~2.2  & -177  &  --- & ~~0.2  & D/B\\       
   NGC 6426   &  -6.69  & -2.36  &   ---      &    --- & ~0.58  & ~14.6  & ~103  &  --- & ~~5.8  & IH	\\     
   Djorg 1    &  -6.26  &-2.025  &   ---      &    --- & ---	& ~~4.1  & ~335  &  --- & ~-0.5  & IH\\        
   Terzan 5   &  -7.87  &  0.16  &  ~0.32(17) &    --- & -1.00  & ~~2.4  & ~138  &  --- & ~~0.3  & D/B  \\     
   NGC 6440   &  -8.75  & -0.20  &   ---      &    --- & -1.00  & ~~1.3  & ~256  &  --- & ~~0.6  & D/B\\       
   NGC 6441   &  -9.64  & -0.44  &   ~0.21(1) &   0.83 & -0.76  & ~~3.9  &~~-64  &  --- & ~-1.0  & D/B \\      
   Terzan 6   &  -7.67  & -0.40  &  ---       &    --- & -1.00  & ~~1.6  & -168  &  --- & ~-0.4  & D/B    \\   
   NGC 6453   &  -6.88  & -1.48  &  ---       &    --- &  ---	& ~~1.8  & ~-13  &  --- & ~-0.6  & D/B \\      
   UKS-1      &  -6.88  & -0.40  &  ~0.33(15) &    --- & -1.00  & ~~0.8  & ---   &  --- & ~~0.1  & D/B    \\   
   NGC 6496   &  -7.23  & -0.46  &  ---       &   0.87 & -1.00  & ~~4.3  & ~~43  &  --- & ~-2.0  & D/B \\      
   Terzan 9   &  -3.85  & -2.07  &  ---       &    --- &  ---	& ~~1.6  & ~-29  &  --- & ~-0.2  & D/B \\      
   Djorg 2    &  -6.98  &-0.525  &  ---       &    --- & -1.00  & ~~1.4  & ---   &  --- & ~-0.3  & D/B  \\     
   NGC 6517   &  -8.28  & -1.24  &  ---       &    --- &  ---	& ~~4.3  & ~110  &  --- & ~~1.3  & D/B \\      
   Terzan 10  &  -6.31  &-0.725  &  ---       &    --- & -1.00  & ~~2.4  &  ---  &  --- & ~-0.2  & D/B    \\   
   NGC 6522   &  -7.67  & -1.45  &  ---       &    --- & ~0.71  & ~~0.6  & ~~38  &  --- & ~-0.5  & D/B  \\     
   NGC 6535   &  -4.75  & -1.79  &  ---       &   0.86 & ~1.00  & ~~3.9  & ~310  &  --- & ~~1.2  & IH	\\     
   NGC 6528   &  -6.56  &  0.07  &  ~0.24(20) &    --- & -1.00  & ~~0.6  & -146  &  --- & ~-0.6  & D/B \\      
   NGC 6539   &  -8.30  & -0.53  &  ~0.43(15) &    --- & -1.00  & ~~3.1  &~ 170  &  --- & ~~1.0  & D/B   \\    
   NGC 6540   &  -5.38  &-1.225  &  ---       &    --- &  ---	& ~~4.4  &   16  &  --- & ~-0.2  & D/B   \\    
   NGC 6544   &  -6.66  & -1.47  &  ---       &   0.86 & ~1.00  & ~~5.3  & ~~27  &  --- & ~-0.1  & D/B  \\     
   NGC 6541   &  -8.37  & -1.82  &  ~0.43(13) &   1.05 & ~1.00  & ~~2.2  & ~~28  &  --- & ~-1.4  & D/B  \\     
   NGC 6553   &  -7.77  & -0.16  &  ~0.26(21) &    --- & -1.00  & ~~2.2  & ~~36  &  --- & ~-0.3  & D/B \\      
   NGC 6558   &  -6.46  & -1.37  &  ~0.37(22) &    --- & ~0.70  & ~~1.0  & ~187  &  --- & ~-0.8  & D/B \\      
   IC 1276    &  -6.67  & -0.65  &  ---       &    --- &  ---	& ~~3.7  & ~-80  &  --- & ~~0.5  & D/B  \\     
   Terzan 12  &  -4.14  &-0.525  &   ---      &    --- &  ---	& ~~3.4  & ~-67  &  --- & ~-0.2  & D/B\\       
   NGC 6569   &  -8.30  & -0.72  &  ---       &    --- &  ---	& ~~2.9  & ~~25  &  --- & ~-1.2  & D/B   \\    
   NGC 6584   &  -7.68  & -1.50  &  ---       &   0.89 & -0.15  & ~~7.0  & -255  &  181 & ~-3.8  & IH	\\       
   NGC 6624   &  -7.49  & -0.42  &   ---      &   0.88 & -1.00  & ~~1.2  & ~~44  &  --- & ~-1.1  & D/B\\       
   NGC 6626   &  -8.18  & -1.46  &  ---       &    --- & ~0.90  & ~~2.7  & ~~27  &  ~52 & ~-0.5  & D/B\\       
   NGC 6638   &  -7.13  & -0.99  &   ---      &    --- & -0.30  & ~~2.3  & ~103  &  --- & ~-1.2  & D/B\\       
   NGC 6637   &  -7.64  & -0.59  &  ~0.31(23) &   0.91 & -1.00  & ~~1.9  & ~~72  &  --- & ~-1.6  & D/B    \\   
   NGC 6642   &  -6.77  & -1.19  &  ---       &    --- & ---	& ~~1.7  & ~227  &  --- & ~-0.9  & D/B \\      
   NGC 6652   &  -6.68  & -0.76  &  ---       &   0.91 & -1.00  & ~~2.8  & ~135  &  --- & ~-2.0  & D/B \\      
   NGC 6656   &  -8.50  & -1.70  &  ~0.38(24) &   1.06 & ~0.91  & ~~4.9  & ~156  &  ~42 & ~-0.4  & D/B\\       
   Pal 8      &  -5.52  & -0.37  &  ---       &    --- & -1.00  & ~~5.6  & ~~67  &  --- & ~-1.5  & D/B  \\     
   NGC 6681   &  -7.11  & -1.62  &  ---       &   0.97 & ~0.96  & ~~2.1  & ~-44  &  --- & ~-1.9  & D/B   \\    
   NGC 6712   &  -7.50  & -1.02  &  ---       &    --- & -0.62  & ~~3.5  & ~214  &  183 & ~-0.5  & IH  \\      
   NGC 6715   & -10.01  & -1.44  &  ~0.16(25) &   0.87 & ~0.75  & ~19.2  & ---   &  --- & ~-6.5  & dSph\\      
   NGC 6717   &  -5.66  & -1.26  &  ---       &   1.00 & ~0.98  & ~~2.4  & ~100  &  --- & ~-1.3  & D/B\\       
   NGC 6723   &  -7.84  & -1.10  &  ~0.50(26) &   1.01 & -0.08  & ~~2.6  & ~~83  &  --- & ~-2.6  & D/B   \\    
   NGC 6749   &  -6.70  & -1.62  &  ---       &    --- & ~1.00  & ~~5.0  & ~130  &  --- & ~-0.3  & IH	\\     
   NGC 6752   &  -7.73  & -1.55  &  ~0.43(1)  &   1.02 & ~1.00  & ~~5.2  & ~-24  &  ~21 & ~-1.7  & D/B   \\    
   NGC 6760   &  -7.86  & -0.40  &  ---       &    --- & -1.00  & ~~4.8  & ~100  &  --- & ~-0.5  & D/B \\      
   NGC 6779   &  -7.38  & -2.00  &  ---       &   1.10 & ~0.98  & ~~9.7  & ~~91  &  259 & ~~1.5  & IH	 \\    
   Terzan 7   &  -5.05  & -0.12  & -0.03(27)  &   0.53 & -1.00  & ~16.0  & ---   &  --- & ~-8.0  & dSph  \\    
   Pal 10     &  -5.79  &-0.125  &  ---       &    --- & -1.00  & ~~6.4  & ~~58  &  --- & ~~0.3  & D/B   \\    
   Arp 2      &  -5.29  & -1.74  &  ~0.34(28) &   0.89 & ~0.86  & ~21.4  & ---   &  --- & -10.2  & dSph  \\    
   NGC 6809   &  -7.55  & -1.93  &  ~0.42(1)  &   1.02 & ~0.87  & ~~3.9  &~-129  &  165 & ~-2.1  & D/B  \\     
   Terzan 8   &  -5.05  &-2.025  &  ~0.45(28) &   0.97 & ~1.00  & ~19.1  & ---   &  --- & -10.8  & dSph \\     
   Pal 11     &  -6.86  & -0.45  &  ---       &    --- & ---	& ~~7.9  & ~~77  &  --- & ~-3.5  & IH	 \\    
   NGC 6838   &  -5.60  & -0.82  &  ~0.40(1)  &   0.94 & -1.00  & ~~6.7  &~~ 40  &  ~40 & ~-0.3  & D/B   \\    
   NGC 6864   &  -8.55  & -1.29  &   ---      &    --- & -0.07  & ~14.6  & ~144  &  --- & ~-9.0  & IH	\\     
   NGC 6934   &  -7.46  & -1.56  &  ---       &   0.88 & ~0.25  & ~12.8  & ~328  &  279 & ~-5.1  & IH	\\     
   NGC 6981   &  -7.04  & -1.48  &  ---       &   0.87 & ~0.14  & ~12.9  & ~277  &  --- & ~-9.2  & IH	 \\    
   NGC 7006   &  -7.68  & -1.46  &  ~0.28(29) &    --- & -0.28  & ~38.8  & ---   &  --- & -13.8  & OH	\\     
   NGC 7078   &  -9.17  & -2.33  &  ~0.40(1)  &   1.01 & ~0.67  & ~10.4  & ~~63  &  ~92 & ~-4.7  & IH  \\      
   NGC 7089   &  -9.02  & -1.66  &  ~0.41(30) &   0.96 & ~0.96  & ~10.4  & ~~~5  &  306 & ~-6.7  & IH  \\      
   NGC 7099   &  -7.43  & -2.33  &  ~0.37(1)  &   1.08 & ~0.89  & ~~7.1  & ~214  &  324 & ~-5.9  & IH \\       
   Pal 12     &  -4.48  & -0.81  & -0.01(7)   &   0.64 & -1.00  & ~15.9  & ---   &  --- & -14.1  & dSph  \\    
   Pal 13     &  -3.74  & -1.78  &  ---       &    --- & -0.20  & ~26.7  &  ---  &  --- & -17.5  & OH	\\     
   NGC 7492   &  -5.77  & -1.69  &  ~0.34(31) &   0.93 & ~0.81  & ~24.9  & ---   &  --- & -23.1  & OH  \\      
\hline	
\label{t:class}						  
\end{longtable} 					  
\end{scriptsize}	
\begin{small}				  
\noindent Notes: (1) Carretta et al. (2009b); (2) Shetrone \& Keane (2000); (3) Yong \& Grundahl (2008); (4) McWilliam, Geisler \& Rich (1992); 
(5) Shetrone et al. (2001); (6) Ivans (2009); (7) Brown, Wallerstein \& Zucker (1997); (8) Sneden et al. (2004); (9) McCarthy \& Nemec (1997); 
(10) Lee et al. (2006); (11) Smith, Sneden \& Kraft (2002); (12) Cavallo et al. (2004); (13) Lee \& Carney (2002); (14) Sneden et al. (2000); (15) Origlia et al. (2005); (16) Feltzing et al. (2009); 
(17) Origlia \& Rich (2004); (18) Barbuy et al. (2006, corrrected upward by 0.20 dex); (19) Lee et al. (2004); (20) Carretta et al. (2001); (21) Cohen et al. (1999); 
(22) Barbuy et al. (2007, corrrected upward by 0.20 dex); (23) Lee (2007); (24) Marino et al. (2009); (25) Brown et al. (1999); (26) Fulton \& Carney (1996); (27) Sbordone et al. (2007); 
(28) Mottini et al. (2008); (29) Kraft et al. (1998); (30) Russell (1996); (31) Cohen \& Melendez (2005).
\end{small}					  
\begin{table*}
\centering
\caption{Comparison of our classification with the one by Mackey and van den
Bergh 2005}\label{t:matrix}
\begin{tabular}{c|c|c|c|c}
\hline                                    
            &Disk/Bulge & Old Halo & Young Halo & SGR \\
\hline                                    
Disk/Bulge  &	34      &  36	   &	5  	& --- \\
\hline	 
Inner Halo  &   1       &   23     &	9  	& --- \\
\hline
Outer Halo  &	1       &  11	   &	15 	& --- \\
\hline
 dSph	    &	---     &  ---     &	 1 	& 6   \\
 \hline  
 \end{tabular}
\end{table*}

\begin{table}
\centering
\caption{Globular clusters of LMC, SMC, Fornax dSph and their properties}\label{t:class2}
\begin{tabular}{cccccc}
\hline
\hline
Name & $M_V$ & [Fe/H]       &     [$\alpha$/Fe] & Age & HBR\\
\hline    
     &     	    &	     &  	   &	        &  \\
NGC 1466    &	 -7.26  &    -1.76(1)	   &   ---      &  1.04(5)     & ~0.41(4,6) \\
NGC 1754    &	 -7.09  &    -1.50(1,2)    &   ---	&  0.96(2,3)   & ~0.46(3,6) \\
NGC 1786    &	 -7.70  &    -1.76(1,2,9)  &   0.43(9)  &  1.12(2,3)   & ~0.39(6)   \\
NGC 1835    &	 -8.30  &    -1.72(1,2)    &   ---	&  0.99(2,3)   & ~0.52(3,6) \\
NGC 1841    &	 -6.82  &    -2.02(1)	   &   ---      &  0.89(8)     & ~0.72(4,6) \\
NGC 1898    &	 -7.49  &    -1.32(1,2,10) &   0.06(10) &  0.98(2,3)   & -0.02(3,6) \\ 
NGC 1916    &	 -8.24  &    -2.05(1,2)    &   ---      &  0.93(2,3)   & ~0.97(6)   \\
NGC 1928    &	 -6.06  &    -1.27(6)	   &   ---      &  0.87(13)    & ~0.94(6)   \\
NGC 1939    &	 -6.85  &    -2.02(2,6)    &   ---      &  0.94(2,3,13)& ~0.94(6)   \\
NGC 2005    &	 -7.40  &    -1.74(1,2,10) &   0.17(10) &  1.02(2,3)   & ~0.88(6)   \\
NGC 2019    &	 -7.75  &    -1.56(1,2,10) &   0.28(10) &  1.20(2,3)   & ~0.61(3,6) \\
NGC 2210    &	 -7.51  &    -1.76(1,12,9) &   0.37(9)  &  1.05(12)    & ~0.61(4,6) \\
NGC 2257    &	 -7.25  &    -1.83(1,9)    &    0.34(9) &  1.00(5)     & ~0.46(4,6) \\
Hodge 11    &	 -7.45  &    -2.09(1,10)   &   0.38(10) &  1.14(5)     & ~0.98(4,6) \\
Reticulum   &	 -5.22  &    -1.64(1,6)    &    ---     &  0.81(13)    & -0.02(4,6) \\
ESO121-SC03 &    -4.37  &    -0.93(12,7)   &   ---      &  0.71(14)    & -1.00(6)   \\
\hline
NGC 121     &	 -7.82  &    -1.46(15)     &   0.24(18) &  0.87(15)    & -0.95(6)   \\
Lindsay 1   &	 -5.46  &    -1.14(16)     &    ---     &  0.58(16)    & -1.00(6)   \\
Kron 3      &    -6.86  &    -1.08(16)     &    ---     &  0.51(16)    & -1.00(6)   \\
NGC 339     &	 -5.96  &    -1.12(16)     &	---	&  0.47(16)    & -1.00(6)   \\ 
NGC 361     &	 -6.12  &    -1.45(17)     &	---	&  0.63(17)    & -1.00(6)   \\
NGC 416     &	 -7.48  &    -1.00(16)     &	---	&  0.47(16)    & -1.00(6)   \\
Lindsay 38  &	  ---   &    -1.59(16)     &	---	&  0.51(16)    & -1.00(6)   \\
Lindsay 113 &	 -5.29  &    -1.38(17)     &	---	&  0.31(17)    & ---        \\
\hline
Fornax 1    &	  -5.32 &   -2.50(19)	   &   0.25(19) &  1.04        & -0.30(6)   \\
Fornax 2    &	  -7.03 &   -2.10(19)	   &   0.31(19) &  1.01        & ~0.50(6)   \\
Fornax 3    &	  -7.66 &   -2.40(19)	   &   0.15(19) &  1.09        & ~0.44(6)   \\
Fornax 4    &	  -6.83 &   -1.90(20)	   &   ---	&  0.83        & -0.42(6)   \\
Fornax 5    &	  -6.82 &   -2.20(20)	   &   ---	&  1.08        & ~0.52(6)   \\
\hline 
\end{tabular} 
\begin{footnotesize}       
\begin{list}{}{}
\item Notes: (1) Suntzeff et al. 1992; (2) Beasley et al. 2002; (3) Olsen et al. 1998;(4) Walker 1992b; (5) Johnson et al. 1999; 
(6) Mackey \& Gilmore 2004a; (7) Geisler et al. 1997; (8) Saviane et al. 2002; (9) Mucciarelli et al. 2009; 
(10) Johnson et al. 2006; (11) van den Bergh 2006; (12) Hill et al. 2000; (13) Mackey \& Gilmore 2004b; 
(14) Mackey et al. 2006; (15) Glatt et al. 2008a; (16) Glatt et al. 2008b; (17) Kayser et al. 2006; 
(18) Johnson et al. 2004; (19)  Letarte et al. (2006); (20) Buonanno et al. (1999).                             
\end{list}
\end{footnotesize}
\end{table}

\end{appendix}

\newpage

\Online 

\begin{landscape}
\setcounter{table}{4}
\begin{table}
\centering
\caption{Correlation coefficients for 21 parameters of the 19 GCs.}
\begin{scriptsize}
\setlength{\tabcolsep}{1.5mm}
\begin{tabular}{lccccccccccccccccccccc}
\hline
       &$M_V$   &   HBR & [Fe/H] &  $r_h$     & $\log T_{eff}^{max}$ & relage  
       & IQR   &   \%P &   \%I   &    \%E     & $<$SiCa$>$           & [$\alpha$/Fe] 
       & O$_{min}$  & O$_{max}$  & Mg$_{min}$ & Mg$_{max}$
       & Na$_{min}$ & Na$_{max}$ & Si$_{min}$ & Si$_{max}$           &   Z   \\    
   &1     &2     &3     &4     &5     &6     &7     &8     &9     &10     &11     &12    &13    &14    &15    &16    &17    &18    &19    &20    &21  \\
\hline
~~1&   -- & 0.304&-0.293& 0.113&-0.433& 0.387&-0.463&-0.393& 0.551&-0.614& 0.533& 0.495& 0.495& 0.375& 0.564& 0.336&-0.350& 0.074& 0.349& 0.178& 0.071\\    
~~2& 0.304&   -- &-0.731& 0.131& 0.454& 0.407& 0.198&-0.204& 0.329&-0.420& 0.128& 0.178& 0.121& 0.366& 0.096& 0.224&-0.657&-0.181&-0.244&-0.254&-0.229\\
~~3&-0.293&-0.731&   -- &-0.187&-0.195&-0.549& 0.182& 0.003&-0.277& 0.585&-0.142&-0.252&-0.387&-0.658& 0.045&-0.310& 0.661& 0.271& 0.026&-0.034& 0.070\\
~~4& 0.113& 0.131&-0.187&   -- &-0.157& 0.005& 0.138&-0.214& 0.195&-0.131& 0.439& 0.376&-0.078& 0.197& 0.193& 0.127&-0.146&-0.219& 0.371& 0.078& 0.050\\
~~5&-0.433& 0.454&-0.195&-0.157&   -- &-0.199& 0.693& 0.453&-0.468& 0.361&-0.347&-0.333&-0.471&-0.056&-0.681&-0.272&-0.207&-0.389&-0.530&-0.334&-0.232\\
~~6& 0.387& 0.407&-0.549& 0.005&-0.199&   -- &-0.332&-0.396& 0.593&-0.710& 0.464& 0.592& 0.707& 0.603& 0.425& 0.752&-0.075& 0.306& 0.356& 0.505&-0.167\\
~~7&-0.463& 0.198& 0.182& 0.138& 0.693&-0.332&   -- & 0.248&-0.486& 0.654&-0.188&-0.297&-0.788&-0.228&-0.475&-0.423&-0.030&-0.262&-0.438&-0.465&-0.207\\
~~8&-0.393&-0.204& 0.003&-0.214& 0.453&-0.396& 0.248&	-- &-0.914& 0.561&-0.484&-0.517&-0.372&-0.118&-0.615&-0.551& 0.130&-0.263&-0.321&-0.206&-0.185\\
~~9& 0.551& 0.329&-0.277& 0.195&-0.468& 0.593&-0.486&-0.914&   -- &-0.848& 0.573& 0.661& 0.621& 0.398& 0.671& 0.740&-0.204& 0.211& 0.404& 0.320& 0.187\\
 10&-0.614&-0.420& 0.585&-0.131& 0.361&-0.710& 0.654& 0.561&-0.848&   -- &-0.542&-0.680&-0.779&-0.667&-0.559&-0.797& 0.265&-0.097&-0.397&-0.378&-0.121\\
 11& 0.533& 0.128&-0.142& 0.439&-0.347& 0.464&-0.188&-0.484& 0.573&-0.542&   -- & 0.957& 0.218& 0.368& 0.503& 0.630& 0.012&-0.032& 0.801& 0.473& 0.153\\
 12& 0.495& 0.178&-0.252& 0.376&-0.333& 0.592&-0.297&-0.517& 0.661&-0.680& 0.957&   -- & 0.345& 0.419& 0.485& 0.805& 0.054& 0.053& 0.784& 0.547& 0.060\\
 13& 0.495& 0.121&-0.387&-0.078&-0.471& 0.707&-0.788&-0.372& 0.621&-0.779& 0.218& 0.345&   -- & 0.486& 0.462& 0.598&-0.074& 0.372& 0.336& 0.561& 0.074\\
 14& 0.375& 0.366&-0.658& 0.197&-0.056& 0.603&-0.228&-0.118& 0.398&-0.667& 0.368& 0.419& 0.486&   -- & 0.219& 0.477&-0.376&-0.352& 0.089& 0.157& 0.292\\
 15& 0.564& 0.096& 0.045& 0.193&-0.681& 0.425&-0.475&-0.615& 0.671&-0.559& 0.503& 0.485& 0.462& 0.219&   -- & 0.486& 0.036& 0.350& 0.386& 0.224& 0.156\\
 16& 0.336& 0.224&-0.310& 0.127&-0.272& 0.752&-0.423&-0.551& 0.740&-0.797& 0.630& 0.805& 0.598& 0.477& 0.486&	-- & 0.108& 0.277& 0.476& 0.599&-0.007\\
 17&-0.350&-0.657& 0.661&-0.146&-0.207&-0.075&-0.030& 0.130&-0.204& 0.265& 0.012& 0.054&-0.074&-0.379& 0.036& 0.108&   -- & 0.539& 0.310& 0.308&-0.349\\
 18& 0.074&-0.181& 0.271&-0.219&-0.389& 0.306&-0.262&-0.263& 0.211&-0.097&-0.032& 0.053& 0.372&-0.352& 0.350& 0.277& 0.539&   -- & 0.176& 0.330&-0.536\\
 19& 0.349&-0.244& 0.026& 0.371&-0.530& 0.356&-0.438&-0.321& 0.404&-0.397& 0.801& 0.784& 0.336& 0.089& 0.386& 0.476& 0.310& 0.176&   -- & 0.721& 0.061\\
 20& 0.178&-0.254&-0.034& 0.078&-0.334& 0.505&-0.465&-0.206& 0.320&-0.378& 0.473& 0.547& 0.561& 0.157& 0.224& 0.599& 0.308& 0.330& 0.721&   -- & 0.120\\
 21& 0.071&-0.229& 0.070& 0.050&-0.232&-0.167&-0.207&-0.185& 0.187&-0.121& 0.153& 0.060& 0.074& 0.292& 0.156&-0.007&-0.349&-0.536& 0.061& 0.120&  --  \\
\hline													      	      	      	      
\end{tabular}													      	      	      
\end{scriptsize}													      	      
\label{t:pca2}															      
\end{table}
\end{landscape}

\begin{table*}
\setcounter{table}{6}
\centering
\caption{Correlations, with number of clusters, degrees of liberty, Pearson 
coefficients (without sign), and significance in percent.}
\begin{tabular}{lcccc}
\hline
 &$M_V$ & [Fe/H] &age &E$_{tot}$ \\
\hline
\multicolumn{5}{c}{Structural parameters} \\
HBR          &19 17 0.30 $<$90  &19 17 0.73 $>99$  &19 17 0.41 90-95 &17 15 0.01 $<$90\\
c            &19 17 0.35 $<$90  &19 17 0.29 $<$90  &19 17 0.31 $<$90 &17 15 0.03 $<$90\\
ellipticity  &18 16 0.36 $<$90  &18 16 0.05 $<$90  &18 16 0.44 90-95 &16 14 0.61 98-99\\
$r_h$        &19 17 0.11 $<$90  &19 17 0.19 $<$90  &19 17 0.00 $<$90 &17 15 0.33 $<$90\\
$r_t$        &19 17 0.33 $<$90  &19 17 0.41 90-95  &19 17 0.04 $<$90 &17 15 0.59 98-99\\
logTeff      &19 17 0.43 90-95  &19 17 0.20 $<$90  &19 17 0.20 $<$90 &17 15 0.18 $<$90\\
\multicolumn{5}{c}{Orbital and positional parameters} \\
R$_{GC}$     &19 17 0.00 $<$90  &19 17 0.30 $<$90  &19 17 0.24 $<$90 &17 15 0.59 98-99\\
$|Z|$        &19 17 0.02 $<$90  &19 17 0.40 90-95  &19 17 0.07 $<$90 &17 15 0.48 95.0\\
E$_{tot}$    &17 15 0.34 $<$90  &17 15 0.20 $<$90  &17 15 0.64 $>99$ &	       \\
age          &19 17 0.43 90-95  &19 17 0.37 $<$90  & 	             &17 15 0.78 $>99$\\
$M)V$        &		        &19 17 0.29 $<$90  &19 17 0.39 90.0  &17 15 0.34 $<$90\\
${\rm [Fe/H]}$       &19 17 0.29 $<$90  &		   &19 17 0.55 98-99 &17 15 0.20 $<$90\\
\multicolumn{5}{c}{Primordial  abundances} \\
Mg+Al+Si     &18 16 0.42 90-95  &18 16 0.18 $<$90  &18 16 0.77 $>99$ &16 14 0.68 $>99$\\
$\alpha$     &19 17 0.49 95-98  &19 17 0.25 $<$90  &19 17 0.59 $>99$ &17 15 0.64 $>99$\\
\multicolumn{5}{c}{First generation abundances} \\
Mg$_{max}$   &19 17 0.34 $<$90  &19 17 0.31 $<$90  &19 17 0.75 $>99$ &17 15 0.61 99.0\\
Si$_{min}$   &19 17 0.35 $<$90  &19 17 0.03 $<$90  &19 17 0.36 $<$90 &17 15 0.37 $<$90\\
Al$_{min}$   &18 16 0.19 $<$90  &18 16 0.23 $<$90  &18 16 0.23 $<$90 &16 14 0.24 $<$90\\
Na$_{min}$   &19 17 0.35 $<$90  &19 17 0.66 $>99$  &19 17 0.08 $<$90 &17 15 0.26 $<$90\\
O$_{max }$   &19 17 0.37 $<$90  &19 17 0.66 $>99$  &19 17 0.60 $>99$ &17 15 0.00 $<$90\\
P            &19 17 0.39 90.0   &19 17 0.00 $<$90  &19 17 0.40 90-95 &17 15 0.33 $<$90\\
\multicolumn{5}{c}{Second generation abundances} \\
Mg$_{min}$   &19 17 0.56 98-99  &19 17 0.04 $<$90  &19 17 0.42 90-95 &17 15 0.35 $<$90\\
Si$_{max}$   &19 17 0.18 $<$90  &19 17 0.03 $<$90  &19 17 0.51 95-98 &17 15 0.46 90-95\\
Al$_{max}$   &18 16 0.26 $<$90  &18 16 0.40 90.0   &18 16 0.15 $<$90 &16 14 0.13 $<$90\\
O$_{min}$    &19 17 0.50 95-98  &19 17 0.39 90.0   &19 17 0.71 $>99$ &17 15 0.40 $<$90\\
Na$_{max}$   &19 17 0.07 $<$90  &19 17 0.27 $<$90  &19 17 0.31 $<$90 &17 15 0.46 90-95\\
I            &19 17 0.55 98-99  &19 17 0.28 $<$90  &19 17 0.59 $>99$ &17 15 0.37 $<$90\\
E            &19 17 0.61 $>99$  &19 17 0.58 99.0   &19 17 0.71 $>99$ &17 15 0.34 $<$90\\
\multicolumn{5}{c}{Link first-second generation} \\
slope[Na/Al] &18 16 0.46 95.0  &18 16 0.08 $<$90   &18 16 0.35 $<$90 &16 14 0.34 $<$90\\
IQR[O/Na]    &19 17 0.46 95-98  &19 17 0.18 $<$90  &19 17 0.33 $<$90 &17 15 0.18 $<$90\\
\hline
\end{tabular}
\label{t:c}
\end{table*}

\end{document}